\documentclass[11pt]{article}

\usepackage{mathrsfs}
\usepackage{amsmath,amssymb}
\usepackage{epsfig}
\usepackage{relsize}
\usepackage{graphicx}
\usepackage{bbm}
\usepackage[shortlabels]{enumitem}
\usepackage{xcolor}
\usepackage{bm}
\usepackage{float}
\usepackage{cite}
\usepackage{tcolorbox}

\definecolor{lightred}{RGB}{255,127,127}
\definecolor{lightgreen}{RGB}{127,255,127}
\definecolor{lightblue}{RGB}{127,127,255}
\definecolor{linkcolor}{rgb}{0,0,0.6}
\usepackage[ pdftex,colorlinks=true,
pdfstartview=FitV,
linkcolor= linkcolor,
citecolor= linkcolor,
urlcolor= linkcolor,
hyperindex=true,
hyperfigures=false]
{hyperref}

\numberwithin{equation}{section}

\newcommand{\noi}{\noindent}
\newcommand{\dd}{\text{d}}
\newcommand{\p}{\partial}
\newcommand{\g}{\mathfrak{g}}
\newcommand{\R}{\mathbb{R}}
\newcommand{\D}{\mathbb{D}}
\newcommand{\C}{\mathbb{C}}
\newcommand{\Z}{\mathbb{Z}}
\newcommand{\CP}{\mathbb{CP}^1}
\newcommand{\Lc}{\mathcal{L}}

\newcommand{\Tr}{\text{Tr}}
\newcommand{\CS}{\text{CS}}
\newcommand{\zb}{{\bar{z}}}
\newcommand{\vp}{\varphi}
\newcommand{\Zc}{\bm{\mathcal{Z}}}
\newcommand{\Ze}{\bm{\zeta}}
\newcommand{\ze}{\zeta}
\newcommand{\gh}{\widehat{g}}
\newcommand{\Dc}{\mathcal{D}}
\newcommand{\Id}{\text{Id}}
\newcommand{\Rc}{\mathcal{R}}
\newcommand{\tp}{\null^{t}\null}
\newcommand{\Ab}{\mathbb{A}}
\newcommand{\At}{\widetilde{\mathbb{A}}}
\newcommand{\kf}{\mathfrak{k}}
\newcommand{\ff}{\mathfrak{f}}
\newcommand{\diag}{\text{diag}}
\newcommand{\Ad}{\text{Ad}}

\DeclareFontEncoding{LS1}{}{}
\DeclareFontSubstitution{LS1}{stix}{m}{n}
\DeclareSymbolFont{stixsymbols}{LS1}{stixscr}{m}{n}
\SetSymbolFont{stixsymbols}{bold}{LS1}{stixscr}{b}{n}
\DeclareMathSymbol{\kay}{\mathalpha}{stixsymbols}{"6B}
\DeclareMathSymbol{\hay}{\mathalpha}{stixsymbols}{"68}

\DeclareMathAlphabet{\mathdsl}{U}{bbm}{m}{sl}
\newcommand{\ub}{\mathdsl{u}}
\newcommand{\ut}{\widetilde{\mathdsl{u}}}
\newcommand{\gb}{\mathdsl{g}}
\newcommand{\gt}{\widetilde{\mathdsl{g}}}

\newcommand{\df}{\mathfrak{d}}
\newcommand{\Lb}{\mathbb{L}}
\newcommand{\Lt}{\widetilde{\mathbb{L}}}

\def\res{\mathop{\text{res}\,}}

\makeatletter
\DeclareFontFamily{OMX}{MnSymbolE}{}
\DeclareSymbolFont{MnLargeSymbols}{OMX}{MnSymbolE}{m}{n}
\SetSymbolFont{MnLargeSymbols}{bold}{OMX}{MnSymbolE}{b}{n}
\DeclareFontShape{OMX}{MnSymbolE}{m}{n}{
    <-6>  MnSymbolE5
   <6-7>  MnSymbolE6
   <7-8>  MnSymbolE7
   <8-9>  MnSymbolE8
   <9-10> MnSymbolE9
  <10-12> MnSymbolE10
  <12->   MnSymbolE12
}{}
\DeclareFontShape{OMX}{MnSymbolE}{b}{n}{
    <-6>  MnSymbolE-Bold5
   <6-7>  MnSymbolE-Bold6
   <7-8>  MnSymbolE-Bold7
   <8-9>  MnSymbolE-Bold8
   <9-10> MnSymbolE-Bold9
  <10-12> MnSymbolE-Bold10
  <12->   MnSymbolE-Bold12
}{}

\let\llangle\@undefined
\let\rrangle\@undefined
\DeclareMathDelimiter{\llangle}{\mathopen}%
                     {MnLargeSymbols}{'164}{MnLargeSymbols}{'164}
\DeclareMathDelimiter{\rrangle}{\mathclose}%
                     {MnLargeSymbols}{'171}{MnLargeSymbols}{'171}
\makeatother

\begin{document}

\thispagestyle{empty}

\begin{center}
\huge \textbf{4-dimensional Chern-Simons theory \\
and integrable field theories} \\~\\
{\Large Sylvain Lacroix~\Large} \vspace{12pt}\\
\normalsize II. Institut fur Theoretische Physik, Universitat Hamburg, \textit{Luruper Chaussee 149, 22761 Hamburg}\vspace{2pt}\\
\normalsize Zentrum fur Mathematische Physik, Universitat Hamburg, \textit{Bundesstrasse 55, 20146 Hamburg} \vspace{12pt}\\
\large \begingroup\ttfamily\href{mailto:sylvain.lacroix@desy.de}{\color{black}sylvain.lacroix@desy.de}\endgroup \vspace{25pt}
\end{center}

\noi \textsc{Abstract:} These lecture notes concern the semi-holomorphic 4d Chern-Simons theory and its applications to classical integrable field theories in 2d and in particular integrable sigma-models. After introducing the main properties of the Chern-Simons theory in 3d, we will define its 4d analogue and explain how it is naturally related to the Lax formalism of integrable 2d theories. Moreover, we will explain how varying the boundary conditions imposed on this 4d theory allows to recover various occurences of integrable sigma-models through this construction, in particular illustrating this on two simple examples: the Principal Chiral Model and its Yang-Baxter deformation.\vspace{7pt}

\noi These notes were written for the lectures delivered at the school ``Integrability, Dualities and Deformations'', that ran from 23 to 27 August 2021 in Santiago de Compostela and virtually.\vspace{15pt}

\newpage

\renewcommand{\baselinestretch}{0.98}\normalsize
\tableofcontents
\renewcommand{\baselinestretch}{1.0}\normalsize

\newpage

\section{Introduction}

\paragraph{Motivations.} Integrable systems form an important class of theories, for which the presence of a high number of symmetries allows the development of exact methods in the computation of physical observables. As described in Ana L. Retore's lectures \textit{``Introduction to classical and quantum integrability''}~\cite{LectureRetore}, integrable systems appear in various domains of physics, from mechanical systems to spin chains and 2-dimensional field theories. Important examples belonging to the latter class are integrable $\sigma$-models, with the prototypical example being the Principal Chiral Model. The panorama of integrable $\sigma$-models have been shown to be quite rich, for instance through the discovery of continuous deformations of some of these models that preserve their integrability. We refer to Ben Hoare's lectures \textit{``Integrable deformations of sigma models''}~\cite{LectureHoare} for more details and references. Moreover, integrable $\sigma$-models (or more precisely string theories) have played an important role in the context of the AdS/CFT correspondence, for instance through the example of the Green-Schwarz superstring on AdS$_5\times$S$^5$, which is dual to the maximally supersymmetric Yang-Mills in 4 dimensions (see for instance the review~\cite{Beisert:2010jr} for more details on the uses of integrability in both sides of the duality).

As explained in Ana L. Retore's lectures \textit{``Introduction to classical and quantum integrability''}~\cite{LectureRetore}, for 2-dimensional classical field theories, integrability is generally proven by exhibiting a so-called Lax connection, which ensures the existence of an infinite number of conserved charges. Without an underlying guiding principle, finding this Lax connection is often a difficult task, which makes the question of determining whether a given 2-dimensional field theory is integrable or not a challenging problem. Faced with this difficulty, it is natural to search for systematic ways of constructing integrable field theories, in which the existence of a Lax connection is automatically ensured. In addition of providing a guiding method for exploring the panorama of integrable field theories, such a formalism would offer a general insight into the structure of these systems and their properties. The goal of these lectures is to review one of these approaches, proposed by K. Costello and M. Yamazaki in~\cite{Costello:2019tri} just 2 years ago and still actively studied currently, which is based on a 4-dimensional gauge theory called the semi-holomorphic 4-dimensional Chern-Simons theory. In particular, we will explain how this approach allows to recover many integrable $\sigma$-models, including the Principal Chiral Model and its deformations\footnote{For completeness, let us note that another unifying formalism for integrable 2-dimensional field theories (and in particular integrable $\sigma$-models) has been developed in parallel, based on the so-called affine Gaudin models~\cite{Feigin:2007mr,Vicedo:2017cge}. This complementary approach, which is rooted in algebraic structures and in particular the representation theory of affine Kac-Moody algebras, is in fact deeply related to the 4-dimensional Chern-Simons theory, as shown in~\cite{Vicedo:2019dej} by B. Vicedo.}.

The 4-dimensional semi-holomorphic Chern-Simons theory was initially defined by N. Nekrasov in~\cite{NekrasovThesis} and was rediscovered independently by K. Costello in~\cite{Costello:2013zra}. It was proposed in~\cite{NekrasovThesis} that this 4-dimensional theory can provide a geometric origin to quantum affine algebras, which are algebraic structures underlying the theory of integrable systems. In the reference~\cite{Costello:2013zra} and in subsequent works~\cite{Costello:2013sla,Witten:2016spx,Costello:2017dso,Costello:2018gyb} of K. Costello, M. Yamazaki and E. Witten, it was shown that integrable lattice models (spin chains) can be naturally obtained using this 4-dimensional theory (see also~\cite{Yagi:2015lha,Yagi:2016oum,Ashwinkumar:2018tmm,Costello:2018txb,Saidi:2018mdg,Bittleston:2019gkq,Bittleston:2019mol,Ashwinkumar:2020gxt,Ashwinkumar:2020krt,Costello:2021zcl} for further developments)\footnote{For a more general discussion of the relations between gauge theories and integrability and their history, see for example the introduction of~\cite{Dedushenko:2021mds} and references therein.}. The extension of this approach to also generate integrable 2-dimensional field theories was put forward by K. Costello and M. Yamazaki in~\cite{Costello:2019tri} and has been the subject of many subsequent works~\cite{Delduc:2019whp,Bykov:2019vkf,Bassi:2019aaf,Schmidtt:2019otc,Gaiotto:2020fdr,Fukushima:2020kta,Bykov:2020llx,Fukushima:2020dcp,Costello:2020lpi,Tian:2020ryu,Bykov:2020nal,Tian:2020meg,Benini:2020skc,Hoare:2020mpv,Bykov:2020tao,Gaiotto:2020dhf,Bittleston:2020hfv,Penna:2020uky,Lacroix:2020flf,Caudrelier:2020xtn,Fukushima:2020tqv,Affleck:2021ypq,Chen:2021qto,Fukushima:2021eni,Derryberry:2021rne,Schmidtt:2021rhw,Stedman:2021wrw}. This extension is based on the introduction of 2-dimensional defects in the 4-dimensional theory, which can be either order defects or disorder defects. In these lectures, we will focus on disorder defects.

In the rest of this introduction, our goal will be to sketch how the 4-dimensional Chern-Simons theory is related to integrable systems and in particular integrable 2-dimensional field theories. To do so, we will adopt from now on a slightly more technical point of view, with the objective of presenting a more concrete overview of the ideas underlying this approach.

\paragraph{Integrable systems \textit{vs} 3-dimensional Chern-Simons theory.} Let us first recall some of the main features of integrable systems, focusing mostly on field theories (we refer to Ana L. Retore's lectures \textit{``Introduction to classical and quantum integrability''}~\cite{LectureRetore} for a detailed review of these concepts). We consider a classical field theory defined on a 2-dimensional space-time $\Sigma$ (with a time coordinate $t$ and a space coordinate $x$) and described by fields $\phi_i(t,x)$. As mentioned above, the integrability of such a theory is in general characterised by the existence of a Lax connection. The latter consists of a pair of matrices $\Lc_t(z,t,x)$ and $\Lc_x(z,t,x)$, built from the fields of the theory $\phi_i(t,x)$ and depending meromorphically on an auxiliary complex parameter $z$ that we call the spectral parameter, and that we will generally take here as belonging to the Riemann sphere $\CP=\C \cup \lbrace \infty \rbrace$ (\textit{i.e.} the complex plane plus the point at infinity). The crucial property of the Lax connection is that, in an integrable field theory, the equations of motion can be recast as the zero curvature equation for $\Lc_\mu$:
\begin{equation}\label{Eq:LaxIntro}
\p_t \Lc_x(z,t,x) - \p_x \Lc_t(z,t,x) + \bigl[ \Lc_t(z,t,x), \Lc_x(z,t,x) \bigr] = 0, \qquad \forall \, z\in\CP,
\end{equation}
which we also call the Lax equation. This equation implies that the theory possesses conserved charges (constructed from the so-called monodromy matrix of $\Lc_x$). The arbitrary dependence on the auxiliary spectral parameter $z\in\CP$ ensures that there are infinitely many such charges, which then characterise the integrable structure of the theory. In particular, the presence of a spectral parameter plays a crucial role in establishing the integrability of the theory.

Before discussing its semi-holomorphic 4-dimensional variant, let us briefly describe the standard 3-dimensional Chern-Simons theory, which belongs to the class of so-called topological field theories. It was developed by E. Witten in his seminal paper~\cite{Witten:1988hf}, building on the geometrical works of S-S. Chern and J. Simons. It is a 3-dimensional gauge theory, described by a gauge field $A_\mu$ ($\mu=1,2,3$) valued in a (potentially non-abelian) matrix Lie algebra. The action of this theory is given by
\begin{equation}
S[A] = \frac{k}{4\pi}  \iiint_M \dd x^1 \, \dd x^2 \, \dd x^2 \; \epsilon^{\mu\nu\rho} \, \Tr \left( A_\mu \p_\nu A_\rho + \frac{1}{3} A_\mu [A_\nu, A_\rho] \right),
\end{equation}
where $k$ is a constant, $x^\mu$ are coordinates on $M$, $\p_\mu$ are the corresponding derivatives and $\epsilon^{\mu\nu\rho}$ is the Levi-Civita tensor. At the classical level, one of the key characteristics of this theory is that its equation of motion is the zero curvature equation of $A_\mu$, \textit{i.e.} the vanishing of $\p_\mu A_\nu - \p_\nu A_\mu + [A_\mu,A_\nu]$. A second important property of this theory is its invariance under the gauge transformations $A_\mu \mapsto u A_\mu u^{-1} - (\p_\mu u)u^{-1}$, where $u$ is a field valued in a Lie group corresponding to the choice of Lie algebra (similar to the gauge transformations of a Yang-Mills theory). At the quantum level, the gauge-invariant observables of the theory measure topological properties of the underlying 3-dimensional space-time and in particular are related to knot invariants such as Jones polynomials~\cite{Witten:1988hf}. In these lectures, we will be mostly interested with classical aspects: we will therefore not discuss the quantisation of the Chern-Simons theory and refer to the extensive literature on the subject for more details.

The 3-dimensional Chern-Simons theory shares some formal similarities with the general formalism of integrable systems~\cite{Witten:1989wf}. For instance, the knot invariants, mentioned above in relation with the quantised Chern-Simons theory, satisfy a certain number of invariance properties called Reidemeister moves, one of which resembles the so-called Yang-Baxter equation, characteristic of the theory of quantum integrable systems (see Ana L. Retore's lectures \textit{``Introduction to classical and quantum integrability''}~\cite{LectureRetore}). Moreover, the equation of motion of the Chern-Simons theory takes the form of the vanishing of the curvature of the gauge field $A_\mu$ and therefore resembles the Lax equation of 2-dimensional integrable field theories: in particular, if we denote by $(t,x,\xi)$ the three coordinates of the Chern-Simons theory and restrict our attention to gauge fields with vanishing $A_\xi$-component, the two remaining components $A_t$ and $A_x$ form a pair of matrices independent of $\xi$ and satisfying a 2-dimensional zero curvature equation $\p_t A_x - \p_x A_t + [A_x,A_t] = 0$, similar to the Lax equation. In both these examples, the main missing ingredient to make a more concrete link between the 3-dimensional Chern-Simons theory and the formalism of integrable systems is the spectral parameter $z\in\CP$. Indeed, the Reidemeister move for knot invariants is schematically a Yang-Baxter equation without spectral parameter and in the above discussion the gauge fields components $A_t$ and $A_x$ do not depend on an auxiliary complex parameter $z$. This absence of the spectral parameter is what is solved by the passage from the standard 3-dimensional Chern-Simons theory to its semi-holomorphic 4-dimensional variant, introduced in~\cite{NekrasovThesis,Costello:2013zra}.

\paragraph{The semi-holomorphic 4-dimensional Chern-Simons theory.} In order to take into account the spectral parameter, this 4-dimensional theory incorporates it as part of the space-time on which it is defined. More precisely, this 4-dimensional space-time is defined as $M=\Sigma \times \CP$, where the coordinates $(t,x)$ on $\Sigma$ will play the role of the space-time coordinates of the resulting 2-dimensional integrable model and the complex coordinate $z$ on $\CP$ will play the role of its spectral parameter (see Figure \ref{Fig:Space} below). The theory depends on a gauge field with components $A_t$, $A_x$ and $A_{\bar z}$ but with no $z$-component (where $\bar z$ denotes the complex conjugate of $z$). The action of the theory is taken to be
\begin{equation}
S[A] = \frac{i}{4\pi}  \iint_\Sigma \dd t\,\dd x \iint_{\CP} \dd z\,\dd \zb \; \vp(z) \sum_{\mu,\nu,\rho \in\lbrace t,x,\zb \rbrace} \epsilon^{\mu\nu\rho} \, \Tr \left( A_\mu \p_\nu A_\rho + \frac{1}{3} A_\mu [A_\nu, A_\rho] \right).
\end{equation}
Schematically, the corresponding Lagrangian density is thus defined as the Lagrangian density for a 3-dimensional Chern-Simons theory along the $(t,x,\bar z)$-directions, multiplied by a meromorphic 1-form $\omega=\vp(z)\dd z$ along the $z$-direction (where $\vp(z)$ is a rational function of $z$)\footnote{Another example of a gauge theory multiplied by a meromorphic 1-form and related to integrability was discussed in~\cite{Gorsky:1994dj}. Moreover, gauge theories obtained by multiplying Chern-Simons terms with non-dynamical differential forms were also considered in~\cite{Nair:1990aa,Losev:1995cr} for the case of Kähler forms. The author would like to thank N. Nekrasov for useful comments and discussions on these aspects and related ones.}. This 1-form is a non-dynamical object and is thus part of the input defining the theory.

In the case where this 1-form $\omega$ is simply equal to $\dd z$, it was shown in the works~\cite{Costello:2013zra,Costello:2013sla,Witten:2016spx,Costello:2017dso,Costello:2018gyb} of K. Costello, M. Yamazaki and E. Witten (see also~\cite{NekrasovThesis} for an earlier similar proposal) that this 4-dimensional theory can be used to generate integrable lattice models (spin chains). In this framework, one obtains quantities which satisfy the Yang-Baxter equation with spectral parameter $z$, as a generalisation of the Reidemeister move of the 3-dimensional theory. In particular, with this choice of 1-form $\omega=\dd z$ on $\CP$, one recovers integrable lattice systems associated with so-called rational solutions of the Yang-Baxter equation. This approach can be generalised by replacing $\CP$ with any compact Riemann surface $C$ equipped with a meromorphic 1-form $\omega$ without zeroes. In addition to the Riemann sphere with $\omega=\dd z$, there exist (up to equivalence) two other possibilities for $C$ which admit such a nowhere-vanishing 1-form $\omega$, namely the sphere $\CP$ with $\omega=\dd z/z$ and the torus\footnote{By removing the poles of $\omega$ from $C$, one can see this setup as choosing a (non-compact) Riemann surface with a holomorphic 1-form with no zeroes. For instance, $\omega=\dd z$ on $\CP$ defines a non-vanishing holomorphic 1-form on the complex plane $\C=\CP\setminus\lbrace \infty\rbrace$ and $\omega=\dd z/z$ on $\CP$ defines a non-vanishing holomorphic 1-form on $\C^\times=\CP\setminus\lbrace 0,\infty \rbrace$, which is isomorphic to the cylinder. The non-vanishing 1-form $\omega$ on the torus already has no poles.}: these two cases yield integrable lattice systems associated respectively with trigonometric and elliptic solutions of the Yang-Baxter equation. Remarkably, this geometrisation of the spectral parameter $z\in C$ then recovers the classification of solutions of the Yang-Baxter equation studied by A. Belavin and V. Drinfeld in the context of integrable systems~\cite{BelavinDrinfeld}, where these solutions also belong to these three classes: rational, trigonometric and elliptic.

In these lectures, we will not focus on integrable lattice models but rather on integrable 2-dimensional field theories. The idea that such theories can also arise from the 4-dimensional Chern-Simons setup was initially proposed in the work~\cite{Costello:2019tri} of K. Costello and M. Yamazaki. There are in fact two main possibilities to generate these models. The first one still considers the case where $C$ is either the Riemann sphere or the torus, equipped with a meromorphic 1-form $\omega$ with no zeroes, similarly to the framework yielding lattice models. One then couples the 4-dimensional Chern-Simons theory with 2-dimensional terms called order defects, yielding in the end an integrable field theory on $\Sigma$. This procedure allows to recover \textit{e.g.} the Gross-Neuveu and Thirring models~\cite{Costello:2019tri}, $\sigma$-models on Kälher manifolds~\cite{Costello:2019tri,Bykov:2019vkf,Bykov:2020llx,Bykov:2020nal,Bykov:2020tao,Affleck:2021ypq} and the Zakharov-Mikhailov~\cite{Caudrelier:2020xtn} and Faddeev-Reshetikhin~\cite{Fukushima:2020tqv} models. We will not discuss order defects in these lectures and refer to these references for more details.

Instead we will focus on the second approach proposed in~\cite{Costello:2019tri} to generate integrable 2-dimensional models from 4-dimensional Chern-Simons theory, which is based on so-called disorder defects and which occurs in the case where the 1-form $\omega$ has zeroes. This approach was further developed in~\cite{
Delduc:2019whp,Bassi:2019aaf,Schmidtt:2019otc,Gaiotto:2020fdr,Fukushima:2020kta,Fukushima:2020dcp,Costello:2020lpi,Tian:2020ryu,Tian:2020meg,Benini:2020skc,Hoare:2020mpv,Gaiotto:2020dhf,Bittleston:2020hfv,Penna:2020uky,Lacroix:2020flf,Chen:2021qto,Fukushima:2021eni,Derryberry:2021rne,Schmidtt:2021rhw,Stedman:2021wrw}: in these lectures, we will mainly review the results of~\cite{Costello:2019tri} and~\cite{Delduc:2019whp}.

\begin{figure}[H]
\begin{center}
\includegraphics[scale=1.1]{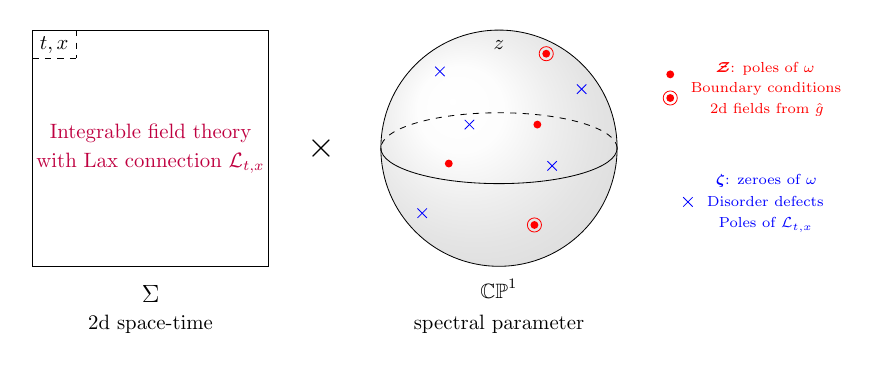}\vspace{-10pt}
\caption{Space-time of the 4-dimensional Chern-Simons theory.}\label{Fig:Space}\vspace{-5pt}
\end{center}
\end{figure}

\begin{figure}[H]
\begin{center}
\includegraphics[scale=1.2]{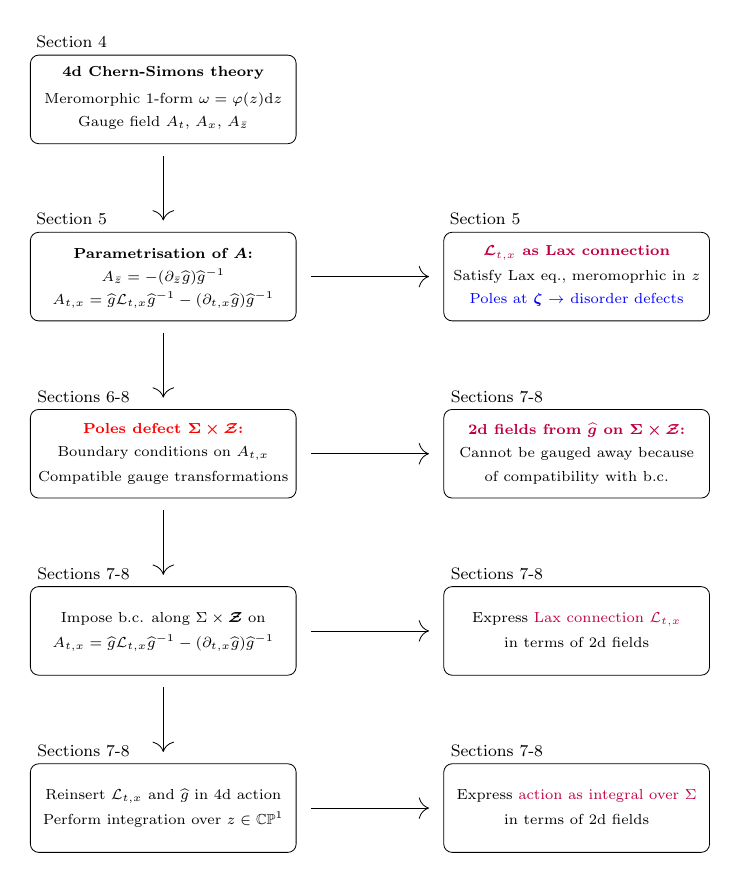}
\caption{Summary of the 4-dimensional Chern-Simons approach with disorder defects.}\label{Fig:Summary}
\end{center}
\end{figure}

\paragraph{Overview.} In the rest of this introduction, we give an overview of the main concepts and results described in these lectures, which we also summarise in a more schematic way in the diagrams \ref{Fig:Space} and \ref{Fig:Summary}. Let us warn here the reader that this overview does not aim at providing an instantaneous real grasp of these aspects and should instead be seen more as a general guide in which these main concepts and their relations are sketched. Our main task in these lectures will then be to introduce these with more details and explanations and to complete step by step the schematic overview sketched in the diagrams \ref{Fig:Space} and \ref{Fig:Summary}. We thus invite the reader to come back to these diagrams throughout the lectures.

We will consider the 4-dimensional Chern-Simons theory with an arbitrary meromorphic 1-form $\omega$ on $\CP$ (we will not consider the case of more general Riemann surfaces $C$ instead of $\CP$), which in particular can have zeroes $\Ze \subset \CP$. The equations of motion of the 4-dimensional theory take the form of a zero curvature equation $\p_\mu A_\nu - \p_\nu A_\mu + [A_\mu,A_\nu] = 0$ along the components $\mu,\nu \in \lbrace t,x,\bar z\rbrace$, at least for $z \in \CP\setminus\Ze$. To make the link with integrable systems, we focus on solutions of these equations for which there exists a gauge where the gauge field has no $\zb$-component: suggestively, we then call the two other components of the gauge field in this gauge $\Lc_t$ and $\Lc_x$. Part of the equations of motion of the theory then tells us that $\p_{\zb}\Lc_{t,x}=0$ for $z\in\CP\setminus \Ze$. This implies that $\Lc_{t,x}$ is holomorphic in $z$ on $\CP\setminus \Ze$: more precisely, one shows that it is meromorphic on $\CP$ with poles at the zeroes $\Ze$ of $\omega$. These singularities of the gauge field at the zeroes of $\omega$ are the so-called disorder defects. The remaining part of the equations of motion then implies that $\Lc_{t,x}$ satisfy a zero curvature equation along the directions $(t,x)$ of $\Sigma$, \textit{i.e.} the Lax equation \eqref{Eq:LaxIntro}. Thus, this 4-dimensional setup exactly generates a Lax connection on $\Sigma$ depending meromorphically on an auxiliary complex coordinate $z\in\CP$, which then confirms the identification of this coordinate with the spectral parameter.

The next step is to extract from this 4-dimensional theory the integrable 2-dimensional model which corresponds to this Lax connection $\Lc_{t,x}$. This procedure has to do with the poles $\Zc\subset \CP$ of $\omega$. When varying the action with respect to the gauge field to derive the equations of motion, one also obtains terms on the 2-dimensional space $\Sigma \times \Zc$, which we call the poles defect. To make sure that the action principle is well-posed, one needs to ensure that these terms vanish: this is done by imposing boundary conditions on the gauge field components $A_t$ and $A_x$ along $\Sigma \times \Zc$. An important consequence of these boundary conditions is that one needs to restrict the set of true, physical, gauge transformations of the theory to the ones that preserve these boundary conditions. To obtain the physical degrees of freedom of the theory, one should determine which are the degrees of freedom contained in $A_\mu$ that cannot be eliminated by a gauge transformation. Schematically, almost all of these degrees of freedom can be gauged away, except for certain 2-dimensional fields located on the poles defect $\Sigma \times \Zc$, which cannot be eliminated because of the boundary conditions imposed on the gauge transformations on this defect. These degrees of freedom on $\Sigma \times \Zc$ form the fundamental fields of the resulting integrable 2-dimensional model. In particular, one shows that it is possible to perform the integration over $z\in\CP$ in the 4-dimensional action, obtaining a 2-dimensional action on $\Sigma$ depending on these fields.

For the sake of concreteness, let us be a bit more precise about the extraction of the 2-dimensional fields. The assumption made earlier that there exists a gauge transformation of the gauge field such that its $\zb$-component vanishes implies the existence of a group-valued field $\gh$ such that $A_{\zb}=-(\p_{\zb}\gh)\gh^{-1}$. Gauge transformations $A_\mu \mapsto u A_\mu u^{-1} - (\p_\mu u)u^{-1}$ act on this field $\gh$ as $\gh\mapsto u\gh$. In particular, the gauge transformation by $u=\gh^{-1}$ would completely gauge away all degrees of freedom in $\gh$: this transformation is however not a true physical gauge transformation as it does not respect the boundary conditions imposed on $\Sigma \times \Zc$. Taking into account these boundary conditions, one finds that the degrees of freedom that cannot be gauged away in $\gh$ are its evaluations on the poles defect $\Sigma \times \Zc$ or more generally the evaluations of its derivatives $\p_z^p \gh$ on this defect (where the possible values of $p$ depend on the order of the poles $\Zc$ and the nature of the boundary conditions). These form the degrees of freedom of the 2-dimensional integrable model. To end the description of this model, let us briefly discuss its Lax connection. The latter was defined as the gauge field after the (unphysical) gauge transformation by $\gh$ which eliminates the $\zb$-component: thus we have $A_{t,x} =  \gh \Lc_{t,x} \gh^{-1} - (\p_{t,x} \gh)\gh^{-1}$. Imposing the boundary conditions on $A_{t,x}$ at the poles defect $\Sigma \times \Zc$ allows one to express the Lax connection $\Lc_{t,x}$ in terms of the fundamental fields of the 2-dimensional model, extracted from $\gh$ on this defect.

The integrable 2-dimensional model obtained by this approach depends on the choice of 1-form $\omega$ and of boundary conditions at the poles of $\omega$: in particular, different choices yield different integrable models. In these lectures, we will discuss in detail the simplest example of an integrable 2-dimensional model obtained by this approach, the Principal Chiral Model, and will describe the choice of $\omega$ and of boundary conditions that leads to this model, following~\cite{Costello:2019tri}. As a second example, we will treat the case of the so-called Yang-Baxter deformation of the Principal Chiral Model, which is treated in~\cite{Delduc:2019whp} by considering a deformation of the 1-form $\omega$ and boundary conditions corresponding to the latter. Finally, we will mention other types of boundary conditions in exercises and as opening perspectives, following mainly~\cite{Delduc:2019whp,Benini:2020skc,Lacroix:2020flf}.

\paragraph{Plan.} The plan of these lectures is the following. In Section \ref{Sec:Preliminaries}, we will recall some useful facts about the formalism of differential forms and in particular how to formulate the Lax equation of integrable field theories using this formalism.

In Section \ref{Sec:3d}, we will introduce the standard 3-dimensional Chern-Simons theory and discuss some of its key features, including its equations of motion and gauge invariance. We will focus on the classical aspects of this theory which are useful to prepare our discussion of its 4-dimensional variant and in particular will not address its quantisation.

We then introduce the semi-holomorphic 4-dimensional Chern-Simons theory in Section \ref{Sec:4d}, focusing first on its general properties. In particular, we will introduce the 1-form $\omega$ and the gauge field of this theory, and will discuss its action, its equations of motion and its gauge transformations.

In Section \ref{Sec:4dLax}, we will explain the relation of this 4-dimensional theory with the Lax formalism of integrable 2-dimensional models. Following the ideas sketched above, we will explain how one can naturally extract from the gauge field of the theory a Lax connection on $\Sigma$ satisfying a zero curvature equation and depending meromorphically on the spectral parameter $z\in\CP$. In particular, we will show that the singularities of this connection in $\CP$ are naturally located at the zeroes of the 1-form $\omega$, forming the disorder defects of the theory.

Section \ref{Sec:Poles} will be devoted to the treatment of the defect terms originating from the poles of $\omega$. After discussing the general form of these defect terms, we will discuss on a simple example how they can be treated by imposing appropriate boundary conditions on the gauge field along the pole defect. Moreover, we will see through this example how the compatibility with this boundary condition restricts the set of true physical gauge transformations of the theory.

The simple example of the Principal Chiral Model is the subject of Section \ref{Sec:PCM}. We will start by describing the 4-dimensional setup one needs to start with to obtain this model and in particular the choice of 1-form $\omega$ and of boundary conditions at its poles, using the simplest example of boundary condition treated in Section \ref{Sec:Poles}. We will then discuss in detail how the 2-dimensional Principal Chiral Model can be constructed from this 4-dimensional setup, by describing the extraction of the 2-dimensional degrees of freedom, the computation of the corresponding Lax connection and the reduction of the 4-dimensional action to the 2-dimensional action of the Principal Chiral Model.

In Section \ref{Sec:MoreBC}, we will illustrate the power and generality of the 4-dimensional Chern-Simons approach by discussing another example of integrable field theory, the Yang-Baxter deformation of the Principal Chiral Model. In particular, we will show that this deformation can be obtained by considering another choice of 1-form $\omega$ and a more involved choice of boundary conditions. Based on the experience of this example, we will discuss briefly in the last part of Section \ref{Sec:MoreBC} the treatment of more general 1-forms $\omega$ and boundary conditions.

Finally, we will conclude in Section \ref{Sec:Conclusion} with a brief discussion of some further developments and interesting future perspectives.

Some exercises are proposed throughout the lecture notes, with the goal of providing some concrete training on certain aspects treated in the lectures and to discuss some further developments. These exercises are marked with a certain number of stars which indicate their difficulty, from 1 (easiest) to 4 (hardest).

\section{Differential forms notations and reminder on Lax formulation}
\label{Sec:Preliminaries}

\subsection{Generalities on differential forms}

Let us consider a manifold $M$ of dimension $d$, with local coordinates $x^\mu$, $\mu \in \lbrace 1,\cdots,d \rbrace$. We will denote by $\Omega^1(M)$ the space of \textit{1-forms on $M$}. In the local coordinates $x^\mu$, such a 1-form can be written as
\begin{equation}
\Lambda = \Lambda_\mu \, \dd x^\mu,
\end{equation}
where the coefficients $\Lambda_\mu$ depend on the coordinates $x^\nu$. In this equation, and in the rest of these lecture notes, we use the Einstein convention of summing over repeated indices. More generally, we will denote by $\Omega^p(M)$ the space of $p$-forms on $M$, whose elements can be written locally as
\begin{equation}
\Lambda = \Lambda_{\mu_1 \cdots \mu_p}\, \dd x^{\mu_1} \wedge \cdots \wedge \dd x^{\mu_p},
\end{equation}
where the coefficients $\Lambda_{\mu_1 \cdots \mu_p}$ are completely antisymmetric with respect to permutations of the indices $\mu_i$. Here, we used the basis $\dd x^{\mu_1} \wedge \cdots \wedge \dd x^{\mu_p}$ of $\Omega^p(M)$, labelled by indices $(\mu_1,\cdots,\mu_p)$ up to permutations, which is built from the elementary 1-forms $\dd x^\mu$ by taking the \textit{exterior product} $\wedge$. In particular, we recall that $\dd x^\mu \wedge \dd x^\nu = - \dd x^\nu \wedge \dd x^\mu$. By linearity, the exterior product extends to a bilinear map $\wedge : \Omega^p(M) \times \Omega^q(M) \rightarrow \Omega^{p+q}(M)$, which satisfies the skew-symmetry property
\begin{equation}
\Lambda \wedge \Theta = (-1)^{pq}\, \Theta \wedge \Lambda, \qquad \forall \, \Lambda \in \Omega^p(M), \; \forall \, \Theta \in \Omega^q(M).
\end{equation}

Let us also consider the \textit{exterior derivative} $\dd$, which defines a linear map $\dd: \Omega^p(M) \rightarrow \Omega^{p+1}(M)$. It acts on the $p$-form $\Lambda = \Lambda_{\mu_1 \cdots \mu_p}\, \dd x^{\mu_1} \wedge \cdots \wedge \dd x^{\mu_p}$ as
\begin{equation}
\dd\Lambda = \frac{\p\Lambda_{\mu_2\cdots\mu_{p+1}}}{\p x^{\mu_1}} \, \dd x^{\mu_1} \wedge \cdots \wedge \dd x^{\mu_{p+1}}.
\end{equation}
Note that for a 1-form $\Lambda=\Lambda_\mu\,\dd x^\mu$, we have
\begin{equation}\label{Eq:d1Form}
\dd\Lambda = \frac{\p \Lambda_\nu}{\p x^\mu} \, \dd x^\mu \wedge \dd x^\nu = \frac{1}{2} \left( \p_\mu \Lambda_\nu - \p_\nu \Lambda_\mu \right) \, \dd x^\mu \wedge \dd x^\nu.
\end{equation}
The exterior derivative satisfies $\dd^2=0$, or more explicitly
\begin{equation}
\dd (\dd \Lambda) = 0, \qquad \forall\, \Lambda\in\Omega^p(M).
\end{equation}
Finally, it satisfies the product rule
\begin{equation}\label{Eq:ProductRule}
\dd(\Lambda\wedge\Theta) = \dd\Lambda\wedge\Theta + (-1)^p \, \Lambda\wedge \dd\Theta, \qquad \forall \, \Lambda \in \Omega^p(M), \; \forall \, \Theta \in \Omega^q(M).
\end{equation}

\subsection{Differential forms valued in a Lie algebra}

Let us consider a matrix Lie algebra $\g$, with commutator $[\cdot,\cdot]$, such as for instance $\mathfrak{su}(N)$ or $\mathfrak{so}(N)$. We will denote by $\Omega^p(M,\g)$ the space of \textit{$p$-forms on $M$ valued in $\g$}. If $A$ and $B$ are $\g$-valued forms, we define the exterior product $A\wedge B$ through the standard formula but using the matrix product to define the multiplication of the components $A_\mu B_\nu$. Note that with this definition, there is in general no (skew-)symmetry relation between $A \wedge B$ and $B \wedge A$, since the matrix product is non-commutative. In particular, this means that the product $A \wedge A$ is in general non-vanishing. For instance, if $A = A_\mu \, \dd x^\mu$ is a $\g$-valued 1-form, then $A\wedge A$ is the $\g$-valued 2-form
\begin{equation}\label{Eq:AwedgeA}
A \wedge A = A_\mu A_\nu \, \dd x^\mu \wedge \dd x^\nu = \frac{1}{2} [ A_\mu, A_\nu ]\, \dd x^\mu \wedge \dd x^\nu,
\end{equation}
where the second equality is obtained by permuting the repeated indices $\mu,\nu$ and using the skew-symmetry of $\dd x^\mu \wedge \dd x^\nu$. In these lectures, we will extensively use the \textit{curvature} of $\g$-valued 1-forms, which is defined as
\begin{equation}\label{Eq:FA}
F(A) = \dd A + A \wedge A = \frac{1}{2} \Bigl( \p_\mu A_\nu - \p_\nu A_\mu + [A_\mu, A_\nu] \Bigr) \, \dd x^\mu \wedge \dd x^\nu,
\end{equation}
where we used Equations \eqref{Eq:d1Form} and \eqref{Eq:AwedgeA} to get the second equality. In particular, we recognise in the term in brackets the standard notion of curvature $F_{\mu\nu} = \p_\mu A_\nu - \p_\nu A_\mu + [A_\mu, A_\nu]$ used in gauge theory. If a 1-form $A$ has vanishing curvature, \textit{i.e.} $F(A)=0$, we will say that it is \textit{flat}.

\subsection{Lax formulation of 2-dimensional integrable field theories}
\label{Sec:Lax}

Let us finally recall briefly some basic facts about the Lax formulation of 2-dimensional integrable field theories. We refer to Ana L. Retore's lectures \textit{``Introduction to classical and quantum integrability''}~\cite{LectureRetore} for details. We consider a 2-dimensional field theory with time coordinate $t\in \R$ and space coordinate $x \in \D$, where $\mathbb{D}$ is taken to be the real line $\R$ or the circle $S^1$, with the appropriate choice of boundary conditions. We will denote by $\Sigma = \R \times \D$ the 2-dimensional space-time on which the theory is defined: having in mind applications to $\sigma$-models, we will sometimes refer to this space-time as the worldsheet.

The integrability of such a 2-dimensional theory is generally based on the existence of a so-called \textit{Lax connection}. The latter consists of two matrices $\Lc_t(z,t,x)$ and $\Lc_x(z,t,x)$, valued in a (complex) Lie algebra $\g^{\C}$, constructed from the fundamental fields of the theory and depending meromorphically on an auxiliary complex parameter $z$, called the \textit{spectral parameter}. These matrices should be such that the equations of motion of the theory can be recast as the following \textit{Lax equation}
\begin{equation}\label{Eq:LaxEq}
\p_t \Lc_x(z,t,x) - \p_x \Lc_t(z,t,x) + \bigl[ \Lc_t(z,t,x), \Lc_x(z,t,x) \bigr] = 0, \qquad \forall\,z.
\end{equation}
Indeed, this equation ensures the existence of an infinite number of conserved charges for this theory.

The Lax equation \eqref{Eq:LaxEq} is sometimes called the zero curvature equation. Indeed, the formalism of the previous subsection suggests a rewriting of this equation in terms of differential forms. More concretely, we introduce the $\g^{\C}$-valued 1-form
\begin{equation}
\Lc(z) = \Lc_t(z)\, \dd t + \Lc_x(z) \, \dd x
\end{equation}
on $\Sigma$, where we kept the $z$-dependence in the notation to stress the fact that this form also depends on the auxiliary spectral parameter. Using the language of the previous subsection, and in particular Equation \eqref{Eq:FA}, the Lax equation \eqref{Eq:LaxEq} can then be recast as the vanishing of the curvature of $\Lc(z)$:
\begin{equation}
F\bigl( \Lc(z) \bigr) = \dd\Lc(z) + \Lc(z) \wedge \Lc(z) = 0, \qquad \forall \, z.
\end{equation}

So far, we did not make precise the ``space'' to which the complex spectral parameter $z$ belongs. In the general theory of integrable systems, this parameter can belong to any Riemann surface, \textit{i.e.} any one-dimensional complex manifold. For instance, for some systems, one can consider $z$ as belonging to a torus, in which case the Lax connection of the theory typically depends on $z$ through elliptic functions. Although the 4d Chern-Simons theory that is the subject of the present lectures can in principle be defined for any choice of Riemann surface for the spectral parameter, we will restrict our attention in these lectures to the simplest case, where the Lax connection is simply a rational function of $z$. In this case, we will see $z$ as simply belonging to the complex plane $\C$, or more precisely to the Riemann sphere $\CP = \C \cup \lbrace \infty \rbrace$ obtained by adding to the complex plane the point at infinity.

Let us finally observe that, in general, the definition of integrability requires not only the existence of an infinite number of conserved charges in the theory but also the involution of these charges, \textit{i.e.} the vanishing of their Poisson brackets in the Hamiltonian formulation. We will not discuss such Hamiltonian properties in detail in these lectures, although we will briefly mention related aspects in Subsection \ref{Sec:CommentsLax}.

To summarise this subsection, in the rest of these lecture notes we will see a 2-dimensional integrable field theory on a worldsheet $\Sigma$ as a theory possessing a Lax connection $\Lc(z)$ satisfying the following key features:\vspace{-2pt}
\begin{enumerate}[(i)]\setlength\itemsep{0.2em}
\item $\Lc(z)$ is a $\g^{\C}$-valued 1-form on $\Sigma$, depending rationally on the spectral parameter $z \in \CP$ ;
\item the equations of motion are equivalent to the Lax equation $\dd \Lc(z) + \Lc(z) \wedge \Lc(z) = 0, \forall \, z\in \CP$. 
\end{enumerate}

\section{3-dimensional Chern-Simons theory}
\label{Sec:3d}

Before studying its 4-dimensional variant, let us first introduce the standard 3-dimensional Chern-Simons theory and discuss some of its main features. We will use the abbreviation 3d-CS to designate this theory. The quantisation of 3d-CS has attracted a lot of attention in the physics and mathematics literature, has been the subject of many works and has led to many important developments. However, for these lectures, we will only need its classical properties and thus restrict our attention to these aspects in the notes for simplicity.

\subsection{Gauge field, Chern-Simons 3-form and action}

\paragraph{Gauge field and Chern-Simons 3-form.} We consider a 3-dimensional manifold $M$, with coordinates $x^\mu$, $\mu\in\lbrace 1,2,3 \rbrace$, which defines the space-time of the 3d-CS theory. The fundamental field of this theory is a \textit{gauge field} $A_\mu(x)$, valued in a matrix Lie algebra $\g$, which we see as a 1-form $A = A_\mu \, \dd x^\mu \, \in \, \Omega^1(M,\g)$ (see Section \ref{Sec:Preliminaries} for notations).

An important object in the theory is the so-called \textit{Chern-Simons 3-form}, defined as
\begin{equation}\label{Eq:CSForm}
\CS(A) = \Tr \left( A \wedge \dd A + \frac{2}{3} A \wedge A \wedge A \right),
\end{equation}
and which is thus an element of $\Omega^3(M)$. For completeness, let us also describe the expression of this form in coordinates. Since $M$ is 3-dimensional, it possesses a completely skew-symmetric tensor $\epsilon^{\mu\nu\rho}$, which we normalise by $\epsilon^{123}=1$. Using Equations \eqref{Eq:d1Form} and \eqref{Eq:AwedgeA}, we then find that the Chern-Simons 3-form reads
\begin{equation}\label{Eq:CSFormCoord}
\CS(A) = \epsilon^{\mu\nu\rho} \, \Tr \left( A_\mu \p_\nu A_\rho + \frac{1}{3} A_\mu [A_\nu, A_\rho] \right) \, \dd x^1 \wedge \dd x^2 \wedge \dd x^3 ,
\end{equation}
where $\dd x^1 \wedge \dd x^2 \wedge \dd x^3$ is the volume form of $M$.

In the above definition, we have used the trace $\Tr$ in the matrix representation of $\g$. This trace defines a bilinear form\footnote{If the chosen matrix representation is the adjoint one, then this bilinear form is the standard Killing form of $\g$.}
\begin{equation}
\begin{array}{rccc}
\langle \cdot, \cdot \rangle : 
& \g \times \g &\longrightarrow & \R \\
& (X,Y) & \longmapsto & \Tr(XY)
\end{array}
\end{equation}
on $\g$, which is ad-invariant, in the sense that
\begin{equation}
\bigl\langle X,[Y,Z] \bigr\rangle = \bigl\langle [X,Y],Z \bigr\rangle, \qquad \forall\,X,Y,Z\in\g.
\end{equation}
From now on, we will suppose that $\langle \cdot, \cdot \rangle$ is non-degenerate\footnote{This is the case for instance when $\g$ is semi-simple.}, \textit{i.e.} that $\langle X,Y \rangle=0$ for all $Y\in\g$ implies $X=0$. We extend this bilinear form on $\g$ to a bilinear form on $\g$-valued forms by considering exterior products. The Chern-Simons form \eqref{Eq:CSForm} can then be rewritten as
\begin{equation}
\CS(A) = \left\langle A, \dd A + \frac{2}{3} A \wedge A \right\rangle,
\end{equation}
which is a form also often found in the literature.

\paragraph{Action.} Recall that $p$-forms can be naturally integrated over $p$-manifolds. In terms of the 3-form $\CS(A)$, the \textit{3d-CS action} is then simply defined as the following integral~\cite{Witten:1988hf}:
\begin{equation}\label{Eq:Action3d}
S[A] = \frac{k}{4\pi} \iiint_M \; \CS(A),
\end{equation}
where $k$ is a constant parameter called the Chern-Simons level. This action then defines a functional on the space of $\g$-valued 1-forms $\Omega^1(M,\g)$. For completeness, let us also give the expression of the action $S[A]$ as a standard integral over the coordinates $x^\mu$. Using the coordinate expression \eqref{Eq:CSFormCoord} of $\CS(A)$ and defining the orientation of $M$ such that its positive volume form is $\dd x^1 \wedge \dd x^2 \wedge \dd x^3$, we get
\begin{equation}\label{Eq:Action3dCoord}
S[A] = \frac{k}{4\pi}  \iiint_M \dd x^1 \, \dd x^2 \, \dd x^2 \; \epsilon^{\mu\nu\rho} \, \Tr \left( A_\mu \p_\nu A_\rho + \frac{1}{3} A_\mu [A_\nu, A_\rho] \right).
\end{equation}

\subsection{Equations of motion}
\label{Sec:EoM3d}

\paragraph{Variation of the Chern-Simons 3-form.} Having defined the action \eqref{Eq:Action3d} of the Chern-Simons theory, we can now derive its equations of motion. For this, let us consider an infinitesimal variation $\delta A$ of the gauge field and compute the corresponding variation of the Chern-Simons 3-form $\CS(A)$. We have
\begin{equation}\label{Eq:dCS1}
\delta\, \CS(A) = \Tr \bigl( \delta A \wedge \dd A + A \wedge \dd(\delta A) + 2 \, A \wedge A \wedge \delta A \bigr),
\end{equation}
where the last part is obtained by observing that the three terms arising from the variation of $\Tr(A\wedge A \wedge A)$ are equal due to the cyclicity of the trace and the invariance of $\dd x^\mu \wedge \dd x^\nu \wedge \dd x^\rho$ under cyclic permutations. To treat the second term, we will use the product rule \eqref{Eq:ProductRule} to write
\begin{equation}
A \wedge \dd(\delta A) = \dd A \wedge \delta A - \dd \bigl( A \wedge \delta A ).
\end{equation}
Reinserting in \eqref{Eq:dCS1} and using the cyclicity of the trace to exchange the position of $\dd A$ and $\delta A$ (note that we do not pick any minus sign by doing so as $\dd A$ is a 2-form), we get the following important result:
\begin{equation}\label{Eq:dCS}
\delta\, \CS(A) = 2 \, \Tr \bigl( F(A) \wedge \delta A \bigr) - \dd \, \Tr\bigl( A \wedge \delta A \bigr),
\end{equation}
where we recall that $F(A)=\dd A + A\wedge A$ is the curvature of $A$, as defined in Equation \eqref{Eq:FA}.

\paragraph{Variation of the action and equation of motion.} From the above expression of the variation of the Chern-Simons 3-form, we readily obtain the variation of the action \eqref{Eq:Action3d} by integrating:
\begin{equation}
\delta S[A] = \frac{k}{2\pi} \iiint_M  \Tr \bigl( F(A) \wedge \delta A  \bigr) - \frac{k}{4\pi} \iint_{\partial M} \Tr(A\wedge \delta A),
\end{equation}
where the second term results from the application of Stokes's theorem and $\p M$ denotes the boundary of the manifold $M$. In order for this boundary term to vanish, we either suppose that $M$ has no boundary or that one of the gauge field component vanishes\footnote{Note that in this case we also restrict to variations $\delta A$ that preserve this boundary condition, \textit{i.e.} $\delta A_\mu|_{\p M}=0$.} on the boundary $\p M$, \textit{i.e.} $A_\mu|_{\p M}=0$ where $\mu$ is a direction tangential to $\p M$.

We are then left with the first term only in the above variation, which is a ``bulk'' term (in opposition to a boundary one). By stationarity of the action, this bulk term should vanish for any variation $\delta A \in \Omega^1(M,\g)$. Since we suppose that the trace bilinear form is non-degenerate (see above), this is equivalent to
\begin{equation}\label{Eq:EoM3d}
F(A) = \dd A + A \wedge A = 0.
\end{equation}
The equation of motion of the 3d Chern-Simons action is thus \textit{the zero curvature equation for the gauge field $A$}, or in other words the flatness of $A$. This is one of the key feature of the Chern-Simons theory.\\

\begin{tcolorbox} \textit{\underline{Exercise 1:} Equations of motion in components} {\Large$\;\;\star$} \vspace{4pt}\\
For the readers less familiar with differential forms computations, we leave as an exercise to rederive the equation of motion $F(A)=0$ working directly in components. For that, start with the expression \eqref{Eq:Action3dCoord} of the action and compute its variation when introducing a variation $\delta A_\mu$ of the gauge field, discarding boundary terms when performing integration by parts. As an alternative check, one can also directly compute the Euler-Lagrange equations of the action \eqref{Eq:Action3dCoord}.
\end{tcolorbox}~

Let us quickly discuss a standard class of solutions of these equations of motion. Let $G$ be a connected and simply-connected Lie group with Lie algebra $\g$ and $g: M \rightarrow G$ be a smooth map from $M$ to $G$. The 1-form $g^{-1}\dd g$ is then valued in the Lie algebra $\g$. It is a standard result that this 1-form satisfies the so-called Maurer-Cartan identity\footnote{This is a familiar identity in the context of non-abelian gauge theories, which states that ``pure gauge'' configurations $A=g^{-1}\dd g$ have vanishing curvature $F(A)=0$.}
\begin{equation}\label{Eq:MC}
d\bigl( g^{-1} \dd g \bigr) + g^{-1} \dd g \wedge g^{-1} \dd g = 0.
\end{equation}
Thus, 
\begin{equation} \label{Eq:PureGauge}
A=g^{-1}\dd g
\end{equation}
is a solution of the equation of motion \eqref{Eq:EoM3d}. A natural question at this point is whether all solutions of the theory are of this form: this is in general not the case, depending on the topology of $M$. More precisely, it is true when $M$ is simply connected. For more general topologies, it is only true locally.

\subsection{Gauge symmetry}
\label{Sec:Gauge3d}

\paragraph{Gauge transformations.} Another important property of the Chern-Simons theory is its gauge invariance. Let us consider the \textit{local transformation} of the gauge field $A \mapsto A^u$ where
\begin{equation}
A^u = u A u^{-1} - (\dd u)u^{-1},
\end{equation}
and $u$ is a smooth function on $M$ valued in the (connected simply-connected) group $G$ with Lie algebra $\g$. We note that the adjoint action $X \mapsto uXu^{-1}$ preserves the Lie algebra $\g$ and that the 1-form $(\dd u) u^{-1}$ is naturally valued in $\g$: thus $A^u$ is also a $\g$-valued 1-form. In components, the above definition of $A^u$ reads
\begin{equation}
A^u_\mu = uA_\mu u^{-1} - (\p_\mu u) u^{-1},
\end{equation}
which we recognise as the standard law of transformation of a gauge field in a non-abelian gauge theory.

Let us note that in the case where the manifold $M$ has a boundary $\p M$ and thus where we imposed a boundary condition $A_\mu|_{\p M}=0$ (see above), one has to restrict the gauge transformations $A \mapsto A^u$ that one considers to the ones that preserve this boundary condition. This requires the term $(\p_\mu u)u^{-1}$ to vanish on the boundary: thus, we restrict to gauge transformations which do not depend on the coordinate $x^\mu$ on the boundary\footnote{More precisely, one standardly defines the true physical gauge transformations by imposing the stronger condition that $u$ is equal to the identity on the boundary. The remaining transformations that satisfy $(\p_\mu u)u^{-1}|_{\p M}=0$ but not $u|_{\p M}=\text{Id}$ are then interpreted as symmetries of the boundary term itself.}. Similar considerations will play an important role in the study of the 4-dimensional Cherns-Simons theory in the next sections.

\paragraph{Invariance of the equations of motion.} Let us now discuss in which sense the Chern-Simons theory is invariant under the gauge transformations $A \mapsto A^u$. It is a standard result from gauge theory that the curvature of $A$ is a covariant quantity under such transformations, \textit{i.e.} that
\begin{equation}\label{Eq:CovF}
F(A^u) = u \,F(A)\,u^{-1}.
\end{equation}
The equations of motion $F(A)=0$ are thus clearly invariant under these gauge transformations.\\

\begin{tcolorbox}
\textit{\underline{Exercise 2:} Covariance of the curvature} {\Large$\;\;\star$} \vspace{4pt}\\
Check Equation \eqref{Eq:CovF} using differential forms, the product rule \eqref{Eq:ProductRule} and the Maurer-Cartan identity \eqref{Eq:MC} for $u$.
\end{tcolorbox}~

Recall that a standard class of solutions of the equation of motion $F(A)=0$ is given by $A=g^{-1}\dd g$, where $g: M \rightarrow G$ is a smooth $G$-valued function -- see Equation \eqref{Eq:PureGauge}. One easily checks that such a solution is gauge equivalent to a vanishing gauge field, since $A^g = 0$. For this reason, we will call these solutions \textit{pure gauge}.

\paragraph{Gauge transformation of the Chern-Simons 3-form.} A natural question at this point is to check whether the Chern-Simons action itself is invariant under gauge transformations (which would in particular be consistent with the invariance of the equations of motion). A first step for that is to compute the transformation of the density of the action, \textit{i.e.} the Cherns-Simons 3-form. From the expression \eqref{Eq:CSForm} of the latter, one finds that this transformation is given by
\begin{equation}\label{Eq:GaugeCS}
\CS(A^u) = \CS(A) + \dd \, \Tr\bigl( u^{-1} \dd u \wedge A \bigr) + \frac{1}{3} \Tr\bigl( u^{-1} \dd u \wedge u^{-1} \dd u \wedge u^{-1} \dd u \bigr).\vspace{2pt}
\end{equation}

\begin{tcolorbox} \textit{\underline{Exercise 3:} Gauge transformation of the Chern-Simons 3-form} {\Large$\;\;\star\star$} \vspace{4pt}\\
Check Equation \eqref{Eq:GaugeCS}. To do so, use the invariance of the trace under conjugacy and the Maurer-Cartan identity \eqref{Eq:MC} for $u$.
\end{tcolorbox}

\paragraph{Gauge invariance of the action.} Let us finally consider the gauge transformation of the action \eqref{Eq:Action3d}, using the transformation \eqref{Eq:GaugeCS} of its density. The second term in \eqref{Eq:GaugeCS} contributes to a boundary term in $S[A^u]$, which vanishes due to the boundary conditions imposed on the gauge field and the transformation parameter $u$. We then find
\begin{equation}
S[A^u] = S[A] + \frac{k}{12\pi} \iiint_M \Tr\bigl( u^{-1} \dd u \wedge u^{-1} \dd u \wedge u^{-1} \dd u \bigr).
\end{equation}
We recognise in the second integral the Wess-Zumino term of $u$, which we also encountered in Ben Hoare's lectures \textit{``Integrable deformations of sigma models''}~\cite{LectureHoare}. As also explained in these lectures, the 3-form
\begin{equation}
\Lambda(u) = \frac{1}{12\pi} \Tr\bigl( u^{-1} \dd u \wedge u^{-1} \dd u \wedge u^{-1} \dd u \bigr)
\end{equation}
is closed (\textit{i.e.} $\dd \Lambda(u)=0$) but in general not exact (\textit{i.e.} not equal to $\dd$ of some 2-form). Thus, this contribution is in general not a boundary term and the action is strictly speaking not gauge invariant. However, at the classical level, this closed term does not affect the equation of motion, consistently with the observation above. Moreover, at least for $G$ a connected and simply-connected compact group, one has
\begin{equation}
\iiint_M \Lambda(u) = 2\pi n, \qquad \text{ where } \qquad n\in \mathbb{Z}
\end{equation}
is an integer called the winding number of $u$. In particular, we find that the coefficient $e^{i S[A]}$ appearing in the path-integral formulation of the theory is invariant under gauge transformations, \textit{i.e.} $e^{i S[A^u]}=e^{i S[A]}$, if the level $k$ is an integer. Thus, under this quantisation condition on $k$, the theory is \textit{gauge invariant} at the quantum level.

\subsection{Comments on the 3-dimensional Chern-Simons theory}

Let us end this section by a few general comments on the properties of the 3d-CS theory. Some of these properties are at the basis of deep and important further developments in the study of this model: however, since they will not be necessary to discuss the main topics of these lectures, we will not present an exhaustive description of these aspects here and will instead briefly sketch some of the main ideas. We refer to the extensive literature on these subjects for more details.

\paragraph{Topological field theory.} The formulation of the 3d-CS theory in this section was based on a description in terms of differential forms and integration thereof. By construction, this makes this formulation independent of the choice of coordinates on the manifold $M$. The 3d-CS theory is thus invariant under diffeomorphisms of the space-time $M$, if we also transform the gauge field $A$ as a 1-form. An important observation here is that this diffeomorphisms invariance was achieved without having to introduce a dynamical metric on the space-time manifold $M$, as one would do for instance in General Relativity. Models with this property of being independent of a background metric are called \textit{topologial field theories}\footnote{More precisely, the 3d-CS theory is a topological field theory of Schwarz-type.}. They take their name from the fact that physical observables of these theories typically measure topological invariants of the underlying space-time manifold. For instance, in his seminal work~\cite{Witten:1988hf}, Witten showed that the correlation functions of Wilson loops in the 3d-CS theory compute knot invariants of the manifold $M$, such as Jones polynomials.

\paragraph{Absence of bulk local degrees of freedom.} As discussed in Subsections \ref{Sec:EoM3d} and \ref{Sec:Gauge3d}, every solution of the equation of motion $F(A)=0$ of the 3d-CS theory can be locally written as a pure gauge solution $A=g^{-1}\dd g$. By a gauge transformation, such solutions can always be brought back to a trivial field $A^g=0$. This shows that the theory does not possess any local degrees of freedom in the bulk (or in other words that there are no propagating fields in the bulk). However, the theory is in general still non-trivial. For instance, it can possess global degrees of freedom, such as Wilson loops, which in particular can encode topological informations on the manifold $M$.

\paragraph{Boundaries and WZW model.} The 3d-CS theory can also possess boundary degrees of freedom. In the previous subsections, we have discussed briefly some aspects related to the boundary $\p M$ of the space-time manifold $M$, if this boundary is non-trivial. In particular, we have sketched in Subsection \ref{Sec:EoM3d} that the well-posedness of the action principle requires the introduction of appropriate boundary conditions on the gauge field on $\p M$. Moreover, we have mentioned in Subsection \ref{Sec:Gauge3d} that the gauge transformations $A \mapsto A^u$ of the theory have to be restricted to transformations that preserve these boundary conditions on $A$. As a consequence, there can exist degrees of freedom on the boundary $\p M$ that cannot be gauged away, contrarily to the bulk degrees of freedom discussed in the previous paragraph. In this case, the theory is then described by an effective 2-dimensional model on $\p M$. The typical example of such a phenomena was also put forward by Witten in his seminal work~\cite{Witten:1988hf}, where the resulting 2-dimensional theory is a chiral Wess-Zumino-Witten model on $\p M$ (in fact, the analysis of~\cite{Witten:1988hf} also extends this correspondence to the quantum level and in particular to the description of the Hilbert space of the Chern-Simons theory). As we will see in more details in Sections \ref{Sec:Poles}, \ref{Sec:PCM} and \ref{Sec:MoreBC}, similar mechanisms will appear in the study of the 4-dimensional Chern-Simons theory.

\paragraph{Towards a relation with integrability.} The attentive reader might have observed that the equation of motion of the 3d-CS theory, namely the zero curvature equation $\dd A + A \wedge A = 0$ of the gauge field, shares some formal similarity with the Lax equation $\dd \Lc(z) + \Lc(z) \wedge \Lc(z) = 0$ characterising integrable 2-dimensional field theories (see the brief reminder in Subsection \ref{Sec:Lax}). One can thus wonder whether there exists a deeper relation and whether one could for instance use the 3d-CS theory to generate 2-dimensional integrable systems. There are however some important differences in the two zero curvature equations under consideration. For instance, the Lax connection $\Lc(z)$ is a 1-form on the 2-dimensional worldsheet $\Sigma$, whereas the Chern-Simons gauge field $A$ is a 1-form on the 3-dimensional manifold $M$: relating the two would then require understanding a passage from 3 to 2 dimensions. This can be acheived by restricting our attention to gauge fields which have a vanishing component along one direction (up to a gauge transformation). A more serious difference is that the Lax connection $\Lc(z)$ depends rationally on the auxiliary complex parameter $z\in \CP$, the spectral parameter. It is unclear how this complex parameter would appear in the context of the 3d-CS theory. As we shall see in the rest of these lectures, the passage from the 3d-CS theory to its ``semi-holomorphic'' 4-dimensional variant is exactly what will allow us to make this relation concrete, by incorporating the spectral parameter $z\in\CP$ into the space-time of the theory.

\section{The 4-dimensional Chern-Simons theory}
\label{Sec:4d}

In this section, we introduce the semi-holomorphic 4-dimensional Chern-Simons theory, which we abbreviate as 4d-CS from now on.

\subsection{Space-time, gauge field and action}
\label{Sec:Action4d}

\paragraph{Space-time.} We define the space-time of the theory as
\begin{equation}
M = \Sigma \times \CP,
\end{equation}
where $\Sigma = \R \times \D$, with $\D=\R$ or $\D=S^1$. We will describe $M$ through the coordinates $(t,x,z,\zb)$, where $(t,x)$ are real coordinates on $\Sigma = \R \times \D$, $z$ is a complex coordinate on $\CP$ and $\zb$ is its complex conjugate. We will later interpret $(t,x)$ as the space-time coordinates of an integrable 2-dimensional field theory on $\Sigma$ and $z$ as the corresponding auxiliary spectral parameter. In particular, this construction gives a geometrical interpretation of the spectral parameter.

\paragraph{Gauge field.} The fundamental field of the model is a gauge field $A\in \Omega(M,\g^{\C})$, defined as a $\g^{\C}$-valued 1-form on $M$:
\begin{equation}
A = A_t \, \dd t + A_x \, \dd x + A_z \, \dd z + A_\zb \, \dd \zb.
\end{equation}
In this definition, we take $\g^{\C}$ to be a semi-simple complex Lie algebra (in a matrix representation). In general, one has to supplement this definition with an appropriate choice of reality conditions on $A$, ensuring that the physical quantities that we will study in this theory are real. For simplicity, we will postpone the discussion of these reality conditions to Subsection \ref{Sec:CommentsLax}, to avoid having to describe these conditions each time we introduce a new object.

\paragraph{Meromorphic 1-form $\bm{\omega}$ and action.} Let us consider the Chern-Simons 3-form
\begin{equation}
\CS(A) = \Tr \left( A \wedge \dd A + \frac{2}{3} A \wedge A \wedge A \right),
\end{equation}
defined from $A$ as in the 3d-CS theory. In the present case, this is a 3-form on the 4-dimensional manifold $M$. Thus, we cannot integrate $\CS(A)$ over $M$ and therefore cannot define the action of the 4-dimensional theory exactly as we defined the 3-dimensional one. Instead, we will define the action as~\cite{NekrasovThesis,Costello:2013zra,Costello:2019tri}
\begin{equation}\label{Eq:Action4d}
S[A] = \frac{i}{4\pi} \iiiint_M \omega \wedge \CS(A),
\end{equation} 
where
\begin{equation}
\omega = \vp(z) \dd z
\end{equation}
is a meromorphic 1-form in the complex parameter $z\in\CP$. This 1-form should be seen as a non-dynamical quantity and thus as part of the data defining the theory.

We will suppose that $\omega$ has a finite number of poles in $\CP$, so that $\vp(z)$ is a rational function of $z$. We will refer to $\vp(z)$ as the \textit{twist function}. We denote by $z_1,\cdots,z_N$ the finite poles of $\omega$ and by $m_1,\cdots,m_N \in \Z_{\geq 1}$ their order\footnote{Note that, when $\omega$ possesses poles of order strictly greater than 1, one needs to regularise the action \eqref{Eq:Action4d} to define it rigorously. We refer to~\cite{Benini:2020skc} for more details about this aspect.}. Moreover, we denote by $m_\infty\in \Z_{\geq 0}$ the order of the pole of $\omega$ at $z=\infty$ (with $m_\infty=0$ if $\omega$ is regular at infinity). Finally, we define $\Zc \subset \CP$ as the set of poles of $\omega$.

\paragraph{Decoupling of $\bm{A_z}$.} Let us consider the action \eqref{Eq:Action4d}. Since $\omega=\vp(z)\dd z$ is a 1-form in the $z$-direction, it is clear that any term containing the component $A_z$ in the Chern-Simons form $\CS(A)$ will vanish in $\omega \wedge \CS(A)$. Thus the field $A_z$ does not appear at all in the action $S[A]$, or in other words the theory is invariant under a shift of $A_z$ by any function. The component $A_z$ therefore completely decouples from the theory and for simplicity we can then put it to zero. We thus write the gauge field $A$ as
\begin{equation}
A = A_t \, \dd t + A_x \, \dd x + A_\zb \, \dd \zb.
\end{equation}
Note that, by the same argument, the derivatives of $A$ with respect to $z$ do not appear in the action \eqref{Eq:Action4d}. Working explicitly in coordinates, we can then write the action as
\begin{equation}
S[A] = \frac{i}{4\pi}  \iint_\Sigma \dd t\,\dd x \iint_{\CP} \dd z\,\dd \zb \; \vp(z) \sum_{\mu,\nu,\rho \in\lbrace t,x,\zb \rbrace} \epsilon^{\mu\nu\rho} \, \Tr \left( A_\mu \p_\nu A_\rho + \frac{1}{3} A_\mu [A_\nu, A_\rho] \right),
\end{equation}
where we write explicitly the sum over the indices $\mu,\nu,\rho$ to stress the fact that this sum can be taken over $t$, $x$ and $\zb$ only.

\subsection{Equation of motion and formal gauge transformations}
\label{Sec:EoM4d}

\paragraph{Bulk equation of motion.} Let us consider an infinitesimal variation $\delta A$ of the gauge field. We have computed in Equation \eqref{Eq:dCS} the corresponding variation of the Chern-Simons 3-form. Taking the exterior product with $\omega$ and using the product rule \eqref{Eq:ProductRule}, we get
\begin{equation}
\delta\, \bigl( \omega \wedge \CS(A) \bigr) = 2 \, \Tr \bigl( \omega \wedge F(A) \wedge \delta A \bigr) - \dd \omega \wedge \Tr\bigl( A \wedge \delta A \bigr) + \dd \Bigl( \omega \wedge \Tr\bigl( A \wedge \delta A \bigr) \Bigr).
\end{equation}
Integrating over $M$, we
get the variation of the action \eqref{Eq:Action4d}. The last term is a total derivative and thus does not contribute to the integral\footnote{Recall that we have $M=\Sigma \times \CP$. $\CP$ does not have boundaries, while $\Sigma$ has a boundary at space-time infinity: we thus ensure that the total derivative term does not contribute by requiring that the gauge field $A$ vanishes at this space-time infinity.}. We thus get
\begin{equation}\label{Eq:VariationS}
\delta S[A] = \frac{i}{2\pi} \iiiint_M \Tr \bigl( \omega \wedge F(A) \wedge \delta A \bigr) - \frac{i}{4\pi} \iiiint_M \dd \omega \wedge \Tr\bigl( A \wedge \delta A \bigr).
\end{equation}
Let us consider the second term. Recall that $\omega = \vp(z) \dd z$ is a 1-form on the $\CP$ part of $M$. Thus, $\vp(z)$ has no derivatives with respect to the coordinates $(t,x)$ of $\Sigma$. Moreover, since $\omega$ is proportional to $\dd z$, we simply have $\dd \omega = \p_{\bar z} \, \vp(z) \dd \zb \wedge \dd z$. As $\omega=\vp(z) \dd z$ is meromorphic in $z$, one can naively expect that its derivative with respect to $\zb$ vanishes and thus that $\dd \omega=0$: this is in fact true only on $\CP\setminus \Zc$, \textit{i.e.} away from the poles $\Zc$ of $\omega$. In fact, as we shall explained in more details below and in Appendix \ref{App:Delta}, the $\zb$-derivative of a function with a pole at a point $z_0 \in \CP$ creates a distribution localised at $z_0$. Thus, $\dd \omega$ is a distribution localised on $\Sigma \times \Zc \subset \Sigma \times \CP$. The second term in \eqref{Eq:VariationS} is then a 2-dimensional term on the space $\Sigma \times \Zc$ associated with the poles of $\omega$, which we will call the \textit{poles defect}. The detailed treatment of this defect term requires the introduction of appropriate boundary conditions on $A$ along $\Sigma \times \Zc$ and will play an important role in the description of the theory: these aspects will be the main subject of Section \ref{Sec:Poles}.

For the moment, let us focus on the first term in the variation \eqref{Eq:VariationS} of the action. This is a bulk term in $M$ and it should then vanish independently of the defect term when we impose the stationarity of the action. We thus obtain the following \textit{bulk equation of motion}:
\begin{equation}\label{Eq:EoM4d}
\omega \wedge F(A) = \omega \wedge ( \dd A + A \wedge A ) = 0.
\end{equation}
As in the 3-dimensional case, this equation of motion takes the form of the zero curvature of the gauge field $A$ (at least, in the present 4-dimensional case, on the points where $\omega$ is non-zero). Note however that this is a zero curvature with respect to the coordinates $(t,x,\zb)$ only: indeed, the presence of $\omega$ removes any $z$-derivative and the $z$-component of the gauge field $A$.

\paragraph{Formal gauge transformations.} Let us consider the \textit{``formal'' gauge transformations}
\begin{equation}
A^u = u A u^{-1} - (\dd u)u^{-1}
\end{equation}
of the gauge field, where $u: M \rightarrow G^{\C}$ is a smooth function on $M$ valued in the connected and simply-connected group $G^{\C}$ with Lie algebra $\g^{\C}$. The terminology ``formal'' introduced here will be justified in more details in Sections \ref{Sec:Poles} to \ref{Sec:MoreBC} and has to do with the fact that the true physical gauge symmetries of the theory are only the ones preserving the boundary conditions that we will impose on $A$ on the poles defect $\Sigma \times \Zc$.

Recall from Equation \eqref{Eq:CovF} that the curvature $F(A)$ is covariant under such transformations, \textit{i.e.} $F(A^u)=u F(A) u^{-1}$. Thus, the equation of motion \eqref{Eq:EoM4d} of the theory is invariant under these transformations. The gauge invariance of the action \eqref{Eq:Action4d} itself is a more subtle question and also requires an appropriate treatment of boundary conditions.

\subsection{Comments on the 4-dimensional Chern-Simons theory}

Let us make a few final comments on the content of this section. We note that other higher-dimensional variants of the 3d-CS theory have been considered before. They naturally generalise the 3-dimensional setting as they define gauge theories on $d$-dimensional manifolds, depending on a $d$-dimensional gauge field $A$ and whose equations of motion are the vanishing of the curvature of $A$. However, such theories exist only for odd dimensions $d$. The 4-dimensional variant considered here is of a different nature: in particular, although the equation of motion of this theory is also the flatness of a gauge field $A$, the presence of the non-dynamical 1-form $\omega=\vp(z)\dd z$ in the action makes the $z$-derivatives and the component $A_z$ of this gauge field decouple and results in a vanishing curvature along the components $(t,x,\zb)$ only.

Recall moreover that the 3d-CS theory is a topological field theory, which is invariant under diffeomorphisms of the underlying space-time manifold without the introduction of a background metric. In the case of the 4d-CS theory, the presence of the 1-form $\omega = \vp(z) \dd z$ in the action \eqref{Eq:Action4d} makes the theory non-topological. In fact, this 4-dimensional theory is topological along the direction $\Sigma$ and holomorphic along the direction $\CP$ (indeed, the solutions $A$ define holomorphic bundles over $\CP$ as they are flat connections with no $z$-component). For this reason, this theory is sometimes refered to as the \textit{semi-holomorphic} 4-dimensional Chern-Simons theory.

\section{Relation with the Lax formalism of 2d integrable field theories}
\label{Sec:4dLax}

In this section, we explain how the 4d-CS theory defined in the previous section is naturally related to the Lax formalism of 2d integrable field theories.

\subsection{Fields \texorpdfstring{$\bm{\gh}$}{gh} and \texorpdfstring{$\bm{\Lc}$}{L}}
\label{Sec:gL}

Let us consider the gauge field $A$ of the model, satisfying the bulk equation of motion \eqref{Eq:EoM4d}. We will restrict to the solutions $A$ for which there exists a formal gauge transformation $A \mapsto A^{\gh^{-1}}$, for some function\footnote{We follow here the conventions of~\cite{Delduc:2019whp}. To ease the comparison with other works in the literature, let us note that $\gh$ coincides with $\widehat{\sigma}^{-1}$ in the notation of~\cite{Costello:2019tri}.} $\gh : M \rightarrow G^{\C}$, \textit{such that $A^{\gh^{-1}}$ has no $\zb$-component}\footnote{Let us note that this formal gauge is similar to the Weyl gauge or axial gauge in electrodynamics and Yang-Mills theories, where one of the component of the gauge field is set to vanish.}. We will then denote by $\Lc$ the corresponding field, which takes the form
\begin{equation}
A^{\gh^{-1}} = \Lc = \Lc_t \, \dd t + \Lc_x\,\dd x.
\end{equation}
As we shall see below, we will interpret $\Lc$ as the Lax connection of an integrable 2d field theory on $\Sigma$. This is the reason why we restrict to this particular class of solutions, as such a relation with integrable field theories does not hold for solutions outside of this class. Concretely, the assumption that such a formal gauge transformation exists is equivalent to writing the component $A_{\zb}$ of the field in the form
\begin{equation}\label{Eq:Azbar}
A_{\zb} = - (\p_{\zb} \gh) \gh^{-1}.
\end{equation}
The two remaining components of $A$ are then given by
\begin{equation}\label{Eq:ALax}
A_\mu = \gh \Lc_\mu \gh^{-1} - (\p_\mu \gh) \gh^{-1}, \qquad \mu = t,x.
\end{equation}
We can thererefore see the above assumption as a parametrisation of the gauge field $A$ in terms of a $G^{\C}$-valued field $\gh$ and two $\g^{\C}$-valued fields $\Lc_t$ and $\Lc_x$.

\subsection{Lax connection}
\label{Sec:Lax4Db}

\paragraph{Interpreting $\bm{\Lc}$ as a Lax connection.} Since the formal gauge transformation $A \mapsto A^{\gh^{-1}}$ preserves the bulk equation of motion \eqref{Eq:EoM4d}, the field $A^{\gh^{-1}}=\Lc$ also satisfies this equation, \textit{i.e.}
\begin{equation}\label{Eq:EoMLax}
\omega \wedge F(\Lc) = \omega \wedge ( \dd \Lc + \Lc \wedge \Lc ) = 0.
\end{equation}
Let us decompose this equation of motion in two parts. We first project it on $\dd z \wedge \dd \zb \wedge \dd t$ and $\dd z \wedge \dd \zb \wedge \dd x$. Since $\Lc$ has no $\zb$-component, this part simply reads
\begin{equation}\label{Eq:EoMLZbar}
\omega\, \p_{\zb} \Lc_\mu = 0, \qquad \text{ for } \mu = t,x.
\end{equation}
This proves that $\Lc$ is holomorphic in $z$, at least on $\Sigma \times \CP$ from which we removed the zeroes of $\omega$. More precisely, as we will see in more details in the next paragraph, this implies that $\Lc$ is meromorphic in $z$, with poles at the zeroes of $\omega$.

Let us now consider the projection of the equation of motion \eqref{Eq:EoMLax} on $\dd z \wedge \dd t \wedge \dd x$. Away from the zeroes of $\omega$, it yields
\begin{equation}
\p_t \Lc_x - \p_x \Lc_t + [\Lc_t, \Lc_x] = 0.
\end{equation}
This is nothing but the zero curvature equation of $\Lc$ on the two-dimensional space-time $\Sigma$. To summarise, we have that $\Lc$ is a 1-form along $\Sigma$, \textit{i.e.} with components only in $\dd t$ and $\dd x$ and
\begin{enumerate}[(i)]\setlength\itemsep{0.2em}
\item $\Lc$ depends meromorphically in $z$,
\item $\Lc$ satisfies $d_\Sigma \Lc + \Lc \wedge \Lc = 0$,
\end{enumerate}
where $\dd_\Sigma$ denotes the exterior derivative of $\Sigma$, along the coordinates $t$ and $x$ only. These are exactly the properties characterising the \textit{Lax connection of an integrable 2d field theory on $\Sigma$}. In particular, we see that the complex coordinate $z$ plays the role of the \textit{spectral parameter} of this 2d integrable field theory. Let us note that the meromorphicity of $\Lc$ is a direct consequence of working in a formal gauge where the $\bar z$-component of the gauge field vanishes: this shows that the assumption, made in Subsection \ref{Sec:gL}, that such a formal gauge exists is indeed necessary to establish a relation with integrable field theories.

\paragraph{Zeroes of $\bm{\omega}$ and disorder defects.} Let us denote by $\Ze$ the set of zeroes of $\omega$. For simplicity, we will suppose here that these zeroes are all finite, \textit{i.e.} $\Ze \subset \C$, and simple. We can then rewrite $\omega$ as
\begin{equation}
\omega = -K \frac{\prod_{y\in \Ze} (z-y)}{\prod_{r=1}^N (z-z_r)^{m_r}},
\end{equation}
for some constant number $K$.

As mentioned in the previous paragraph, the equation of motion \eqref{Eq:EoMLZbar} for $\Lc$ along the $\zb$-component implies that $\Lc$ is meromorphic in $z$, with poles in $\Ze$. To explain this statement more precisely, let us consider the meromorphic function $z \mapsto 1/(z-y)$ with $y$ a fixed point in $\C$. On $\CP\setminus\lbrace y \rbrace$, this function is holomorphic and thus has a vanishing derivative with respect to $\zb$. This $\zb$-derivative can be extended on the full Riemann sphere $\CP$ as a distribution with support $\lbrace y \rbrace$. More precisely, it is a standard result from complex analysis that
\begin{equation}\label{Eq:DerPoleDirac}
\p_{\zb} \left( \frac{1}{z-y} \right) = -2i\pi \, \delta^{(2)}(z-y),
\end{equation}
where $\delta^{(2)}(z-y)$ is the Dirac-distribution on $\CP$ centred in $y$, defined with respect to the volume form $\dd z \wedge \dd \zb$, which then satisfies
\begin{equation}
\iint_{\CP} f(z,\zb)\, \delta^{(2)}(z-y)\; \dd z \wedge \dd \zb = f(y,\bar y)
\end{equation}
for any function $f$. For completeness, we include a derivation of this identity in Appendix \ref{App:Delta}. Using this result, we see that for $\Lc_\mu$ to have a simple pole at a point $y\in\C$, the equation $\omega\,\p_{\zb}\Lc_\mu=0$ requires that $\omega$ has a zero at $y$. More generally, the $\zb$-derivative of $1/(z-y)^p$, for $p\in\Z_{\geq 1}$, is proportional to the $p$-th derivative of $\delta^{(2)}(z-y)$ with respect to $z$. A pole of order $p$ at $y$ in $\Lc_\mu$ would thus require that $\omega$ has a zero of order at least $p$ at $y$. Since we suppose that $\omega$ has only simple zeroes here, we get that $\Lc$ has simple poles at these points, hence
\begin{equation}
\Lc = \sum_{y\in \Ze} \frac{\Lc^{(y)}}{z-y} + V,
\end{equation}
where the $\Lc^{(y)}$'s and $V$ are 1-forms on $\Sigma$, independent of $z$. These singularities of $\Lc$, and thus ultimately of the gauge field $A=\Lc^{\gh}$, at the points $\Ze$ are called \textit{disorder defects}.

Recall that $\Lc$ will satisfy the Lax equation $\dd_\Sigma \Lc + \Lc \wedge \Lc = 0$. The term $\Lc \wedge \Lc$ in this equation contains a singular term $\Lc^{(y)} \wedge \Lc^{(y)} / (z-y)^2$ at each point $y\in \Ze$. Since $\dd_\Sigma \Lc$ has only a pole of order 1 at $y$, the term $\Lc^{(y)} \wedge \Lc^{(y)}$ must vanish on its own. We will ensure that this is the case by imposing\footnote{We refer to~\cite{Costello:2019tri,Delduc:2019whp,Benini:2020skc} for more details on these assumptions and in particular to the concept of admissible solution in~\cite{Benini:2020skc}.} that $\Lc^{(y)}$ is along a single direction in $\Sigma$, i.e. proportional to some $\dd(\alpha_y t + \beta_y x)$, so that the exterior product of $\Lc^{(y)}$ with itself vanishes. The coefficients $\alpha_y$ and $\beta_y$ defining this direction can in principle be taken as arbitrary: different choices will result into different integrable 2d models. However, for these models to be relativistic, one has to restrict (see~\cite{Costello:2019tri,Delduc:2019whp,Lacroix:2020flf}) the directions $\alpha_y t + \beta_y x$ of the disorder defects to be along either one of the two light-cone coordinates
\begin{equation}
x^\pm = \frac{t\pm x}{2}.
\end{equation}
Following the terminology of~\cite{Costello:2019tri}, we then call these \textit{chiral and anti-chiral disorder defects}. This effectively separates the zeroes $\Ze$ of $\omega$ into two disjoint subsets $\Ze_\pm$, such that the chiral disorder defects $\Lc^{(y)}$ for $y\in \Ze_+$ are along $\dd x^+$ and the anti-chiral disorder defects $\Lc^{(y)}$ for $y\in \Ze_-$ are along $\dd x^-$. In this case, the $z$-dependence of the Lax connection is completely fixed and reads
\begin{equation}\label{Eq:Lz}
\Lc = \left( \sum_{y \in \Ze_+} \frac{U_+^{(y)}}{z-y} + V_+ \right) \dd x^+ + \left( \sum_{y \in \Ze_-} \frac{U_-^{(y)}}{z-y} + V_- \right) \dd x^-,
\end{equation}
where $U_\pm^{(y)}$ and $V_\pm$ are $\g^{\C}$-valued fields on $\Sigma$ (independent of $z$).

\paragraph{Comments on defects.} Let us make a quick comment on terminology. In this section and the previous one, we have considered two different concepts that we designated under the name of defects, as they are located on 2-dimensional subspaces of the 4-dimensional manifold $M=\Sigma \times \CP$. We insist however on the different nature and role of these two objects. The first is the poles defect $\Sigma \times \Zc$ associated with the poles of $\omega$, which, as we shall see later, will be directly related to the fundamental fields of the resulting 2d field theory. The second one is the disorder defect $\Sigma \times \Ze$, associated with the zeroes of $\omega$, which as we just saw is related to the singularities of the Lax connection $\Lc$ and thus of the gauge field $A$.

\subsection{Freedom in the choice of \texorpdfstring{$\bm{\gh}$}{gh}}
\label{Sec:FreedomGh}

The field $\gh:M\rightarrow G^{\C}$ that parametrises $A_{\zb}$ as \eqref{Eq:Azbar} plays an important role in this subsection and the following ones. It is however not uniquely defined. Let us then discuss the freedom that one has in defining it. By construction, the parametrisation $A_{\zb} = - (\p_{\zb} \gh) \gh^{-1}$ is invariant under the redefinition
\begin{equation}
\gh \longmapsto \gh v, \qquad \text{ where } \qquad v: \Sigma \longrightarrow G^{\C}
\end{equation}
is a $G^{\C}$-valued function on $\Sigma$ and thus is independent of $z$ and $\zb$. Indeed, we clearly also have $A_{\zb} = - (\p_{\zb} \gh v) (\gh v)^{-1}$. This freedom can for instance be used to fix the value of $\gh$ at any point $z\in\CP$. This redifinition does not change the other components \eqref{Eq:ALax} of the gauge field $A$ if one also changes accordingly the Lax connection $\Lc$ as (see exercise below)
\begin{equation}
\Lc \longmapsto \Lc^v = v^{-1}\Lc v + v^{-1} \dd_\Sigma v.
\end{equation}
This transformation of $\Lc$ is a 2d gauge transformation on $\Sigma$ and reflects the non-uniqueness of the Lax connection of an integrable field theory: indeed, the Lax equation $F(\Lc)=0$ is clearly invariant under this transformation.

The transformation $(\gh,\Lc) \mapsto (\gh v,\Lc^v)$ considered above corresponds to a different choice of parametrisation of the same gauge field $A$. Recall however that the 4d-CS theory is invariant under gauge transformations $A \mapsto A^u$ of the gauge field (if these transformations preserve the boundary conditions imposed on $A$, as we shall discuss in Sections \ref{Sec:Poles}, \ref{Sec:PCM} and \ref{Sec:MoreBC}). One checks (see exercise below) that such a gauge transformation of $A$ amounts to the following redefinition of $\gh$:
\begin{equation}
\gh \longmapsto u\gh, \qquad \text{ where } \qquad u: M \longrightarrow G^{\C},
\end{equation}
while keeping the Lax connection $\Lc$ invariant. In conclusion, one can redefine the field $\gh$ and the Lax connection $\Lc$ by
\begin{equation}
(\gh,\Lc) \longmapsto (u \gh v, \Lc^v),
\end{equation}
where $v:\Sigma\rightarrow G^{\C}$ is independent of $z$ and $u:M \rightarrow G^{\C}$ can depend on $z$ (but should preserve the boundary conditions imposed on $A$).\vspace{7pt}

\begin{tcolorbox} \textit{\underline{Exercise 4:} Freedom in the choice of $\gh$.} {\Large$\;\;\star$} \vspace{4pt}\\
Explicitly check that the transformation $(\gh,\Lc) \mapsto \bigl( \gh v,\Lc^v \bigr)$, with $v: \Sigma \rightarrow G^C$ independent of $z$ and $\zb$, preserves the gauge field $A$.\\
Explicitly check that the transformation $(\gh,\Lc) \mapsto \bigl( u\gh,\Lc \bigr)$, with $u: M \rightarrow G^{\C}$, amounts to a gauge transformation $A\mapsto A^u$ of the gauge field.
\end{tcolorbox}

\subsection{Various comments}
\label{Sec:CommentsLax}

\paragraph{Reality conditions.} Let us quickly discuss the reality conditions that one should impose on the theory, starting with the 1-form $\omega=\vp(z)\dd z$. We require that it is equivariant under complex conjugation, \textit{i.e.} that
\begin{equation}
\overline{\vp(z)} = \vp\bigl( \zb ).
\end{equation}
More concretely, this condition is equivalent to the following requirements. If $z_r \in \Zc$ is a pole of $\omega$, it is either real or coming in a pair of complex conjugate poles, \textit{i.e.} $\overline{z_r}$ also belongs to $\Zc$ and thus $\overline{z_r}=z_{\bar r}$ for some $\bar r \in \lbrace 1,\cdots,N\rbrace$, with the same order $m_{\bar r}=m_r$. We note that these reality conditions also imply that the zeroes $\Ze$ of $\omega$ are either real or in pairs of complex conjugate numbers. For simplicity, we will suppose in the rest of these lecture notes that they are all real.

Let us now discuss the reality conditions imposed on the gauge field $A$. For that, we choose a real form $\g$ of the complex Lie algebra $\g^{\C}$, \textit{i.e.} a real Lie algebra whose complexification is $\g^{\C}$ (for instance $\g=\mathfrak{su}(N)$ for $\g^{\C}=\mathfrak{sl}(N,\C)$). There exists an involutive antilinear automorphism $\tau: \g^{\C} \rightarrow \g^{\C}$ whose fixed points form the real algebra $\g$, \textit{i.e.} $\g=\bigl\lbrace X\in\g^{\C} \, \bigl| \, \tau(X)=X \bigr\rbrace$ (for instance, in the example $\g=\mathfrak{su}(N)$ and $\g^{\C}=\mathfrak{sl}(N,\C)$ above, we have $\tau(X)= -X^\dagger$). We then require that the components $A_\mu$, $\mu=t,x,\zb$, of $A$ are equivariant functions of $z$ under the complex conjugation on $\CP$ and the action of $\tau$ on $\g^C$:
\begin{equation}
\tau\bigl( A_\mu(t,x,z) \bigr) = A_\mu(t,x, \zb ).
\end{equation}
Together with the above condition on $\omega$, this property ensures that the action \eqref{Eq:Action4d} is real (see~\cite{Delduc:2019whp}). We note that one should now also restrict the gauge transformations $A\mapsto A^u$ to the ones preserving these reality conditions, which require that $\tau\bigl(u(t,x,z)\bigr) = u(t,x,\zb)$, where $\tau$ now also denotes the lift of the antilinear automorphism to the group $G^{\C}$.

Let us finally comment on the reality conditions of the field $\gh$ and the Lax connection $\Lc$, which were introduced in this section to parametrise $A$. These quantities are also equivariant, \textit{i.e.} satisfy
\begin{equation}
\tau\bigl( \gh(t,x,z) \bigr) = \gh(t,x, \zb ) \qquad \text{ and } \qquad \tau\bigl( \Lc_\mu(t,x,z) \bigr) = \Lc_\mu(t,x, \zb ),
\end{equation}
where $\tau$ denotes the automorphism of the group $G^{\C}$ in the first equation and the automorphism of the Lie algebra $\g^{\C}$ in the second one. In particular, combining the reality condition of $\Lc$ with the assumption made above that the zeroes $y\in\Ze$ of $\omega$ are real, we find that the quantities $U_\pm$ and $V_\pm^{(y)}$ appearing in the partial fraction decomposition \eqref{Eq:Lz} of $\Lc$ are invariant under $\tau$ and thus belong to the real form $\g$.

\paragraph{Towards an integrable 2d model.} As we have argued above, the quantity $\Lc$ obtained by considering formal gauge transformations eliminating the $\zb$-component of the gauge field possesses the right properties to be the Lax connection of a 2d integrable field theory on $\Sigma$. The main question at this point is then ``what is this integrable theory''? In particular, what are its dynamical fields and its action and how is the Lax connection related to these dynamical fields. As we shall see, these 2d degrees of freedom are related to the defect $\Sigma \times \Zc$ originating from the poles of $\omega$ and their description requires an appropriate treatment of this defect and of the corresponding boundary conditions on the gauge field $A$. This is the subject of the next sections.

\paragraph{Hamiltonian aspects.} As briefly mentioned in the synopsis of integrable 2d field theories in Subsection \ref{Sec:Lax}, the existence of a Lax connection $\Lc(z)$ whose flatness is equivalent to the equations of motion of a model is only a first step in the proof of the integrability of this model, which allows the construction of an infinite number of conserved charges. Indeed, one also has to prove that these charges are in involution, \textit{i.e.} that their pairwise Poisson brackets are vanishing in the Hamiltonian formulation of the model. This is generally done by studying the Poisson bracket of the Lax matrix, \textit{i.e.} the spatial component $\Lc_x$ of the Lax connection. In particular, if this bracket takes the form of a Sklyanin bracket~\cite{Sklyanin:1982tf} or the more general non-ultralocal Maillet bracket~\cite{Maillet:1985fn,Maillet:1985ek}, then this property of involution is ensured.

For the Lax connection obtained from 4d-CS theory as in this section, such an Hamiltonian analysis has been performed in the reference~\cite{Vicedo:2019dej}. In particular, it was shown in this reference that the Poisson bracket of the corresponding Lax matrix is a Maillet bracket. Moreover, this bracket takes a specific form, which is controlled by the 1-form $\omega=\vp(z)\dd z$ defining the 4d-CS setup one starts with. More precisely, as shown in~\cite{Vicedo:2019dej}, the rational function $\vp(z)$ defining $\omega$ is identified with the so-called \textit{twist function}, which was previously shown to control the Hamiltonian structure of many integrable 2d field theories and in particular $\sigma$-models. The structure of integrable field theories with twist function has led to the reintrepretation of these models as so-called \textit{affine Gaudin models}~\cite{Feigin:2007mr,Vicedo:2017cge}. This approach also provides a systematic way to study and construct integrable field theories. In this context, the reference~\cite{Vicedo:2019dej} then shows that this formalism and the 4d-CS approach can be seen as two faces of the same formulation, one based on the Hamiltonian picture and the other on the Lagrangian one.

\section{Poles defect in 4-dimensional Chern-Simons theory}
\label{Sec:Poles}

\subsection{Defect terms in the variation of the action}
\label{Sec:Defect}

In this section, we discuss the treatment of defect terms in the derivation of the equations of motion of the 4d-CS theory. For that, we start by introducing necessary notations related to the poles of the meromorphic 1-form $\omega$.

\paragraph{Poles of $\bm{\omega}$.} Recall from Subsection \ref{Sec:Action4d} that the meromorphic 1-form $\omega$ has finite poles $z_r \in \C$, $r\in\lbrace 1,\cdots,N\rbrace$, with multiplicities $m_r\in \Z_{\geq 1}$ and a potential pole at infinity with multiplicity $m_\infty$ (with $m_\infty=0$ if $\omega$ is regular at infinity). It then takes the general form
\begin{equation}\label{Eq:OmegaLevels}
\omega = \left( \sum_{r=1}^N \sum_{p=0}^{m_r-1} \frac{\ell_{r,p}}{(z-z_r)^{p+1}} - \sum_{p=1}^{m_\infty-1} \ell_{\infty,p}\, z^{p-1} \right) \dd z, 
\end{equation}
where
\begin{equation}
\ell_{r,p}, \quad r\in\lbrace 1,\cdots,N \rbrace, \quad p \in \lbrace 0,\cdots,m_r-1 \rbrace \qquad \text{ and } \qquad \ell_{\infty,p}, \quad p \in \lbrace 1,\cdots,m_\infty-1 \rbrace
\end{equation}
are complex numbers, which we call the \textit{levels} of the theory. Recall here that we consider $\omega$ as a 1-form on $\CP$ and not as a function: in particular, this has important consequences on the pole structure of $\omega$ at $z=\infty$. Indeed, to study the 1-form $\omega$ around $z=\infty$, we perform the change of variable $\xi=1/z$. The presence of $\dd z$ in $\omega$ then introduces stronger divergences in $\xi$, since
\begin{equation}\label{Eq:dOverz}
\dd z = - \frac{\dd \xi}{\xi^2}.
\end{equation}
In particular, one checks (see exercise below) that the behaviour of $\omega$ around $z=\infty$, \textit{i.e.} around $\xi=0$, is given by
\begin{equation}\label{Eq:OmegaInfty}
\omega = \left( \sum_{p=0}^{m_\infty-1} \frac{\ell_{\infty,p}}{\xi^{p+1}} + O(\xi^0) \right) \dd \xi,
\end{equation}
where the numbers $\ell_{\infty,1}, \cdots, \ell_{\infty,m_\infty-1}$ are the ones entering the polynomial part of \eqref{Eq:OmegaLevels} and where $\ell_{\infty,0}$ is defined as
\begin{equation}
\ell_{\infty,0} = - \sum_{r=1}^N \ell_{r,0}.
\end{equation}
In particular, we see that the parametrisation of the polynomial part in \eqref{Eq:OmegaLevels} indeed corresponds to a pole at infinity in $\omega$ of order $m_\infty$.

Recall that we denoted $\Zc \subset \CP$ the set of poles of $\omega$. We also define $\Zc' = \lbrace z_1,\cdots,z_N \rbrace \subset \C$ as the set of its finite poles, so that $\Zc=\Zc'$ if $\omega$ is regular at $z=\infty$ and $\Zc=\Zc' \cup \lbrace \infty \rbrace$ otherwise.

Recall finally that reality conditions introduced in Subsection \ref{Sec:CommentsLax} imply that the poles $z_r$ are either real or come in pairs of complex conjugate numbers $(z_r,z_{\bar r}=\overline{z_r})$. In the first case, the levels $\ell_{r,p}$ appearing in Equation \eqref{Eq:OmegaLevels} as the coefficients of the poles of $\omega$ at $z=z_r$ are themselves real. In the second case, the levels associated with the poles $z_r$ and $z_{\bar r}=\overline{z_r}$ are complex conjugate, \textit{i.e.} $\overline{\ell_{r,p}}=\ell_{\bar r,p}$.\\

\begin{tcolorbox} \textit{\underline{Exercise 5:} Pole structure of $\omega$ at infinity.} {\Large$\;\;\star$} \vspace{4pt}\\
Explicitly check Equations \eqref{Eq:dOverz} and \eqref{Eq:OmegaInfty}.
\end{tcolorbox}

\paragraph{Defect term.} Recall that the variation of the 4d-CS action under an infinitesimal change $\delta A$ of the gauge field is given by Equation \eqref{Eq:VariationS}. In the previous sections, we have focused on the first term in this variation, which is a bulk term in $M=\Sigma \times \CP$, and postponed the treatment of the second term, which is a 2-dimensional defect term. For convenience, we recall this term here (removing overall constants):
\begin{equation}\label{Eq:Defect}
\iiiint_M \dd \omega \wedge \Tr\bigl( A \wedge \delta A \bigr).
\end{equation}
As mentioned in Section \ref{Sec:4d}, the derivative $\dd\omega=\p_{\zb} \vp(z)\, \dd \zb \wedge \dd z$ of $\omega=\vp(z)\dd z$ is a distribution located on the \textit{poles defect} $\Sigma \times \Zc$, where we recall that $\Zc$ is the set of poles of $\omega$. The above term is thus a 2-dimensional integral on $\Sigma \times \Zc$. The consistency of the action principle requires that this defect term also vanishes on its own. To explain how this term is treated, let us start by the simplest example, focusing on simple poles of $\omega$.

\paragraph{The case of simple poles.} Let us then consider a pole $z_r \in \Zc'$, $r\in\lbrace 1,\cdots,N \rbrace$, which we suppose is simple. Recall that $\Zc'$ denotes the set of finite poles of $\omega$: for simplicity, we will restrict our attention here to the case where the pole under investigation is finite, the case of a pole at infinity being treated in a similar way after performing the change of variable $\xi=1/z$. We thus have
\begin{equation}
\omega = \frac{\ell_{r,0}}{z-z_r}\dd z + \text{``regular part at } z=z_r \text{''}.
\end{equation}
Recall the identity \eqref{Eq:DerPoleDirac}, expressing the $\zb$-derivative of $1/(z-y)$ in terms of the Dirac distribution $\delta^{(2)}(z-y)$. Applying this identity in the present case, we get
\begin{equation}
\dd \omega = 2i\pi \,\ell_{r,0} \, \delta^{(2)}(z-z_r) \, \dd z \wedge \dd \zb + \text{``distribution with support away from } z=z_r \text{''}.
\end{equation}
Thus, the contribution of $z_r$ to the defect term \eqref{Eq:Defect} is given, up to a global factor, by
\begin{equation}
\Dc_r=\ell_{r,0} \, \iint_\Sigma \,\Tr\bigl( A \wedge \delta A \bigr) \bigr|_{z=z_r},
\end{equation}
where $|_{z=z_r}$ denotes the evaluation at $z=z_r$. To rewrite this more explicitly, let us introduce the skew-symmetric tensor $\epsilon^{\mu\nu}$ on $\Sigma$, normalised as $\epsilon^{tx}=1$. The contribution of the pole $z_r$ is thus
\begin{equation}\label{Eq:DefectSimple}
\Dc_r=\ell_{r,0}\,\iint_\Sigma \dd t \, \dd x \; \epsilon^{\mu\nu} \,  \Tr\bigl( A_\mu \,\delta A_\nu \bigr) \bigr|_{z=z_r}.
\end{equation}

\paragraph{The case of higher-order poles.} Let us now consider a pole $z_r \in \Zc'$ with arbitrary order $m_r\in\Z_{\geq 1}$. We then have
\begin{equation}
\omega = \sum_{p=0}^{m_r-1} \frac{\ell_{r,p}}{(z-z_r)^{p+1}}\dd z + \text{``regular part at } z=z_r \text{''}.
\end{equation}
By derivating the identity \eqref{Eq:DerPoleDirac} $p$ times with respect to $z$, we find that
\begin{equation}\label{Eq:DerPoleDiracHigher}
\p_{\zb} \left( \frac{1}{(z-y)^{p+1}} \right) = \frac{(-1)^{p+1} 2i\pi}{p!} \, \p_z^p\delta^{(2)}(z-y).
\end{equation}
Thus, we get
\begin{equation}
\dd \omega = 2i\pi \sum_{p=0}^{m_r-1} \frac{(-1)^p\ell_{r,p}}{p!} \, \p_z^p \delta^{(2)}(z-z_r) \, \dd z \wedge \dd \zb + \text{``distribution with support away from } z=z_r \text{''}.
\end{equation}
In particular, the contribution of the pole $z_r$ to the defect term \eqref{Eq:Defect} can be computed, by using the rule
\begin{equation}\label{Eq:dDelta}
\iint_{\CP} f\,\p_z^p \delta^{(2)}(z-z_r) \, \dd z \wedge \dd \zb = (-1)^p \p_z^p f \bigl|_{z=z_r}.
\end{equation}
More precisely, we find that this contribution, up to a global factor, is given by
\begin{equation}\label{Eq:DefectHigherOrder}
\Dc_r = \sum_{p=0}^{m_r-1} \frac{\ell_{r,p}}{p!}\,\iint_\Sigma \dd t \, \dd x \; \epsilon^{\mu\nu} \, \p_z^p \Tr\bigl( A_\mu \,\delta A_\nu \bigr) \bigr|_{z=z_r}.
\end{equation}
This generalises the Equation \eqref{Eq:DefectSimple} for arbitrary $m_r$.\\

\begin{tcolorbox} \textit{\underline{Exercise 6:} Defect terms for higher order poles.} {\Large$\;\;\star$} \vspace{4pt}\\
Explicitly check the equations \eqref{Eq:DerPoleDiracHigher} to \eqref{Eq:DefectHigherOrder}.
\end{tcolorbox}

\paragraph{Total defect term.} Let us denote by $P$ the set of labels $\lbrace 1,\cdots,N \rbrace$ if $\infty$ is not a pole of $\omega$ and $\lbrace 1,\cdots,N,\infty\rbrace$ otherwise. In the second case, we also denote $z_\infty=\infty$ to uniformise the notations. In every case, the poles of $\omega$ are thus $\Zc=\lbrace z_r, \, r\in P \rbrace$, with multiplicities $m_r$. One then expresses the full defect term \eqref{Eq:Defect} by summing the contributions $\Dc_r$ over $r\in P$. As the consistency of the action principle requires that this defect term vanishes, we thus need
\begin{equation}\label{Eq:DefectZero}
\sum_{r\in P} \Dc_r = 0.
\end{equation}
As we will see, this is ensured by imposing appropriate boundary conditions on the gauge field $A$ along the defect $\Sigma \times \Zc$. Different boundary conditions will in the end result in different integrable 2d field theories. To keep things pedagogical, we will not discuss directly the systematic treatment of boundary conditions in the general case. Instead, we will start with a simple example of boundary condition, which we will describe in the next subsection. In Section \ref{Sec:PCM}, we will then study an explicit example of a 4d-CS theory that uses this boundary condition and will describe in detail how to extract the corresponding 2-dimensional integrable field theory, obtaining in this way the Principal Chiral Model. We will come back to the discussion of more general boundary conditions after this first example, in Section \ref{Sec:MoreBC}.

\subsection{Boundary conditions: a simple example at a double pole}
\label{Sec:DoublePole}

\paragraph{The boundary condition.} One possibility to ensure the vanishing \eqref{Eq:DefectZero} of the defect term is to impose that all the quantities $\Dc_r$, $r\in P$, vanish independently, treating each pole one by one. With this idea in mind, let us then focus on one pole $z_r$, which we suppose to be real and of order $m_r=2$. Using the general formula \eqref{Eq:DefectHigherOrder}, valid for a pole of arbitrary order, and the Leibniz rule we find that for $m_r=2$, the defect term reads
\begin{equation}\label{Eq:DefectDouble}
\Dc_r = \iint_\Sigma \dd t \, \dd x \;  \epsilon^{\mu\nu} \Bigl( \ell_{r,0} \, \Tr\bigl( A_\mu|_{z=z_r} \,\delta A_\nu|_{z=z_r} \bigr) + \ell_{r,1} \, \Tr\bigl( \p_z A_\mu|_{z=z_r} \,\delta A_\nu|_{z=z_r} \bigr) + \ell_{r,1} \, \Tr\bigl( A_\mu|_{z=z_r} \,\p_z\delta  A_\nu|_{z=z_r} \bigr) \Bigr).
\end{equation}
We want to find a boundary condition on $A_\mu$ ensuring that this term vanishes for every variation $\delta A_\mu$. This can be done simply by imposing
\begin{equation}\label{Eq:BC-PCM}
A_\mu \,\bigl|_{z=z_r} = 0, \qquad \text{ for } \mu=t,x.
\end{equation}
Note that one should then also restrict to variations of $A$ that respect this boundary condition, hence
\begin{equation}
\delta A_\mu \,\bigl|_{z=z_r} = 0, \qquad \text{ for } \mu=t,x.
\end{equation}
It is then clear that this boundary condition ensures that the defect term \eqref{Eq:DefectDouble} vanishes.

\paragraph{Compatible gauge transformations.} An important consequence of the boundary condition on the pole defect $\Sigma \times \lbrace z_r \rbrace$ is that one should restrict the gauge transformations $A \mapsto A^u$ of the theory to the ones that preserve this condition. Recall that
\begin{equation}
A^u_\mu = u A_\mu u^{-1} - \bigl( \p_\mu u \bigr) u^{-1}.
\end{equation}
For a gauge field satisfying the boundary condition \eqref{Eq:BC-PCM}, we then have
\begin{equation}
A^u_\mu \,\bigl|_{z=z_r} = - \bigl( \p_\mu u \bigr) u^{-1}\bigl|_{z=z_r}, \qquad \text{ for } \mu=t,x.
\end{equation}
For the gauge transformation to be compatible with the boundary condition, we must ensure that $A^u_\mu \,\bigl|_{z=z_r} = 0$. This requires that $\p_\mu u \bigl|_{z=z_r} = 0$ for $\mu=t,x$ and thus that the restriction of $u$ on the defect $\Sigma \times \lbrace z_r \rbrace$ is a constant, independent of the coordinates $(t,x)$ of $\Sigma$. In what follows, we will define the true physical gauge transformations $A \mapsto A^u$ by requiring
\begin{equation}
u |_{z=z_r} = \Id.
\end{equation}
These are not the most general transformations that preserve the boundary condition, as one could have taken $u|_{z=z_r}$ to be any constant element of the group $G$. This results in an additional symmetry group of the theory, which can be seen as the quotient of formal gauge transformations with constant $u|_{z=z_r}$ by the true gauge transformations satisfying $u|_{z=z_r}=\Id$ only. This quotient is a finite dimensional Lie group, isomorphic to $G$ itself, and will be interpreted as a global symmetry of the resulting 2d model, instead of a gauge symmetry. We will discuss this in more details in the example of the Principal Chiral Model in Section \ref{Sec:PCM}.

\section{The Principal Chiral Model from 4-dimensional Chern-Simons}
\label{Sec:PCM}

In this section, we will explain in detail how one recovers the Principal Chiral Model (PCM) from the 4d-CS theory, as was originally done in~\cite{Costello:2019tri}. We will first describe the choice of 4-dimensional theory we start with, in particular the choice of meromorphic 1-form $\omega$ and of associated boundary conditions at its poles. We will then explain how to extract the PCM from it and in particular how to obtain the 2d fields, action and Lax connection of this model.

\subsection{The 4-dimensional setup}
\label{Sec:4dPCM}

\paragraph{Meromorphic 1-form $\bm{\omega}$.} In this section, we will take the meromorphic 1-form $\omega$ to be\footnote{As explained in Section \ref{Sec:CommentsLax}, based on~\cite{Vicedo:2019dej}, the function $\vp(z)$ entering the 1-form $\omega=\vp(z)\dd z$ corresponds in the end to the twist function of the resulting 2d field theory, which controls the Maillet bracket of its Lax matrix. The twist function of the PCM, which we aim to recover here, was computed in~\cite{Maillet:1985ec}. The choice of $\omega$ made here agrees with this twist function.}:
\begin{equation}\label{Eq:OmegaPCM}
\omega = \hay \frac{1-z^2}{z^2}\dd z,
\end{equation}
where $\hay$ is a constant parameter. This form possesses a double pole at $z=0$, as well as a double pole at $z=\infty$ (as can be checked by performing the change of variable $\xi=1/z$). Thus, in the notations of the previous sections, the set of poles of $\omega$ is $\Zc=\lbrace 0,\infty \rbrace$ while its set of finite poles is simply $\Zc'=\lbrace 0 \rbrace$. Moreover, $\omega$ has two real simple zeroes at $z=+1$ and $z=-1$. In the notations of the previous sections, we thus have $\Ze=\lbrace +1,-1 \rbrace$.

\paragraph{Boundary conditions.} Since $\omega$ has double poles at $\Zc=\lbrace 0,\infty \rbrace$, one has to impose boundary conditions on the gauge field along the defect $\Sigma \times \Zc$. We have described an appropriate choice of boundary condition at double poles in Subsection \ref{Sec:DoublePole}. Following these results, we impose in the present case the following boundary conditions:\vspace{-2pt}
\begin{equation}\label{Eq:BC-PCM2}
A_\pm \bigl|_{z=0} = A_\pm \bigl|_{z=\infty} = 0.\vspace{-2pt}
\end{equation}
In this equation, we have used the light-cone components $A_\pm = A_t \pm A_x$ of the gauge-field instead of its space-time components $A_t$ and $A_x$. As explained in Subsection \ref{Sec:DoublePole}, we then also restrict the gauge transformations $A \mapsto A^u$ accordingly, by imposing\vspace{-2pt}
\begin{equation}\label{Eq:GaugePCM}
u |_{z=0} = u |_{z=\infty} = \Id.\vspace{-2pt}
\end{equation}
Let us make a small comment on terminology here. In Subsection \ref{Sec:EoM4d}, we have introduced the notion of ``formal gauge transformations'' to denote the general transformations $A \mapsto A^u$, without restrictions on $u$. In contrast, the standard terminology ``gauge transformation'' refers to the transformations with $u$ satisfying the boundary conditions \eqref{Eq:GaugePCM}, which are the ones that really act on the space of physical configurations of the fields of the theory. A natural question related to gauge transformations that we have postponed in the previous sections is whether these physical gauge transformations preserve the action of the theory. This was proven for a general $\omega$ and general boundary conditions in the reference~\cite{Benini:2020skc}, using homotopical methods. In contrast, the action is not invariant under gauge transformations that do not satisfy the boundary conditions, as one can expect.

\subsection{The 2-dimensional fields}
\label{Sec:FieldPCM}

We are now in a position to extract the integrable 2-dimensional model from the above 4-dimensional setting. Let us start by explaining what are the fundamental fields of this 2-dimensional model. Recall from Equation \eqref{Eq:Azbar} that we parametrise the $\zb$-component of the gauge field as $A_{\zb} = -(\p_{\zb}\gh)\gh^{-1}$, in terms of a group valued field $\gh: M \rightarrow G^{\C}$. Recall also from Subsection \ref{Sec:FreedomGh} that under a gauge transformation $A\mapsto A^u$, this field simply transforms as $\gh \mapsto u\gh$. Such a transformation $\gh \mapsto u\gh$ can in principle be used to completely eliminate $\gh$: however, in the present case, we must restrict the allowed gauged transformations to the ones preserving the boundary conditions and thus satisfying the condition \eqref{Eq:GaugePCM}, \textit{i.e.} $u |_{z=0} = u |_{z=\infty} = \Id$. Thus, we can gauge away all the degrees of freedom in the field $\gh$, except for its evaluations $\gh |_{z=0}$ and $\gh |_{z=\infty}$ on the poles defect $\Sigma \times \lbrace 0,\infty \rbrace$. These are the degrees of freedom that will form the fundamental fields of our 2d field theory.

Recall from Section \ref{Sec:FreedomGh} that there exists another freedom in the choice of $\gh$. Indeed, the gauge field $A$ is unchanged under the transformation $(\gh,\Lc)\mapsto (\gh v, \Lc^v)$, where $v: \Sigma \rightarrow G$ is independent of $z$ (note that because these redefinitions of $(\gh,\Lc)$ do not change the gauge field $A$, they are always compatible with the boundary conditions). This freedom can be used to fix the evaluation of $\gh$ at any point $z\in\CP$ to an arbitrary element of $G^{\C}$, say the identity for instance. In the present case, we will fix the evaluation at infinity to the identity, thus eliminating one of the degrees of freedom $\gh |_{z=\infty}$ that was left after using gauge transformations. We are then left with only one physical 2d degree of freedom $\gh |_{z=0}$, which we will denote by $g$. To summarise, we now have\vspace{-2pt}
\begin{equation}\label{Eq:FieldPCM}
\gh |_{z=0}=g \qquad \text{ and } \qquad \gh |_{z=\infty}=\Id,\vspace{-2pt}
\end{equation}
where $g:\Sigma \rightarrow G$ is a 2d group-valued field, which is the only physical degree of freedom contained in $\gh$ (note that, because of the reality conditions imposed in Subsection \ref{Sec:CommentsLax}, the field $g=\gh |_{z=0}$ is real, in the sense that it belongs to the real form $G$ in $G^{\C}$).\\

Let us use this example to illustrate the general mechanism that allows the extraction of 2d fields from the 4d-CS theory. In a model with an arbitrary meromorphic 1-form $\omega$, one has to eliminate defect terms in the variation of the action by imposing appropriate boundary conditions on $A$ along the poles defect $\Sigma \times \Zc$. This requires at the same time to restrict the type of gauge transformations that one considers, keeping only the ones compatible with these boundary conditions. As a consequence, some of the degrees of freedom in $\gh$ cannot be gauged away: these form the 2-dimensional degrees of freedom of the resulting theory. In the example studied in this subsection, these degrees of freedom are the evaluations of $\gh$ on the defect $\Sigma \times \Zc$. For more general poles and boundary conditions, one can encounter cases where the surviving degrees of freedom also involve derivatives $\p_z^p\gh$ evaluated on the defects $\Sigma \times \Zc$. In all cases, these 2d degrees of freedom are naturally associated with the poles defect $\Sigma \times \Zc$.

\subsection{Lax connection}
\label{Sec:LaxPCM}

In the previous subsection, we have shown that the only physical degree of freedom contained in $\gh$ is a group valued field $g:\Sigma \rightarrow G$. The quantity $\gh$ parametrises the component $A_{\zb}$ of the gauge field. Let us now turn our attention to the other two components $A_\pm$ (the light-cone components along $\Sigma$). As explained in Section \ref{Sec:4dLax}, we parametrise these components in terms of the field $\gh$ and two other Lie algebra valued fields $\Lc_\pm$ as
\begin{equation}\label{Eq:ApmPCM}
A_\pm = \gh \Lc_\pm \gh^{-1} - (\p_\pm \gh) \gh^{-1}.
\end{equation}
A natural question is thus whether there are additional physical degrees of freedom contained in these fields $\Lc_\pm$. In particular, we have argued in Section \ref{Sec:4dLax} that $\Lc_\pm$ plays the role of the Lax connection of the resulting integrable 2d field theory. To really be able to make such an interpretation, $\Lc_\pm$ should not be independent fields and should in fact be expressed in terms of the fundamental fields of this theory, here $g$. This is what we will prove in this subsection.\\

In Subsection \ref{Sec:Lax4Db}, we have studied parts of the equations of the motion of the model and have shown that they fix the $z$-dependence of the fields $\Lc_\pm$. More precisely, we have shown that these fields are meromorphic in $z$, with simple poles at the zeroes $\Ze$ of $\omega$ and that these zeroes are separated into two subsets $\Ze_\pm$, corresponding to the poles of $\Lc_\pm$ respectively, following the general formula \eqref{Eq:Lz}. In the present case, we have $\Ze=\lbrace +1,-1\rbrace$: we will then take $\Ze_\pm = \lbrace \pm 1 \rbrace$.  The $z$-dependence of the fields $\Lc_\pm$ then takes the form
\begin{equation}\label{Eq:LaxPCM}
\Lc_\pm = \frac{U_\pm}{z \mp 1} + V_\pm,
\end{equation}
where $U_\pm$ and $V_\pm$ are $\g$-valued fields on $\Sigma$ (independent of $z$).\\

Let us now impose the boundary conditions \eqref{Eq:BC-PCM2} on the gauge field $A_\pm$. We first focus on the one at infinity, which we recall is $A_\pm|_{z=\infty}=0$. From Equation \eqref{Eq:LaxPCM}, we simply have
\begin{equation}
\Lc_\pm\bigl|_{z=\infty} = V_\pm.
\end{equation}
Evaluating Equation \eqref{Eq:ApmPCM} at $z=\infty$ and recalling from Equation \eqref{Eq:FieldPCM} that $\gh|_{z=\infty}=\Id$, we then get
\begin{equation}
A_\pm \bigl|_{z=\infty} = \Lc_\pm\bigl|_{z=\infty} = V_\pm.
\end{equation}
The boundary condition $A_\pm|_{z=\infty}=0$ then simply implies that $V_\pm$ vanishes.

We now turn our attention to the boundary condition $A_\pm|_{z=0}=0$. From Equation \eqref{Eq:LaxPCM} and the fact that $V_\pm=0$, we get
\begin{equation}
\Lc_\pm\bigl|_{z=0} = \mp U_\pm.
\end{equation}
Evaluating Equation \eqref{Eq:ApmPCM} at $z=0$ and recalling from Equation \eqref{Eq:FieldPCM} that $\gh|_{z=0}=g$, we then get
\begin{equation}
A_\pm \bigl|_{z=0} = \mp g \,U_\pm g^{-1} - (\p_\pm g)g^{-1}.
\end{equation}
The boundary condition $A_\pm|_{z=0}=0$ then yields $U_\pm = \mp g^{-1}\p_\pm g$. Reinserting in Equation \eqref{Eq:LaxPCM}, we finally express $\Lc_\pm$ as
\begin{equation}\label{Eq:LaxPCM2}
\Lc_\pm = \frac{g^{-1}\p_\pm g}{1 \mp z}.
\end{equation}
We recognise here the Lax connection the PCM, as seen in Ben Hoare's lectures \textit{``Integrable deformations of sigma models''}~\cite{LectureHoare}. In particular, we have shown, as announced, that $\Lc_\pm$ does not contain any additional physical degrees of freedom and is expressed purely in terms of $g$.\\

As in the previous subsection, let us use this example to explain the general strategy that one follows to determine the Lax connection of a 2d model in the framework of 4d-CS, after having extracted the 2d fields from $\gh$ at the poles defect $\Sigma \times \Zc$. The first step is to choose a separation of the zeroes $\Ze$ of $\omega$ into the subsets $\Ze_\pm$, thus fixing the $z$-dependence \eqref{Eq:Lz} of $\Lc_\pm$ in terms of fields $U_\pm^{(y)}$ ($y\in\Ze_\pm$) and $V_\pm$. We then impose the chosen boundary conditions on $A_\pm$ at the poles defect $\Sigma \times \Zc$: these conditions translate to a system of equations on $U_\pm^{(y)}$ and $V_\pm$, allowing us to solve for these quantities in terms of the fundamental fields of the model.

\subsection{Action}
\label{Sec:ActionPCM}

Let us finally explain how to recover the action of the PCM from 4d-CS. For that, we will show that replacing the gauge field $A$ by its expression in terms of $\gh$ and $\Lc_\pm$ in the 4-dimensional action \eqref{Eq:Action4d} and further reinserting the expression \eqref{Eq:LaxPCM2} of $\Lc_\pm$, we will be able to explicitly perform the integration over $z\in\CP$ and obtain an integral over $\Sigma$ in terms of the 2d-field $g$.

\begin{figure}[H]
\begin{center}
\includegraphics[scale=1.3]{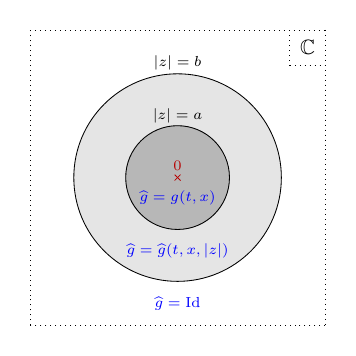}\vspace{-17pt}
\caption{Choice of gauge for the field $\widehat{g}$.}\label{Fig:Island}
\end{center}
\end{figure}

\paragraph{Preparing the field $\bm{\gh}$.} To make this computation easier, it will be useful to use the gauge transformations to first simplify the field $\gh$. Recall that a gauge transformation by $u$ acts on $\gh$ as $\gh \mapsto u\gh$. Moreover, we have to restrict to gauge parameters $u$ satisfying the boundary conditions \eqref{Eq:GaugePCM}. As explained in Subsection \ref{Sec:FieldPCM}, such a gauge transformation cannot modify the evaluations $\gh|_{z=0}=g$ and $\gh|_{z=\infty}=\Id$. However, as $u$ is completely unconstrained outside of $z=0$ and $z=\infty$, we can always fine tune it to bring $\gh$ to any smooth function on $M$ with these fixed evaluations. We will choose to take $\gh$ satisfying the following conditions (see Figure \ref{Fig:Island}):
\begin{itemize}\setlength\itemsep{0.2em}
\item there exists $a>0$ such that $\gh$ is equal to $g$ for $|z| \leq a$ ;
\item there exists $b>a$ such that $\gh$ is equal to $\Id$ for $|z| \geq b$ ;
\item for $a \leq |z| \leq b$, $\gh$ depends only on $t$, $x$ and the modulus $|z|$ and interpolates smoothly between $g(t,x)$ at $|z|=a$ and $\Id$ at $|z|=b$.
\end{itemize}

\paragraph{Expressing the action in terms of $\bm{\gh}$ and $\bm{\Lc}$.} Recall the expression \eqref{Eq:Action4d} of the 4-dimensional action in terms of the Chern-Simons 3-form $\CS(A)$. By the general construction of Section \ref{Sec:4dLax}, the gauge field $A$ is related to $\Lc$ by the formal gauge transformation by $\gh^{-1}$, \textit{i.e.} $A=\Lc^{\gh}$. According to our terminology, this transformation is only formal, since the gauge parameter $\gh^{-1}$ does not satisfy the boundary conditions \eqref{Eq:GaugePCM}: thus, the 4-dimensional action is not invariant under this transformation and we cannot compute $S[A]$ simply by replacing $A$ with $\Lc$. However, we have determined in Equation \eqref{Eq:GaugeCS} the general transformation rule of the Cherns-Simons 3-form $\CS(A)$ under such a formal gauge transformation, independently of whether it preserves the boundary conditions or not\footnote{Although this equation was derived for the 3d-CS theory, it readily applies to the present 4-dimensional case.}. In the present case, this equation tells us that
\begin{equation}\label{Eq:CsGhL}
\CS(A) = \CS\bigl(\Lc^{\gh}\bigr) = \CS(\Lc) + \dd \, \Tr\bigl( \gh^{-1} \dd \gh \wedge \Lc \bigr) + \frac{1}{3} \Tr\bigl( \gh^{-1} \dd \gh \wedge \gh^{-1} \dd \gh \wedge \gh^{-1} \dd \gh \bigr).
\end{equation}
Since $\Lc$ has components only along $\dd t$ and $\dd x$, or equivalently $\dd x^+$ and $\dd x^-$, the term $\Lc \wedge \Lc \wedge \Lc$ vanishes. The Chern-Simons form of $\Lc$ then simply reads
\begin{equation}
\CS(\Lc) = \Lc \wedge \dd \Lc.
\end{equation}
In the action, this form will appear wedged with $\omega$. Since the latter is along $\dd z$ and $\Lc$ is along $\dd x^\pm$, this term $\omega \wedge \CS(\Lc)$ will be proportional to $\omega \wedge \p_{\zb}\Lc \wedge \Lc$. Yet, by virtue of the equations of motion \eqref{Eq:EoMLZbar} along $\zb$, this term vanishes. Thus, the first term in Equation \eqref{Eq:CsGhL} does not contribute to the action \eqref{Eq:Action4d}. The second term is a total derivative: once wedged with $\omega$ in the action, we will use the product rule \eqref{Eq:ProductRule} to bring the exterior derivative on $\omega$, discarding the boundary terms created by the total derivative (recall that we suppose that the fields decrease sufficiently fast at the space-time infinity of $\Sigma \times \CP$). In the end, we thus write the action as
\begin{equation}\label{Eq:ActionPCM1}
S = \frac{i}{4\pi} \iiiint_M \dd \omega \wedge \Tr\bigl( \gh^{-1} \dd \gh \wedge \Lc \bigr) + \frac{i}{12\pi} \iiiint_M \omega \wedge \Tr\bigl( \gh^{-1} \dd \gh \wedge \gh^{-1} \dd \gh \wedge \gh^{-1} \dd \gh \bigr).
\end{equation}
We will now compute the two terms in this equation separately.

\paragraph{Computation of the first term.} Let us start with the first term. Recall that $\omega$ has double poles at $0$ and $\infty$ and thus that $\dd\omega$ is a distribution with support $\lbrace 0,\infty \rbrace$. More precisely, using $\omega=\left(\frac{\hay}{z^2}-\hay \right)\dd z$ and Equation \eqref{Eq:DerPoleDiracHigher}, we find that
\begin{equation}
\dd \omega = -2i\pi \hay\,\p_z \delta^{(2)}(z) \, \dd z \wedge \dd \zb \; + \; \text{ ``distribution with support at } z=\infty \text{''}.
\end{equation}
Reinserting in the first term of Equation \eqref{Eq:ActionPCM1}, one can then explicitly perform the integration over $z\in\CP$, yielding 2-dimensional terms on $\Sigma \times \lbrace 0 \rbrace$ and $\Sigma \times \lbrace \infty \rbrace$. Recall from the previous paragraph that we chose $\gh$ to be constant equal to the identity in a neighbourhood of $z=\infty$. The contribution located on $\Sigma \times \lbrace \infty \rbrace$ then vanishes. Using the identity \eqref{Eq:dDelta}, we thus get
\begin{equation}
\frac{i}{4\pi} \iiiint_M \dd \omega \wedge \Tr\bigl( \gh^{-1} \dd \gh \wedge \Lc \bigr) = -\frac{\hay}{2} \iint_\Sigma \p_z \Tr \bigl( \gh^{-1}\dd_\Sigma \gh \wedge \Lc \bigr)\bigr|_{z=0},
\end{equation}
where we recall that $\dd_\Sigma$ is the exterior derivative along $\Sigma$. In the previous paragraph, we have chosen $\gh$ independent of $z$ and equal to $g$ in a neighbourhood of $z=0$. Thus, $\gh|_{z=0}=g$ and $\p_z\gh|_{z=0}=0$. Developing the $z$-derivative by the Leibniz rule in the above equation, we then find that the term where $\p_z$ acts on $\gh^{-1}\dd_\Sigma \gh$ does not contribute and that
\begin{equation}
\frac{i}{4\pi} \iiiint_M \dd \omega \wedge \Tr\bigl( \gh^{-1} \dd \gh \wedge \Lc \bigr) = -\frac{\hay}{2} \iint_\Sigma \Tr \bigl( g^{-1}\dd_\Sigma g \wedge \p_z \Lc\bigr|_{z=0} \bigr).
\end{equation}
Using light-cone coordinates, we get\footnote{We choose the orientation of $\Sigma$ such that $\dd t \wedge \dd x$ is a positive volume form. In light-cone coordinates, the positive volume form is thus $- \dd x^+ \wedge \dd x^-$.}
\begin{equation}
\frac{i}{4\pi} \iiiint_M \dd \omega \wedge \Tr\bigl( \gh^{-1} \dd \gh \wedge \Lc \bigr) = \frac{\hay}{2} \iint_\Sigma \dd x^+ \, \dd x^- \; \Tr \Bigl( g^{-1}\p_+ g \, \bigl(\p_z \Lc_-\bigr|_{z=0}\bigr) -   g^{-1}\p_- g \,\bigl(\p_z \Lc_+\bigr|_{z=0}\bigr)\Bigr).
\end{equation}
From the explicit expression \eqref{Eq:LaxPCM2} of the light-cone Lax connection $\Lc_\pm$ of the model, we get
\begin{equation}
\p_z \Lc_\pm \bigr|_{z=0} = \pm g^{-1}\p_\pm g.
\end{equation}
Thus, we finally obtain
\begin{equation}
\frac{i}{4\pi} \iiiint_M \dd \omega \wedge \Tr\bigl( \gh^{-1} \dd \gh \wedge \Lc \bigr) = -\hay \iint_\Sigma \dd x^+ \, \dd x^- \; \Tr \bigl( g^{-1}\p_+ g \; g^{-1} \p_- g \bigr).
\end{equation}
We already recognise in this expression the action of the PCM, as seen in Ben Hoare's lectures \textit{``Integrable deformations of sigma models''}\footnote{For completeness and to ease the comparison, let us note that the 2-dimensional measure $\dd^2x$ defined in Ben Hoare's lectures coincides, in the present notations, with $\dd^2x = \dd t\,\dd x = 2\,\dd x^+\,\dd x^-$.}~\cite{LectureHoare}. We thus expect the second term in Equation \eqref{Eq:ActionPCM1} to vanish.

\paragraph{Computation of the second term.} Let us thus turn our attention to this second term. Recall that our choice of $\gh$ is equal to the identity for $|z|>b$, for some $b>0$. The integral over $z$ in the second term of \eqref{Eq:ActionPCM1} then reduces to $D_b \subset \CP$, the disk of radius $b$ centred around $z=0$. Writing things in components, using Equation \eqref{Eq:AwedgeA} and taking into account permutations over the labels $(\zb,t,x)$, we get
\begin{align}
&\frac{i}{12\pi} \iiiint_M \omega \wedge \Tr\bigl( \gh^{-1} \dd \gh \wedge \gh^{-1} \dd \gh \wedge \gh^{-1} \dd \gh \bigr)\\
& \hspace{35pt} = \frac{i}{4\pi} \iiiint_{D_b \times \Sigma} \dd z \wedge \dd \zb \wedge \dd t \wedge \dd x  \; \vp(z)\, \Tr\Bigl( \gh^{-1} \p_{\zb} \gh \; \bigl[ \gh^{-1} \p_t \gh, \gh^{-1} \p_x \gh \bigr] \Bigr), \notag
\end{align}
where we recall that $\vp(z)$ is such that $\omega=\vp(z)\dd z$. Recall moreover that we chose $\gh$ to depend on $z$ only through the modulus $r=|z|$, and not through the argument $\theta = \arg z$. In particular, we then have
\begin{equation}
\p_{\zb} \gh = \frac{1}{2}e^{i\theta} \p_r \gh.
\end{equation}
Let us then perform the change of variable $(z,\zb) \mapsto (r,\theta)$ in the above integral, using $\dd z \wedge \dd \zb = -2ir \, \dd r \wedge \dd \theta$. We choose the orientation on $D_b$ such that the positive surface form is $r\,\dd r \wedge \dd \theta$. We then get
\begin{align}
&\frac{i}{12\pi} \iiiint_M \omega \wedge \Tr\bigl( \gh^{-1} \dd \gh \wedge \gh^{-1} \dd \gh \wedge \gh^{-1} \dd \gh \bigr)\\
& \hspace{35pt} = \frac{1}{4\pi} \iiiint_{[0,2\pi] \times [0,b] \times \Sigma}  \dd \theta \, \dd r \, \dd t \, \dd x \; re^{i\theta}\vp\bigl(r e^{i\theta}\bigr)\, \Tr\Bigl( \gh^{-1} \p_r \gh \; \bigl[ \gh^{-1} \p_t \gh, \gh^{-1} \p_x \gh \bigr] \Bigr). \notag
\end{align}
Let us consider for a moment a general 1-form $\omega = \vp(z)\dd z$, with Laurent series
\begin{equation}
\vp(z) = \sum_{q=-m_0}^{+\infty} a_q z^q
\end{equation}
around $z=0$. We then have
\begin{equation}
\int_0^{2\pi} \dd \theta\,r e^{i\theta} \vp\bigl(r e^{i\theta}\bigr) = \sum_{q=-m_0}^{+\infty} a_q r^{1+q} \underbrace{\int_0^{2\pi} \dd \theta\,e^{i(1+q)\theta}}_{2\pi\,\delta_{q,-1}} = 2\pi a_{-1} = 2\pi \, \res_{z=0} \omega.
\end{equation}
Thus, performing the integral over $\theta$ in the above expression, we get
\begin{align}\label{Eq:WZ}
&\frac{i}{12\pi} \iiiint_M \omega \wedge \Tr\bigl( \gh^{-1} \dd \gh \wedge \gh^{-1} \dd \gh \wedge \gh^{-1} \dd \gh \bigr) \\
& \hspace{35pt} = \frac{1}{2} \res_{z=0} \omega \iiint_{[0,b] \times \Sigma}  \dd r \, \dd t \,  \dd x \; \Tr\Bigl( \gh^{-1} \p_r \gh \; \bigl[ \gh^{-1} \p_t \gh, \gh^{-1} \p_x \gh \bigr] \Bigr). \notag
\end{align}
In the present case, $\omega$ has no residue at $z=0$ and this term thus vanishes, as announced. The action \eqref{Eq:ActionPCM1} is therefore indeed equal to the PCM action
\begin{equation}\label{Eq:ActionPCM}
S[g] = -\hay \iint_\Sigma \dd x^+ \, \dd x^- \; \Tr \bigl( g^{-1}\p_+ g \; g^{-1} \p_- g \bigr).
\end{equation}

\paragraph{Summary.} We have thus proved that the 4-dimensional action \eqref{Eq:Action4d} reduces in the present case to the 2-dimensional action of the PCM, after performing explicitly the integral over $z\in\CP$. In the process, we have used the explicit expression of $\Lc_\pm$ in terms of $g$ and $z$, which we obtained by solving the equations of motion $\omega \p_{\zb} \Lc_\pm = 0$ and imposing the boundary conditions on $A$. Recall that there is one remaining equation of motion, which is the zero curvature equation
\begin{equation}
\p_+ \Lc_- - \p_- \Lc_+ + \bigl[ \Lc_+, \Lc_- \bigr] = 0
\end{equation}
of the Lax connection, which describes the dynamic along $\Sigma$. By construction, this equation of motion coincides with the equation of motion of the 2-dimensional action obtained after integrating over $z$, as can be checked explicitly in the case of the PCM treated here. This ensures that the resulting 2-dimensional model is integrable, in the sense that its equation of motion can indeed be recast as the above Lax equation. This is the general mechanism behind the construction of integrable 2-dimensional field theories from 4d-CS.

In view of extending this construction to more general theories than the PCM, let us note that the expression of the action in terms of $\gh$ and $\Lc$ found in Equation \eqref{Eq:ActionPCM1} was in fact very general and did not depend on any assumptions specific to the PCM.

\paragraph{Comments on symmetries.} Recall from Subsection \ref{Sec:DoublePole} that for the type of boundary condition imposed at the double pole $z=0$ considered for this example, the compatibility of the gauge transformations $A \mapsto A^u$ with this boundary condition requires only $u|_{z=0}$ to be constant along $\Sigma$. However, as mentioned in Subsection \ref{Sec:DoublePole}, we considered as true gauge transformations only the ones with $u|_{z=0}=\Id$, explaining that the remaining symmetries will be interpreted as global symmetries of the resulting 2d theory. Let us illustrate this mechanism explicitly. A gauge transformation by $u$ acts on $\gh$ as $\gh \mapsto u\gh$. Since the Principal Chiral Field $g$ is extracted as $g=\gh|_{z=0}$, a gauge transformation with $u_0=u|_{z=0}$ constant acts on $g$ as $g \mapsto u_0 g$. This is the standard left $G$-isometry of the PCM, which one easily checks is indeed a symmetry of the action \eqref{Eq:ActionPCM}.

The 1-form $\omega$ of the PCM also possesses a double pole at $z=\infty$, associated with the same boundary condition as at $z=0$. Here also one can consider a gauge transformation $A\mapsto A^u$ with $u|_{z=\infty}=u_\infty$ constant but possibly different from the identity (we will suppose here that $u|_{z=0}=\Id$ at the other double pole $z=0$). This transformation naturally acts on $\gh|_{z=\infty}$ by multiplication on the left by $u_\infty$. However, recall from Subsection \ref{Sec:FieldPCM} that we have used the freedom $\gh \mapsto \gh v$ with $v : \Sigma \rightarrow G$ to fix $\gh|_{z=\infty}$ to the identity. Thus, to go back to the same setting, we have to combine the gauge transformation with an appropriate right multiplication by $v=u_\infty^{-1}$ to ensure that the new field $\gh$ still evaluates to the identity at infinity: we are then considering the transformation $\gh \mapsto u\, \gh \,u_\infty^{-1}$. In particular, this acts on the Principal Chiral Field $g=\gh|_{z=0}$ as $g \mapsto g u_{\infty}^{-1}$. This then corresponds to the right $G$-isometry of the PCM, which also preserves the action \eqref{Eq:ActionPCM}.

\subsection{To go further}

\paragraph{Adding a Wess-Zumino term.} We computed the second term in the action \eqref{Eq:ActionPCM1} in Equation \eqref{Eq:WZ}. In particular, for the case of the PCM, this term vanishes since $\omega$ has no residue at the double pole $z=0$. We note however that, up to this vanishing prefactor, the expression found in \eqref{Eq:WZ} corresponds to a Wess-Zumino term for $g$. More precisely, this expression can be rewritten as
\begin{equation}
\frac{i}{12\pi} \iiiint_M \omega \wedge \Tr\bigl( \gh^{-1} \dd \gh \wedge \gh^{-1} \dd \gh \wedge \gh^{-1} \dd \gh \bigr) = - \frac{1}{2} \res_{z=0} \omega \; I_{\text{WZ}}[g],
\end{equation}
where the Wess-Zumino term $I_{\text{WZ}}[g]$ is defined as
\begin{equation}\label{Eq:DefWZ}
I_{\text{WZ}}[g] = - \iiint_{[0,b] \times \Sigma}  \dd r \, \dd t \, \dd x \; \Tr\Bigl( \gh^{-1} \p_r \gh \; \bigl[ \gh^{-1} \p_t \gh, \gh^{-1} \p_x \gh \bigr] \Bigr).
\end{equation}
Following this observation, one can in fact generalise the construction of this section to include a Wess-Zumino term in the action, by considering a 1-form $\omega$ with a double pole which possesses a non-zero residue. We leave this as an exercise for the reader below.

Before stating this exercise, let us briefly comment on the above definition of the Wess-Zumino term compared to the one in Ben Hoare's lectures \textit{``Integrable deformations of sigma models''}~\cite{LectureHoare}. At first sight, one might have the impression that there is a minus sign difference in these definitions: in fact, it is not the case since by construction $\gh$ is a 3-dimensional extension of $g$ which is equal to $g$ on the ``internal'' boundary $\lbrace 0 \rbrace \times \Sigma$ of $[0,b] \times \Sigma$ and not on the ''external'' boundary $\lbrace b \rbrace \times \Sigma$.\\

\begin{tcolorbox} \textit{\underline{Exercise 7:} PCM with Wess-Zumino term} {\Large$\;\;\star\star$} \vspace{4pt}\\
For this exercise, we will consider a deformation of the 1-form \eqref{Eq:OmegaPCM} corresponding to the PCM by shifting the double pole at $z=0$ to a double pole at $z=z_1$, while keeping the zeroes fixed:
\begin{equation}
\omega = \hay \frac{1-z^2}{(z-z_1)^2}\dd z.
\end{equation}
We will parametrise $z_1$ as $z_1=\kay/\hay$. The 1-form $\omega$ has double poles at $z=z_1$ and $z=\infty$. As in the PCM case, we impose the boundary conditions $A_\pm|_{z=z_1}=A_\pm|_{z=\infty}=0$ and restrict to gauge transformations $A\mapsto A^u$ with $u|_{z=z_1}=u|_{z=\infty}=\Id$.
\begin{enumerate}[1.]
\item Check that
\begin{equation}
\omega = \left(\dfrac{\hay^2-\kay^2}{\hay} \dfrac{1}{(z-z_1)^2} - \dfrac{2\kay}{z-z_1} - \hay \right)\dd z.
\end{equation}
In particular, the deformation of the double pole at $z=0$ in the PCM case to a double pole at $z=z_1$ in the present case introduces a non-vanishing residue at this pole.
\item Similarly to the PCM case, show that one can fix $\gh|_\infty=\Id$ and that the only physical degree of freedom left in $\gh$ is $g=\gh|_{z=z_1}$.
\item The Lax connection $\Lc_\pm$ of the model takes the same form \eqref{Eq:LaxPCM} as in the PCM case. Impose the boundary conditions on $A_\pm$ at $z=\infty$ and $z=z_1$ and deduce that
\begin{equation}
\Lc_\pm = \frac{1 \mp z_1}{1 \mp z} g^{-1}\p_\pm g.
\end{equation}
\item Reinsert $\Lc_\pm$ and $\gh$ in the action \eqref{Eq:ActionPCM1} and perform the integration over $z$. Use the PCM case detailed in Subsection \ref{Sec:ActionPCM} as a guide, replacing the pole $z=0$ by $z=z_1$ (in particular, choose an appropriate gauge for $\gh$). In the end, you should find the action of the PCM with Wess-Zumino term (where the later is defined as in Equation \eqref{Eq:DefWZ}):
\begin{equation}
S[g] = -\hay \iint_\Sigma \dd x^+ \, \dd x^- \; \Tr \bigl( g^{-1}\p_+ g \; g^{-1} \p_- g \bigr) + \kay\,I_{\text{WZ}}[g].
\end{equation}
\end{enumerate}
\end{tcolorbox}

\paragraph{Gauged formulation of the PCM.} In Subsection \ref{Sec:FieldPCM}, we have used the freedom $\gh \mapsto \gh v$ to fix the value of $\gh|_{z=\infty}$ to the identity. One can wonder what one would obtain if we did not impose such a condition and kept $\widetilde{g}=\gh|_{z=\infty}$ as a dynamical field. Such a setup in fact results in a gauged formulation of the PCM, possessing two $G$-valued fields $(g,\widetilde{g})$ but invariant under the local transformations $(g,\widetilde{g}) \mapsto (gv,\widetilde{g}v)$ inherited from the freedom $\gh \mapsto \gh v$. The following exercise covers a more detailled treatment of this gauged model.\\

\begin{tcolorbox} \textit{\underline{Exercise 8:} Gauged formulation of the PCM} {\Large$\;\;\star\star$} \vspace{4pt}\\
Repeat the derivation of the PCM discussed in this section but keeping the field at infinity in $\gh$, \textit{i.e.} with two fields $g=\gh|_{z=0}$ and $\widetilde{g}=\gh|_{z=\infty}$, by following the steps below.
\begin{enumerate}
\item Solve the boundary conditions $A_\pm|_{z=0}=A_\pm|_{z=\infty}=0$ to find the expression of the Lax connection:
\begin{equation}
\Lc_\pm = \frac{g^{-1}\p_\pm g - \widetilde{g}^{-1}\p_\pm \widetilde{g}}{1 \mp z} + \widetilde{g}^{-1}\p_\pm \widetilde{g}.
\end{equation}
\item Show that, up to a gauge transformation, $\gh$ can be chosen as a function of $t$, $x$ and $r=|z|$ only, interpolating smoothly between $g$ for $r\in[0,a]$, $\Id$ for $r\in[b,b']$ and $\widetilde{g}$ for $r\in[a',+\infty]$, for some $0<a<b<b'<a'$.
\item Reinsert $\gh$ and $\Lc_\pm$ in the action \eqref{Eq:ActionPCM1} and perform the integration over $z\in\CP$ to obtain
\begin{equation}
S[g,\widetilde{g}] = -\hay \iint_\Sigma \dd x^+ \, \dd x^- \; \Tr \Bigl( \bigl( g^{-1}\p_+ g - \widetilde{g}^{-1}\p_+ \widetilde{g} \bigr) \bigl( g^{-1} \p_- g - \widetilde{g}^{-1}\p_- \widetilde{g} \bigr) \Bigr).
\end{equation}
Treat the term coming from the double pole at $z=\infty$ using the change of variable $z=\dfrac{1}{\xi}$.
\item Check that the above action is invariant under the local symmetry $(g,\widetilde{g}) \mapsto (gv,\widetilde{g}v)$ and that using this gauge symmetry to fix $\widetilde{g}=\Id$ one recovers the action \eqref{Eq:ActionPCM}. How does the Lax connection change under this transformation?
\item Adapting the discussion of the last paragraph of Subsection \ref{Sec:ActionPCM}, discuss the global symmetries of the model in this gauged formulation.
\end{enumerate} 
\end{tcolorbox}

\paragraph{Coupled PCM.} In this section, we have derived the PCM on $G$ as the theory arising from 4d-CS with a meromorphic 1-form $\omega$ possessing 2 double poles. One can naturally wonder which theory one gets starting with a 1-form $\omega$ possessing an arbitrary number $N+1$ of double poles. As shown in~\cite{Costello:2019tri}, this theory is an integrable coupled version of the PCM on $G^N$, which was previously obtained in~\cite{Delduc:2018hty,Delduc:2019bcl} using the approach of affine Gaudin models. We leave as a more advanced exercise the derivation of this coupled model.\\

\begin{tcolorbox} \textit{\underline{Exercise 9:} Integrable coupled PCMs with Wess-Zumino terms}{\Large$\;\;\star\!\star\!\star\star$} \vspace{4pt}\\
We consider the meromorphic 1-form
\begin{equation}
\omega = - K \prod_{r=1}^N  \frac{(z-\ze_r^+)(z-\ze_r^-)}{(z-z_r)^2}\dd z,
\end{equation}
with $N+1$ double poles $\Zc=\lbrace z_1,\cdots,z_N,\infty \rbrace \subset \CP$ and $2N$ simple zeroes separated in subsets $\Ze_\pm = \lbrace \ze_1^\pm,\cdots,\ze_N^\pm \rbrace \subset \C$, following the notations of Sections \ref{Sec:4d} and \ref{Sec:4dLax}.
\end{tcolorbox}
\begin{tcolorbox}
At the double poles, we impose the boundary conditions $A_\pm|_{z=z_r}=A_\pm|_{z=\infty}=0$ and accordingly restrict to gauge transformations $A \mapsto A^u$ with $u|_{z=z_r}=u|_{z=\infty}=\Id$, similarly to the case of the PCM. The degrees of freedom in $\gh$ that cannot be eliminated by a gauge transformation are the fields $g_r = \gh|_{z=z_r}$ and $\gh|_{z=\infty}$. Moreover, the latter can be fixed to $\gh|_{z=\infty}=\Id$ using the freedom $\gh \mapsto \gh v$ with $v:\Sigma \mapsto G$ independent of $z$.
\begin{enumerate}
\item Show that the boundary conditions imply
\begin{equation}
\Lc_\pm|_{z=\infty} = 0 \qquad \text{ and } \qquad  \Lc_\pm|_{z=z_r} = g_r^{-1}\p_\pm g_r.
\end{equation}
\item Following the results of Subsection \ref{Sec:Lax4Db} and in particular Equation \eqref{Eq:Lz}, $\Lc_\pm$ has simple poles at the zeroes $\ze_r^\pm$. Combined with the fact that $\Lc_\pm|_{z=\infty} = 0$ proved in the previous question, we see that $\Lc_\pm$, as a function of $z$, belongs to the space $\mathcal{F}_\pm$ of meromorphic functions with simple poles at $z=\ze_r^\pm$ and vanishing at $z=\infty$. In other words, we have, for some $\g$-valued fields $U_\pm^{(r)}$,
\begin{equation}
\Lc_\pm = \sum_{z=1}^N \frac{U_\pm^{(r)}}{z-\ze_r^\pm}.
\end{equation}
In the above equation, we have used a natural expansion of $\Lc_\pm$ along the fractions $z \mapsto 1/(z-\ze_r^\pm)$, which form an obvious basis of $\mathcal{F}_\pm$. This is however not the most suited basis for our problem. Prove that the functions
\begin{equation}
\alpha_{\pm,r}(z) = \frac{\vp_{\pm,r}(z_r)}{\vp_{\pm,r}(z)}, \qquad \text{ with } \qquad \vp_{\pm,r}(z) = \frac{\prod_{i=1}^N (z-\ze_i^\pm)}{\prod_{\substack{ s=1 \\ s\neq r}}^N (z-z_s)},
\end{equation}
form a basis of $\mathcal{F}_\pm$ and additionally satisfy
\begin{equation}
\alpha_{\pm,r}(z_s) = \delta_{rs}.
\end{equation}
\item Combining the results of questions 1 and 2, conclude that
\begin{equation}
\Lc_\pm = \sum_{r=1}^N \alpha_{\pm,r}(z)\, g_r^{-1}\p_\pm g_r.
\end{equation}
\item Reinsert $\gh$ and $\Lc_\pm$ in the action \eqref{Eq:ActionPCM1} and perform the integration over $z\in\CP$ to obtain the 2-dimensional action of the model. For that, first prepare $\gh$ in a similar way as for the PCM in Subsection \ref{Sec:ActionPCM}: we take it to be equal to the identity outside of disks of radius $b$ around each points $z_r$ and to depend only on $t$, $x$ and $r_r=|z-z_r|$ on these disks (in the language of~\cite{Delduc:2019whp}, we say that $\gh$ is of archipelago type in this gauge). In the end, write the action in the form
\begin{equation}
S[g_1,\cdots,g_N] = -\sum_{r,s=1}^N \hay_{rs}\,\iint_\Sigma \dd x^+ \, \dd x^- \;  \Tr \bigl( g_r^{-1}\p_+ g_r \; g_s^{-1} \p_- g_s \bigr) + \sum_{r=1}^N \kay_r\,I_{\text{WZ}}[g_r],
\end{equation}
with $\kay_r = -\frac{1}{2} \res_{z=z_r} \omega$ and $\hay_{rs}$ some coefficients to determine.
\end{enumerate}
\end{tcolorbox}

\section{The Yang-Baxter model and more general boundary conditions}
\label{Sec:MoreBC}

In this last section, we explore more general boundary conditions than the ones considered in the previous sections for the Principal Chiral Model. This will allow us to construct other integrable $\sigma$-models from 4d-CS theory and for instance to recover the Yang-Baxter model~\cite{Klimcik:2002zj,Klimcik:2008eq}, which we encountered in Ben Hoare's lectures \textit{``Integrable deformations of sigma models''}~\cite{LectureHoare}.

\subsection{The 4-dimensional setup}
\label{Sec:4dYB}

\paragraph{1-form $\bm{\omega}$.} In this subsection, we take the 1-form $\omega$ to be\footnote{This 1-form is equal to $\omega=\vp(z)\dd z$ with $\vp(z)$ the twist function of the Yang-Baxter model as computed in~\cite{Delduc:2013fga}.}
\begin{equation}\label{Eq:OmegaYB}
\omega = \frac{\hay}{1-\eta^2} \frac{1-z^2}{z^2-\eta^2}\dd z,
\end{equation}
where $\hay$ and $\eta$ are real parameters. It has simple zeroes at $\Ze=\lbrace +1,-1 \rbrace$ and poles at $\Zc=\lbrace +\eta,-\eta,\infty \rbrace$, the first two being simple and the third one being of order 2. Note that setting $\eta=0$, we recover the 1-form \eqref{Eq:OmegaPCM} corresponding to the PCM. Accordingly, the model that we will construct in this section will be a deformation of the PCM. Performing the partial fraction decomposition of $\omega$, we rewrite it as
\begin{equation}
\omega = \left( \frac{\hay}{2\eta} \frac{1}{z-\eta} - \frac{\hay}{2\eta} \frac{1}{z+\eta} - \frac{\hay}{1-\eta^2} \right) \dd z.
\end{equation}
In the general terminology of Section \ref{Sec:4d}, we then get that the levels associated to the simple poles $z_1=+\eta$ and $z_2=-\eta$ are given by
\begin{equation}\label{Eq:LevelsYB}
\ell_{1,0} = \frac{\hay}{2\eta} \qquad \text{ and } \qquad \ell_{2,0} = - \frac{\hay}{2\eta}.
\end{equation}

\paragraph{Poles defect and boundary conditions.} Since $z=\infty$ is a double pole of $\omega$, it creates a defect term in the variation of the action, following the general analysis of Section \ref{Sec:Poles}. This defect term takes the exact same form in terms of the gauge field $A$ as in the example discussed in Subsection \ref{Sec:DoublePole} and for the PCM. We will treat this boundary defect in the exact same way as we did there: more precisely, we ensure that it vanishes by requiring that the gauge field satisfies the boundary condition
\begin{equation}
A_\pm \bigl|_{z=\infty} = 0.
\end{equation}

Let us now turn our attention to the simple poles $z_1=+\eta$ and $z_2=-\eta$. They create a defect term in the variation of the action given by Equation \eqref{Eq:DefectSimple}, which depends on the levels $\ell_{r,0}$. Using the expression \eqref{Eq:LevelsYB} of the latter and combining the two contributions together, we thus obtain the defect term
\begin{equation}\label{Eq:DefectYB}
\Dc=\Dc_1+\Dc_2 = \frac{\hay}{2\eta}\,\iint_\Sigma \dd t \, \dd x \; \epsilon^{\mu\nu} \, \Bigl(  \Tr\bigl( A_\mu \,\delta A_\nu \bigr) \bigr|_{z=+\eta} - \Tr\bigl( A_\mu \,\delta A_\nu \bigr) \bigr|_{z=-\eta} \Bigr).
\end{equation}
Following the general strategy described in these notes, we want to ensure that this defect term vanishes by imposing an appropriate boundary condition on the gauge field at $z=\pm \eta$. The main difference with the boundary conditions that we have treated so far will be that this one involves both poles $z_1=+\eta$ and $z_2=-\eta$. In other words, we will not ensure that both defect terms $\Dc_1$ and $\Dc_2$ vanish independently, but instead will only ensure that their sum $\Dc=\Dc_1+\Dc_2$ vanishes.

For that, we will search for the appropriate boundary condition in the form of a linear relation between the evaluations $A_\mu|_{z=+\eta}$ and $A_\mu|_{z=-\eta}$. Let us first note that as $\pm \eta$ are real poles, the reality conditions discussed in Subsection \ref{Sec:CommentsLax} imply that these evaluations $A_\mu|_{z=\pm\eta}$ are valued in the real form $\g$. We will then take the following ansatz for the boundary condition:
\begin{equation}\label{Eq:BC-YB}
(\Rc+\Id)A_\mu|_{z=+\eta} = (\Rc-\Id)A_\mu|_{z=-\eta},
\end{equation}
where $\Rc: \g \rightarrow \g$ is a constant linear operator on $\g$. Let us note that in the limit $\eta \to 0$, this condition simply becomes $A_\mu|_{z=0}=0$ and we thus recover the boundary condition \eqref{Eq:BC-PCM2} of the PCM: we are therefore constructing a continuous deformation of the latter. One checks (see exercise below) that the above condition is equivalent to
\begin{equation}\label{Eq:BC-YB2}
\exists \, X_\mu\in\g\, \text{ s.t. } A_\mu\bigl|_{z=\pm \eta} = (\Rc \mp \Id)X_\mu.
\end{equation}

\begin{tcolorbox} \textit{\underline{Exercise 10:} Yang-Baxter boundary condition.} {\Large$\;\;\star$} \vspace{4pt}\\
Check that Equations \eqref{Eq:BC-YB} and \eqref{Eq:BC-YB2} are equivalent.
\end{tcolorbox}

\noi The variation $\delta A_\mu$ of the gauge field should preserve the boundary condition \eqref{Eq:BC-YB2} and thus satisfies
\begin{equation}
\delta A_\mu|_{z=\pm \eta} = (\Rc \mp \Id)\delta X_\mu.
\end{equation}
Let us reintroduce the above equation in the defect term \eqref{Eq:DefectYB}: after a little bit of algebra, we find
\begin{equation}
\Dc = -\frac{\hay}{\eta}\,\iint_\Sigma \dd t \, \dd x \; \epsilon^{\mu\nu} \, \Tr\bigl( (\Rc+\tp\Rc) X_\mu \,\delta X_\nu \big),
\end{equation}
where $\tp\Rc: \g \rightarrow \g$ denotes the transpose of $\Rc$ with respect to the trace bilinear form, which satisfies:
\begin{equation}
\Tr\bigl(\tp\Rc X\, Y\bigr) = \Tr\bigl(X\, \Rc Y\bigr), \qquad \forall \, X,Y\in\g.
\end{equation}
We then see that this defect term $\Dc$ vanishes if we require the operator $\Rc$ to be skew-symmetric, \textit{i.e.}
\begin{equation}
\tp\Rc = - \Rc.
\end{equation}
In conclusion, at the moment, our choice of boundary condition on the poles defect $\Sigma \times \lbrace +\eta,-\eta \rbrace$ is \eqref{Eq:BC-YB}, or equivalently \eqref{Eq:BC-YB2}, with $\Rc$ a skew-symmetric operator on $\g$. As we will see in the next paragraph, we will find that $\Rc$ should satisfy additional conditions coming from compatibility with the gauge transformations.

\paragraph{Compatible gauge transformations.} We now need to restrict the gauge transformations $A \mapsto A^u$ to the ones that preserve the chosen boundary conditions. For the condition at the pole $z=\infty$, we ask that
\begin{equation}\label{Eq:uInfty-YB}
u|_{z=\infty}=\Id,
\end{equation}
as in the case of the PCM (which shared the same boundary condition at infinity).\\

Let us now turn our attention to the boundary condition at the poles $\lbrace +\eta,-\eta \rbrace$. We define
\begin{equation}
\Ab_\mu = \bigl( A_\mu \bigl|_{z=+\eta}, A_\mu \bigl|_{z=-\eta} \bigr) \; \in \; \g \times \g.
\end{equation}
The boundary condition \eqref{Eq:BC-YB2} can be rephrased as
\begin{equation}\label{Eq:BC-YB3}
\Ab_\mu \in \kf, \qquad \text{ where } \qquad \kf = \bigl\lbrace \bigl( (\Rc-\Id)X, (\Rc+\Id)X \bigr), \; X\in\g \bigr\rbrace \; \subset \; \g \times \g.
\end{equation}
A formal gauge transformation $A \mapsto A^u$ transforms the quantity $\Ab_\mu$ as
\begin{equation}
\Ab_\mu \longmapsto \Ab_\mu^{\ub} = \ub\,\Ab_\mu\,\ub^{-1} - (\p_\mu \ub) \ub^{-1},
\end{equation}
where we defined
\begin{equation}\label{Eq:uYB}
\ub = \bigl( u \bigl|_{z=+\eta}, u \bigl|_{z=-\eta} \bigr) \; \in \; G \times G.
\end{equation}
We can see $\Ab_\mu$ as a 1-form in $\Omega^1(\Sigma,\g \times\g)$, on which functions $\ub : \Sigma \rightarrow G\times  G$ act as $(G \times G)$-gauge transformations. We want to restrict these transformations to the ones preserving the boundary condition \eqref{Eq:BC-YB3}, \textit{i.e.} we want $\Ab^{\ub}_\mu$ valued in the subspace $\kf \subset \g \times \g$.

One can of course ensure this by requiring that $\ub$ is the identity element of $G\times G$: this however would leave us with no residual gauge transformations compatible with the boundary condition and would yield a degenerate theory. Thus, we would like to find a way to ensure that there are non-trivial gauge transformations $\ub$ that are compatible with the boundary condition $\Ab_\mu \in \kf$. This is done by asking $\kf$ to be a subalgebra of $\g \times \g$. Indeed, under this additional assumption, there exists a subgroup $K$ of $G \times G$ with Lie algebra $\kf$ and the gauge transformations such that
\begin{equation}\label{Eq:BCuYB}
\ub \in K
\end{equation}
preserve the boundary condition $\Ab_\mu \in \kf$. We now have to determine under which condition on $\Rc$ the space $\kf$ defined in \eqref{Eq:BC-YB3} is a subalgebra. A sufficient condition for that  is to take $\Rc$ to be a solution of the modified Classical Yang Baxter Equation (mCYBE)
\begin{equation}\label{Eq:mCYBE}
\bigl[ \Rc X, \Rc Y \bigr] - \Rc\bigl[ \Rc X, Y \bigr] - \Rc\bigl[ X, \Rc Y \bigr] + \bigl[ X,Y \bigr] = 0, \qquad \forall \, X,Y\in\g.
\end{equation}
We leave it as an exercise to check this fact and to determine some of the properties of this subalgebra $\kf$.\\

\begin{tcolorbox} \textit{\underline{Exercise 11:} mCYBE and subalgebras of $\g \times\g$.} {\Large$\;\;\star$} \vspace{4pt}\\
Prove that the subspace $\kf$ of $\g\times\g$ is a subalgebra if $\Rc$ satisfies the mCYBE \eqref{Eq:mCYBE}. Moreover, check that we have the following direct sum as vector spaces:
\begin{equation}
\g\times\g = \kf \oplus \g^{\diag},
\end{equation}
where $\g^{\diag} = \bigl\lbrace (X,X), \; X\in\g \bigr\rbrace$ is the diagonal subalgebra of $\g\times\g$ (note that this is not a direct sum of Lie algebras since elements of $\g^{\diag}$ and $\kf$ do not commute). This equips the algebra $\g\times\g$ with the structure of a so-called Drinfel'd double.\end{tcolorbox}

\subsection{The 2-dimensional model}
\label{Sec:2dYB}

\paragraph{2-dimensional fields.} Now that we have determined the boundary conditions at the poles $\Zc = \lbrace +\eta,-\eta,\infty \rbrace$, let us describe the fundamental fields of the resulting 2-dimensional integrable model. Similarly to the example of the PCM treated in Subsection \ref{Sec:FieldPCM}, we will extract them from the 4-dimensional field $\gh$ that parametrises the $\zb$-component of the gauge field $A_{\zb} =  -(\p_{\zb}\gh)\gh^{-1}$. Recall that a gauge transformation $A \mapsto A^u$ acts on $\gh$ as $\gh \mapsto u\gh$. Using this freedom, we can gauge away all degrees of freedom in $\gh$ except for its evaluations at the poles $\Zc = \lbrace +\eta,-\eta,\infty \rbrace$, since $u$ has to satisfy boundary conditions at these points.

For instance, we required in Equation \eqref{Eq:uInfty-YB} that $u|_{z=\infty}=\Id$. We thus cannot gauge away the evaluation $\gh|_{z=\infty}$. Recall however from Subsection \ref{Sec:FreedomGh} that there is another freedom in the definition of $\gh$, which acts as $\gh  \mapsto \gh v$ for $v : \Sigma \rightarrow G$ independent of $z$. We can use this freedom to fix
\begin{equation}
\gh \bigl|_{z=\infty} = \Id.
\end{equation}

Let us now turn our attention to the degrees of freedom associated with the poles $\lbrace +\eta,-\eta \rbrace$. To extract them from $\gh$, we define
\begin{equation}\label{Eq:DefGb}
\gb = \bigl( \gh \bigl|_{z=+\eta}, \gh \bigl|_{z=-\eta} \bigr) \; \in \; G \times G.
\end{equation}
Under a gauge transformation $\gh \mapsto u\gh$, this field transforms as
\begin{equation}
\gb \longmapsto \ub \gb,
\end{equation}
where we recall that $\ub$ is defined as in Equation \eqref{Eq:uYB}. Recall moreover that to ensure compatibility with the boundary condition, we imposed in Equation \eqref{Eq:BCuYB} that $\ub$ is valued in the subgroup $K$ of $G\times G$. Thus, we can gauge away the $K$-valued degrees of freedom in the field $\gb$, or in other words the surviving degrees of freedom left after gauge transformations belong to the quotient $K\! \setminus \!(G\times G)$. Let us describe these degrees of freedom more explicitly. We have seen in Exercise 7 that, at the level of the Lie algebra, we have the vector space decomposition $\g\times\g = \kf \oplus \g^{\diag}$. We will suppose that this decomposition lifts to the group and thus that $G\times G = K \cdot G^{\diag}$. Using the gauge freedom $\gb\mapsto\ub\gb$ with $\ub\in K$, we can thus fix $\gb$ to be valued in the diagonal subgroup $G^{\diag}$. In other words, we can fix $\gb$ to be of the form
\begin{equation}
\gb = (g,g) \in G^{\text{diag}},
\end{equation}
where $g : \Sigma \rightarrow G$ is a 2-dimensional $G$-valued field. This is the fundamental field of the model that we are constructing in this subsection. Recalling the definition \eqref{Eq:DefGb} of $\gb$, we rewrite the above equation as the following condition on $\gh$:
\begin{equation}\label{Eq:gYB}
\gh \bigl|_{z=+\eta} = \gh \bigl|_{z=-\eta} = g.
\end{equation}

\paragraph{Lax connection.} Now that we have extracted the 2-dimensional field $g$ of the model, let us determine its Lax connection $\Lc_\pm$. Following the general results of Subsection \ref{Sec:Lax4Db} and in particular Equation \eqref{Eq:Lz}, $\Lc_\pm$ has poles at the zeroes $\Ze_\pm$ of $\omega$. In the present case, $\omega$ has zeroes $\Ze=\lbrace +1,-1\rbrace$. We choose to separate them into $\Ze_\pm=\lbrace \pm 1 \rbrace$, as in the case of the PCM treated in Subsection \ref{Sec:LaxPCM}. Similarly to this case, we then get that the Lax connection takes the form
\begin{equation}\label{Eq:LaxYB0}
\Lc_\pm = \frac{U_\pm}{z \mp 1} + V_\pm,
\end{equation}
where $U_\pm$ and $V_\pm$ are $\g$-valued fields on $\Sigma$. To determine them, we need to impose the boundary conditions on the gauge field components
\begin{equation}
A_\pm = \gh\,\Lc_\pm\,\gh^{-1} - \bigl( \p_\pm \gh)\gh^{-1}.
\end{equation}
Using the fact that we fixed $\gh|_{z=\infty}=\Id$ in the previous paragraph, we get
\begin{equation}
A_\pm \bigl|_{z=\infty} = \Lc_\pm\bigl|_{z=\infty} = V_\pm.
\end{equation}
Thus the boundary condition $A_\pm|_{z=\infty}=0$ (see above) simply implies that $V_\pm=0$.

We are left to determine the fields $U_\pm$. This is done by imposing the boundary condition \eqref{Eq:BC-YB} at the poles $\lbrace +\eta,-\eta \rbrace$. After a few manipulations, one expresses $U_\pm$ in terms of the field $g=\gh|_{z=+\eta} = \gh|_{z=-\eta}$. Reinserting in the Lax connection, we get in the end
\begin{equation}\label{Eq:LaxYB}
\Lc_\pm = \frac{1-\eta^2}{1 \mp z}  \frac{1}{1 \pm \eta\,\Rc_g} g^{-1}\p_\pm g,
\end{equation}
where $\Rc_g = \Ad_g^{-1} \circ \Rc \circ \Ad_g$. This coincides with the Lax connection of the Yang-Baxter model, as constructed in~\cite{Klimcik:2008eq} and reviewed in Ben Hoare's lectures \textit{``Integrable deformations of sigma models''}~\cite{LectureHoare}. Moreover, taking $\eta=0$, we recover the Lax connection \eqref{Eq:LaxPCM2} of the PCM, justifying that the model that we are currently constructing is a deformation of the latter. We leave as an exercise the derivation of Equation \eqref{Eq:LaxYB}.\vspace{6pt}

\begin{tcolorbox} \textit{\underline{Exercise 12:} Lax connection of the Yang-Baxter model.} {\Large$\;\;\star$} \vspace{4pt}\\
Prove Equation \eqref{Eq:LaxYB} by imposing the boundary condition \eqref{Eq:BC-YB} and using Equation \eqref{Eq:gYB}.
\end{tcolorbox}

\paragraph{Action.} Let us finally discuss the action of the model. Similarly to the case of the PCM treated in Subsection \ref{Sec:ActionPCM}, we start with the expression \eqref{Eq:ActionPCM1} of the 4-dimensional action in terms of the fields $\gh$ and $\Lc_\pm$. We then reinsert the expression of $\gh$ and $\Lc_\pm$ in terms of $g$ in this equation, after using the gauge transformations to bring $\gh$ to a simple form. We then explicitly perform the integral over $z\in\CP$ and obtain in the end a 2-dimensional action depending only on the field $g$. More precisely, we find
\begin{equation}\label{Eq:ActionYB}
S[g] = -\hay \iint_\Sigma \dd x^+ \, \dd x^- \; \Tr \left( g^{-1}\p_+ g \; \frac{1}{1-\eta \,\Rc_g} g^{-1} \p_- g \right).
\end{equation}
As for the Lax connection, this action agrees with the one of the Yang-Baxter model, first introduced in~\cite{Klimcik:2002zj} and discussed in Ben Hoare's lectures \textit{``Integrable deformations of sigma models''}~\cite{LectureHoare}. Moreover, we check that in the limit $\eta=0$, we recover the action of the PCM \eqref{Eq:ActionPCM}. The derivation of this action is the subject of the next exercise.\vspace{6pt}

\begin{tcolorbox} \textit{\underline{Exercise 13:} Action of the Yang-Baxter model.} {\Large$\;\;\star\star$} \vspace{4pt}\\
The goal of this exercise is to prove Equation \eqref{Eq:LaxYB}. We will do it step by step.
\begin{enumerate}
\item We first need to ``prepare'' $\gh$, similarly to what we did for the PCM in Subsection \ref{Sec:ActionPCM}. Show that using the gauge transformations $\gh \mapsto u\gh$, one can always bring $\gh$ to a form satisfying the following properties (following the terminolgy of~\cite{Delduc:2019whp}, we then say that $\gh$ satisfies the archipelago condition):
\begin{itemize}
\item[$\bullet$] there exists $a$, $0<a<\eta$, such that $\gh$ is equal to $g$ on $\Sigma \times D_{\pm \eta,a}$, where $D_{\pm \eta,a} \subset \CP$ are disks of radius $a$ centred around $z=\pm \eta$ ;
\item[$\bullet$] there exists $b$, $a<b<\eta$, such that $\gh$ is constant equal to $\Id$ on $\Sigma \times \bigl( \CP \setminus (D_{+\eta,b} \sqcup D_{-\eta,b}) \bigr)$ ;
\item[$\bullet$] on $\Sigma \times D_{\pm \eta,b}$, $\gh$ depends only on $t$, $x$ and $r_\pm = |z \mp \eta|$ and interpolates between $g$ at $r_\pm = a$ and $\Id$ at $r_\pm = b$.
\end{itemize}

\end{enumerate} 
\end{tcolorbox}
\begin{tcolorbox}
\begin{enumerate}\setcounter{enumi}{1}
\item Use the identity \eqref{Eq:DerPoleDirac} to express the first term in Equation \eqref{Eq:ActionPCM1} as a 2-dimensional term on $\Sigma \times \lbrace \infty,+\eta,-\eta\rbrace$. Reinsert the above form of $\gh$ and the expression \eqref{Eq:LaxYB} of $\Lc_\pm$ and conclude that this term is equal on its own to the action \eqref{Eq:ActionYB}.
\item To conclude, we thus need to prove that the second term in Equation \eqref{Eq:ActionPCM1} vanishes. To do so, first use the form of $\gh$ to write this term as the sum of two integrals over $\Sigma \times D_{+\eta,b}$ and $\Sigma \times D_{-\eta,b}$. Then use the same strategy as in Subsection \ref{Sec:ActionPCM} for the PCM to express these integrals as Wess-Zumino terms for $g|_{z=+\eta}=g$ and $g|_{z=-\eta}=g$, integrating over the angle coordinates $\theta_\pm=\arg(z\mp\eta)$ of the disks $D_{\pm\eta,b}$. Finally prove that these two contributions cancel due to the fact that\vspace{-3pt}
\begin{equation}\label{Eq:OppositeLevelsYB}
\res_{z=+\eta} \omega \; + \; \res_{z=-\eta}\omega = 0.
\end{equation}
\end{enumerate}
\end{tcolorbox}

\subsection{To go further}

\paragraph{Split vs non-split.} In this subsection, we have constructed the Yang-Baxter model based on an $R$-matrix satisfying the mCYBE \eqref{Eq:mCYBE}. This operator is more precisely a split $R$-matrix, since the last term in \eqref{Eq:mCYBE} is $+[X,Y]$. One can also consider non-split solutions, which satisfy the mCYBE with the last term replaced by $-[X,Y]$, and their corresponding Yang-Baxter model (see Ben Hoare's lectures \textit{``Integrable deformations of sigma models''}~\cite{LectureHoare} for more details). This variant can also be obtained from the 4d Chern-Simons theory by changing the nature of the reality conditions on the poles. Instead of two simple real poles $z=\pm \eta$ for the split case, the non-split model is obtained by considering two simple complex poles $z=\pm i\eta$, conjugate to each other. The derivation of the 2-dimensional non-split model from the 4d-CS theory is very similar to the one of the split model described in this section. The main difference is that the real double $\g \!\times\! \g$ is replaced by the complex double $\g^{\C}$ and the diagonal subalgebra $\g^{\diag}$ by the real form $\g$ inside $\g^{\mathbb{C}}$ (see~\cite{Delduc:2019whp} for more details).

\paragraph{Bi-Yang-Baxter model.} In this section, we have rederived the Yang-Baxter model starting from the 4d-CS theory with a 1-form $\omega$ that possesses two simple poles at $z=\pm \eta$ and a double pole at $z=\infty$. The simple poles at $z=\pm \eta$ can be seen as a deformation of the double pole at $z=0$ of the PCM. One can further deform the theory by considering a deformation of the double pole at $z=\infty$ into a pair of simple poles. We obtain in this way a second deformation of the Yang-Baxter model, which is called the Bi-Yang-Baxter model~\cite{Klimcik:2008eq,Klimcik:2014bta}. This is the subject of the following exercise.\vspace{6pt}

\begin{tcolorbox} \textit{\underline{Exercise 14:} Bi-Yang-Baxter model.} {\Large$\;\;\star\star\star$} \vspace{4pt}\\
We consider the meromorphic 1-form\vspace{-2pt}
\begin{equation}
\omega = K\frac{1-z^2}{(z-z_1)(z-z_2)(z-z_3)(z-z_4)}\dd z,\vspace{-2pt}
\end{equation}
with two simple zeroes $\Ze_\pm = \lbrace \pm 1 \rbrace$ and four simple poles $\Zc =\lbrace z_1,z_2,z_3,z_4 \rbrace$ given by\vspace{-2pt}
\begin{equation}
z_1 = \alpha_L, \quad z_2 = -\alpha_L, \quad z_3 = \alpha_R^{-1}, \quad z_4 = -\alpha_R^{-1}.\vspace{-1pt}
\end{equation}
Here $K$, $\alpha_L$ and $\alpha_R$ are real parameters.
\end{tcolorbox}
\begin{tcolorbox}
\begin{enumerate}
\item Check that the levels $\ell_{r,0}=\res_{z=z_r}\omega$ satisfy $\ell_{1,0}+\ell_{2,0}=\ell_{3,0}+\ell_{4,0}=0$. Write the defect terms associated with the pairs of poles $\lbrace z_1,z_2 \rbrace$ and $\lbrace z_3,z_4 \rbrace$ and compare to the defect term considered for the Yang-Baxter model in Subsection \ref{Sec:4dYB}. Deduce that we can make these defect terms vanish by imposing the boundary conditions
\begin{equation}
\bigl( A_\pm\bigl|_{z=z_1},A_\pm\bigl|_{z=z_2} \bigr) \in \kf \qquad \text{ and } \qquad \bigl( A_\pm\bigl|_{z=z_3},A_\pm\bigl|_{z=z_4} \bigr) \in \kf,
\end{equation}
where $\kf = \bigl\lbrace \bigl( (\Rc-\Id)X, (\Rc+\Id)X \bigr), \; X\in\g \bigr\rbrace$ is the subalgebra of $\g \times \g$ introduced in Subsection \ref{Sec:4dYB} and in particular in Equation \eqref{Eq:BC-YB3}.
\item Prove that, acting with gauge transformations compatible with these boundary conditions, we can put $\gh|_{z=z_1}=\gh|_{z=z_2}$ and $\gh|_{z=z_3}=\gh|_{z=z_4}$ (use similar reasoning as in Subsection \ref{Sec:2dYB}). We then use the freedom $\gh \mapsto \gh v$ to fix $\gh|_{z=z_3}=\gh|_{z=z_4}=\Id$ . The only surviving degree of freedom in $\gh$ is then the 2-dimensional field $g=\gh|_{z=z_1}=\gh|_{z=z_2}$.
\item According to the results of Subsection \ref{Sec:Lax4Db} and in particular Equation \eqref{Eq:Lz}, the Lax connection takes the form
\begin{equation}
\Lc_\pm = \frac{U_\pm}{z \mp 1} + V_\pm.
\end{equation}
Impose the boundary conditions on $A_\pm$ at the poles $z_r$ to determine $U_\pm$ and $V_\pm$ in terms of $g$:
\begin{equation}
U_\pm = \mp \frac{(1-\alpha_L^2)(1-\alpha_R^2)}{1-\alpha_L^2\alpha_R^2} \frac{1}{1\pm \eta_L\,\Rc_g \pm\eta_R\,\Rc}g^{-1}\p_\pm  g\qquad \text{ and } \qquad V_\pm = \mp \alpha_R \frac{1 \pm \alpha_R \Rc}{1-\alpha_R^2}U_\pm,
\end{equation}
with
\begin{equation}
\eta_L = \alpha_L \frac{1-\alpha_R^2}{1-\alpha_L^2\alpha_R^2} \qquad \text{ and } \qquad \eta_R = \alpha_R \frac{1-\alpha_L^2}{1-\alpha_L^2\alpha_R^2}.
\end{equation}
\item Reinsert $\Lc_\pm$ and $\gh$ in the action \eqref{Eq:ActionPCM1} and perform the integration over $z$, after choosing an appropriate gauge for $\gh$ (inspire yourself from the previous examples treated in the lectures and the exercises). Show that the 2-dimensional action finally reads
\begin{equation}
S[g] = -\hay \iint_\Sigma \dd x^+ \, \dd x^- \; \Tr \left( g^{-1}\p_+ g \; \frac{1}{1-\eta_L \,\Rc_g - \eta_R\,\Rc} g^{-1} \p_- g \right),
\end{equation}
with
\begin{equation}
\hay = -\frac{K \alpha_R^2(1-\alpha_L^2)(1-\alpha_R^2)}{(1-\alpha_L^2\alpha_R^2)^2}.
\end{equation}
This coincides with the action of the bi-Yang-Baxter model introduced in~\cite{Klimcik:2008eq,Klimcik:2014bta}.
\item Discuss the limit $\alpha_R \to 0$.
\end{enumerate}
\end{tcolorbox}

\paragraph{Yang-Baxter model with Wess-Zumino term.} We have seen in Exercise 9 that the second term in the action \eqref{Eq:ActionPCM1} vanishes for the Yang-Baxter model, due to Equation \eqref{Eq:OppositeLevelsYB} which states that the residues of $\omega$ at $z=\pm \eta$ (the levels of the theory) are opposite. One can consider a more general setting were these levels are not opposite: this results in the Yang-Baxter model with a Wess-Zumino term. This model was first introduced in~\cite{Delduc:2014uaa} and constructed in more generality in~\cite{Hoare:2020mpv} (building also on the results of~\cite{Klimcik:2019kkf}): we refer to Ben Hoare's lectures \textit{``Integrable deformations of sigma models''}~\cite{LectureHoare} for more comments about its construction. Its 4d-CS interpretation was described in~\cite{Delduc:2019whp,Hoare:2020mpv}. This 4-dimensional setup requires a careful treatment of boundary conditions, which is slightly more complicated than the one of the Yang-Baxter model without Wess-Zumino term: for brevity, we will not explore these aspects further in these notes and refer to~\cite{Delduc:2019whp,Hoare:2020mpv} for more details.

\paragraph{The $\bm{\lambda}$-model.} In addition to the Yang-Baxter model, another prototypical example of integrable deformed $\sigma$-model is the $\lambda$-model, introduced by Sfetsos in~\cite{Sfetsos:2013wia}. It can also be obtained from the 4d-CS theory~\cite{Delduc:2019whp} (see also~\cite{Schmidtt:2019otc}), starting from the same 1-form $\omega$ as the (split) Yang-Baxter model, given in Equation \eqref{Eq:OmegaYB}. The crucial difference with the Yang-Baxter model is the choice of a different boundary condition at the pair of poles $\lbrace +\eta,-\eta \rbrace$. The details of this construction are the subject of Exercise 10 below. The fact that the Yang-Baxter and $\lambda$-models are obtained from the same 1-form $\omega$ but different boundary conditions result in the end in a deep relation between these two models~\cite{Vicedo:2015pna,Hoare:2015gda,Sfetsos:2015nya,Klimcik:2015gba}, the so-called Poisson-Lie T-duality~\cite{Klimcik:1995ux,Klimcik:1995jn,Klimcik:1995dy}. We refer to~\cite{Delduc:2019whp,Lacroix:2020flf} for more details about the occurrence of Poisson-Lie T-duality in the context of 4d-CS theory.\\

\begin{tcolorbox} \textit{\underline{Exercise 15:} $\lambda$-model.} {\Large$\;\;\star\star$} \vspace{4pt}\\
We consider the same 1-form $\omega$ as in the (split) Yang-Baxter model, given by Equation \eqref{Eq:OmegaYB}.
\begin{enumerate}
\item Check that $A_\pm|_{z=+\eta}=A_\pm|_{z=-\eta}$ is an appropriate choice of boundary condition to ensure that the defect term \eqref{Eq:DefectYB} vanishes. This choice of boundary condition differs from the one considered for the Yang-Baxter model in Subsection \ref{Sec:4dYB}. For the double pole at infinity, we keep the same boundary condition $A_\pm|_{z=\infty}=0$.
\item Show that the gauge transformations $A \mapsto A^u$ compatible with the boundary conditions are the ones satisfying $u|_{z=+\eta}=u|_{z=-\eta}$.
\item We use the freedom $\gh \mapsto \gh v$ with $v : \Sigma \rightarrow G$ to fix $\gh|_{z=\infty}=\Id$. Argue that the physical degrees of freedom of the theory are then extracted from the evaluations $\gh|_{z=+\eta}$ and $\gh|_{z=-\eta}$ and that the gauge transformations compatible with the boundary conditions result in an additional freedom $(\gh|_{z=+\eta},\gh|_{z=-\eta})\mapsto (u_0\,\gh|_{z=+\eta},u_0\,\gh|_{z=-\eta})$ with $u_0:\Sigma \rightarrow G$. We can use this freedom to fix the field $\gh|_{z=-\eta}$ to the identity and then define the 2-dimensional field of the model $g$ as $\gh|_{z=+\eta}$. We thus have\vspace{-3pt}
\begin{equation}
\gh|_{z=+\eta} = g \qquad \text{ and } \qquad \gh|_{z=-\eta}=\Id.\vspace{-2pt}
\end{equation}
\item The Lax connection takes the same form \eqref{Eq:LaxYB0} as in the Yang-Baxter model. Impose the boundary conditions on $A_\pm$ at the poles $\Zc=\lbrace \infty,+\eta,-\eta \rbrace$ to determine $U_\pm$ and $V_\pm$. Conclude that
\begin{equation*}
\Lc_+ = \frac{1}{1-z}\frac{2\lambda}{1+\lambda} \frac{1}{\lambda-\Ad_g^{-1}}g^{-1}\p_+ g \qquad \text{ and } \qquad \Lc_- = \frac{1}{1+z} \frac{2}{1+\lambda} \frac{1}{\lambda^{-1}-\Ad_g^{-1}}g^{-1}\p_- g,
\end{equation*}
\end{enumerate}
\end{tcolorbox}
\begin{tcolorbox}
\begin{enumerate}\setcounter{enumi}{4}
\item[] where we have introduced
\begin{equation}
\lambda = \frac{1+\eta}{1-\eta}.
\end{equation}
\item Finally compute the 2-dimensional action of the model using a similar strategy as in the previous examples and exercises and prove that
\begin{equation}\label{Eq:ActionLambda}
S[g] =  S_{\text{WZW},\kay}[g] - 2\kay \iint \dd x^+\,\dd x^- \; \Tr\left(\p_+ gg^{-1}\,\frac{1}{\lambda^{-1}-\Ad^{-1}_g} g^{-1}\p_- g \right),
\end{equation}
where
\begin{equation*}
\kay = - \frac{1}{2}\res_{z=+\eta} \omega \qquad \text{ and } \qquad S_{\text{WZW},\kay}[g] = -\kay \iint_\Sigma \dd x^+ \, \dd x^- \; \Tr \bigl( g^{-1}\p_+ g \; g^{-1} \p_- g \bigr) + \kay\,I_{\text{WZ}}[g]
\end{equation*}
is the conformal Wess-Zumino-Witten action. We recognise in Equation \eqref{Eq:ActionLambda} the action of the $\lambda$-model, introduced in~\cite{Sfetsos:2013wia}.
\end{enumerate}
\end{tcolorbox}

\paragraph{Trigonometric formulation.} In this section, we have focused on the rational formulation of the Yang-Baxter model, \textit{i.e.} with a Lax connection depending rationally on the spectral parameter $z\in\CP$. This model also possesses a trigonometric formulation~\cite{Kawaguchi:2012gp}, with a Lax connection expressed in terms of trigonometric functions of the spectral parameter, which is then seen as belonging to the cylinder instead of the Riemann sphere. Based on this observation, it was shown in~\cite{Fukushima:2020kta} that this trigonometric formulation can also be obtained from 4d-CS theory by replacing the Riemann sphere $\CP$ with the cylinder in the choice of 4-dimensional space-time. By considering a variant of the boundary condition in this trigonometric formulation, another type of Yang-Baxter deformation was also constructed in~\cite{Fukushima:2020kta} (we will not develop this aspect further here and refer to this reference for more details). A similar analysis was carried out in~\cite{Tian:2020ryu} for the $\lambda$-model.

\subsection{General boundary conditions}
\label{Sec:GenBC}

\paragraph{Jets and isotropic subspaces.} Let us end this section with a brief sketch on the treatment of more general boundary conditions, following the results of~\cite{Benini:2020skc,Lacroix:2020flf}. We consider a general $\omega$, of the form \eqref{Eq:OmegaLevels}, with poles $z_r$, $r\in P$, of arbitrary order $m_r\in\Z_{\geq 1}$. Following the analysis of Subsection \ref{Sec:Defect} and in particular Equation \eqref{Eq:DefectHigherOrder}, the defect term associated with a pole $z_r$ involves the evaluation of the gauge field $A_\mu$ on the defect $\Sigma \times \lbrace z_r \rbrace$, as well as the evaluation of its derivatives $\p_z^p A_\mu$ for $p$ up to $m_r-1$. We gather these quantities, for all poles, in a unique object
\begin{equation}
\At_\mu = \Bigl( \p_z^p A_\mu\bigl|_{z=z_r} \Bigr)_{r}^{p\in\lbrace 0,\cdots,m_r-1\rbrace} \;,
\end{equation}
which is valued in a vector space $\g^{[\Zc]}$ built from copies of the algebra $\g$ and its complexification $\g^{\C}$ (depending on the reality conditions of the poles $\Zc$). This object is called the \textit{jet of $A$} on the poles defect $\Sigma \times \Zc$. In terms of this object, the vanishing of the defect term in the variation of the 4d-CS action can be rewritten as
\begin{equation}\label{Eq:DefectForm}
\sum_{r\in P} \Dc_r =
\iint_\Sigma \dd t\,\dd x\;\epsilon^{\mu\nu} \llangle \At_\mu, \delta \At_\nu \rrangle = 0,
\end{equation}
where $\llangle \cdot,\cdot\rrangle$ is a well-chosen bilinear form on the vector space $\g^{[\Zc]}$, whose definition depends on the levels $\ell_{r,p}$ associated with the poles.

To ensure that this defect equation is satisfied, we impose the boundary condition 
\begin{equation}\label{Eq:BC-Gen}
\At_\mu \in \ff, \qquad \mu=t,x,
\end{equation}
where $\ff$ is an isotropic subspace of $\g^{[\Zc]}$ with respect to the bilinear form $\llangle \cdot,\cdot\rrangle$. Here, isotropic means that
\begin{equation}
\llangle \mathbb{X},\mathbb{Y}\rrangle = 0, \qquad \forall\,\mathbb{X},\mathbb{Y}\in\ff.
\end{equation}
Indeed, restricting to $\At_\mu$ valued in $\ff$ and thus to variations $\delta\At_\mu$ also valued in $\ff$, we get that $\llangle \At_\mu,\delta\At_\nu \rrangle = 0$ by isotropy of $\ff$, ensuring the defect equation \eqref{Eq:DefectForm}.

Let us quickly comment on how this general formulation can be illustrated through the example of the Yang-Baxter model treated in the previous subsection. For this example, $\omega$ possesses two simple poles $z_1=+\eta$ and $z_2=-\eta$ and a double pole at infinity, hence $\Zc=\lbrace +\eta,-\eta,\infty \rbrace$. Since we treated the pole at infinity separately, we focused in Subsection \ref{Sec:4dYB} on the evaluations $\Ab_\mu=(A_\mu|_{z=+\eta},A_\mu|_{z=-\eta})$, which belonged to the vector space $\df=\g\times\g$. In the general formulation sketched above, this space $\df$ is the subspace of $\g^{[\Zc]}$ that corresponds to the poles $\lbrace +\eta,-\eta \rbrace \subset \Zc$ and $\Ab_\mu$ is the part of $\At_\mu$ corresponding to these poles. In Subsection \ref{Sec:4dYB}, we have imposed the boundary condition $\Ab_\mu \in \kf$, where $\kf$ is the subspace of $\df$ defined in terms of $\Rc$ as in Equation \eqref{Eq:BC-YB3}. We then have found that the vanishing of the defect term required that $\Rc$ is a skew-symmetric operator. In the language developed above, this requirement can be reinterpreted as the condition for $\kf$ to be an isotropic subspace of $\df$. The treatment of the Yang-Baxter model in Subsection \ref{Sec:4dYB} can thus be seen as a particular case of the general formulation sketched above (where we have singled out the poles $\lbrace +\eta,-\eta \rbrace \subset \Zc$).

\paragraph{Compatible gauge transformations.} The next step in the treatment of the defect term is to determine the gauge transformations that are compatible with the boundary condition \eqref{Eq:BC-Gen}. For that, we need to examine which action a formal gauge transformation $A\mapsto A^u$ induces on the jet $\At_\mu$. To describe this action, one needs to introduce some well-chosen algebraic structure on $\g^{[\Zc]}$. In particular, as explained in~\cite{Benini:2020skc}, there exists a natural Lie bracket $[\cdot,\cdot]_{[\Zc]}$ on $\g^{[\Zc]}$, which turns into a Lie algebra that we call the \textit{defect Lie algebra}. The definition of this bracket depends on the order of the poles $\Zc$ and is such that the bilinear form $\llangle\cdot,\cdot\rrangle$ is ad-invariant with respect to this bracket. For instance, if all the poles in $\Zc$ are real and simple, then the Lie algebra $\g^{[\Zc]}$ is simply equal to the direct product $\g^{\times|\Zc|}$. For more general types of poles, this Lie algebra is defined in a slightly more complicated way and involves what are called Takiff or jet algebras. We will not enter into more details on this construction in these lecture notes and refer to~\cite{Benini:2020skc} for more details. Since $\g^{[\Zc]}$ possesses a Lie algebraic structure, one can define the corresponding connected and simply connected Lie group $G^{[\Zc]}$, which we call the \textit{defect group}.

We now have enough material to describe the action of the gauge transformation $A\mapsto A^u$ induced on the jet $\At_\mu \in \g^{[\Zc]}$. First, one checks that this action depends only on the evaluations $\p_z^p u|_{z=z_r}$ for $p\in\lbrace 0,\cdots,m_r-1\rbrace$. These evaluations can be naturally gathered into one unique object $\ut$ valued in the defect group $G^{[\Zc]}$. In terms of this object, the induced gauge transformation on $\At_\mu$ is given by
\begin{equation}\label{Eq:GaugeDefect}
\At_\mu \longmapsto \At_\mu^{\ut} = \ut\,\At_\mu\,\ut^{-1} - (\p_\mu \ut)\ut^{-1}.
\end{equation}
In this equation, the inversion $\ut^{-1}$ is defined with respect to the group structure on $G^{[\Zc]}$. Similarly, the conjugacy of $\At_\mu$ by $\ut$ is defined through the adjoint action of the group $G^{[\Zc]}$ on its Lie algebra $\g^{[\Zc]}$. Thus the Lie algebra and group structures introduced on $\g^{[\Zc]}$ and $G^{[\Zc]}$ naturally capture the form of the gauge transformations of $\At_\mu$.

We now have the proper formalism to discuss the compatibility of the gauge transformations with the boundary condition. Indeed, we want the gauge transformation \eqref{Eq:GaugeDefect} to preserve the boundary condition \eqref{Eq:BC-Gen}, \textit{i.e.} we want $\At_\mu^{\ut} \in\ff$. This is similar to the setup which we encountered in Subsection \ref{Sec:4dYB} for the case of the Yang-Baxter model, where the roles of $\At_\mu$, $\ut$, $\g^{[\Zc]}$, $G^{[\Zc]}$ and $\ff$ were played by $\Ab_\mu$, $\ub$, $\df=\g\times\g$, $D=G\times G$ and $\kf$ (which correspond to removing the objects related to the pole at $\infty$, treated separately). It was argued in this example that one can ensure the existence of non-trivial compatible gauge transformations by requiring that $\kf$ is a subalgebra of $\df$. In the present case, we thus do a similar hypothesis and suppose that $\ff$ is a subalgebra of $\g^{[\Zc]}$, equipped with the bracket $[\cdot,\cdot]_{[\Zc]}$. We then denote by $F$ the corresponding subgroup of $G^{[\Zc]}$ and require that $\ut$ is valued in $F$. With these restrictions, it is clear that the gauge transformation \eqref{Eq:GaugeDefect} preserves the boundary condition $\At_\mu\in\ff$.

Thus, the problem of finding appropriate boundary conditions at the poles defect amounts in the end to finding isotropic subalgebras of $\g^{[\Zc]}$ (subalgebras with respect to the Lie bracket $[\cdot,\cdot]_{[\Zc]}$ and isotropic with respect to the bilinear form $\llangle\cdot,\cdot\rrangle$). In fact, to get well-posed models, one should also require these subalgebras to be maximal isotropic (we refer to~\cite{Benini:2020skc,Lacroix:2020flf} for more details on this aspect).

\paragraph{The 2-dimensional fields.} Given a maximal isotropic subalgebra $\ff$ of $\g^{[\Zc]}$, we have now defined an appropriate choice of boundary conditions on the poles defect $\Sigma \times \Zc$. We then have all the ingredients to extract an integrable 2-dimensional model from this 4d-CS theory. As in the examples considered above, the fundamental fields of this 2d model are extracted from $\gh$. Recall that gauge transformations act on $\gh$ as $\gh \mapsto u\gh$ and should be compatible with the boundary conditions on the poles defect $\Sigma \times \Zc$. The degrees of freedom in $\gh$ at points outside of this defect can thus be gauged away by such a transformation. Similarly, the evaluations $\p_z^p \gh|_{z=z_r}$ for $p\geq m_r$ can also be eliminated by gauge transformations that trivially preserve the boundary conditions. Thus, the 2d fundamental fields can only be extracted from the evaluations $\p_z^p \gh|_{z=z_r}$, for $p\in\lbrace0,\cdots,m_r-1 \rbrace$. These evaluations naturally form a 2d field $\gt: \Sigma \rightarrow G^{[\Zc]}$, valued in the defect group $G^{[\Zc]}$.

The action induced by a gauge transformation $\gh\mapsto u\gh$ on this field $\gt$ is given by $\gt \mapsto \ut \gt$, where $\ut$ is the $G^{[\Zc]}$-valued field considered in the previous paragraph and formed from the evaluations $\p_z^p u|_{z=z_r}$ and where the product of $\ut$ and $\gt$ is defined from the group structure on $G^{[\Zc]}$. Recall that $\ut$ must be valued in the subgroup $F$ of $G^{[\Zc]}$ to be compatible with the boundary condition. Thus, there is a residual gauge transformation $\gt \mapsto \ut \gt$ on $\gt$ by $\ut \in F$. Recall finally from Subsection \ref{Sec:FreedomGh} that the parametrisation of $A_{\bar z}$ in terms of $\gh$ is invariant under a redefinition $\gh \mapsto \gh v$ where $v:\Sigma \mapsto G$ is independent of $z$. This results in another gauge transformation on $\gt$, which takes the form of the multiplication on the right by an element $\Delta(v)$ in $G^{[\Zc]}$, where $\Delta : G \rightarrow G^{[\Zc]}$ is an embedding of $G$ in $G^{[\Zc]}$ (if all the poles $\Zc$ are real and simple, in which case $G^{[\Zc]}=G^{\times|\Zc|}$, $\Delta$ is simply the diagonal embedding $\Delta(v)=(v,v,\cdots,v)$). In the end, we then have a $G^{[\Zc]}$-valued field $\gt$, defined up to gauge transformations
\begin{equation}\label{Eq:GaugeEdge}
\gt \longmapsto \ut\, \gt\, \Delta(v), \qquad \text{ with }\quad \ut\in F, \quad v\in G.
\end{equation}
The physical degrees of freedom of the model are then naturally described by the double coset $F\!\setminus\! G^{[\Zc]} / \Delta(G)$.

\paragraph{Lax connection and action.} The next step in the construction of the 2d integrable model is to describe the Lax connection $\Lc_\pm$ of the model. From the general results of Subsection \ref{Sec:Lax4Db} and in particular Equation \eqref{Eq:Lz}, its $z$-dependence is fixed by the choice of a separation of the zeroes $\Ze$ of $\omega$ into two subsets $\Ze_\pm$. For the reader's convenience, we recall this equation here:
\begin{equation}
\Lc = \left( \sum_{y \in \Ze_+} \frac{U_+^{(y)}}{z-y} + V_+ \right) \dd x^+ + \left( \sum_{y \in \Ze_-} \frac{U_-^{(y)}}{z-y} + V_- \right) \dd x^-,
\end{equation}
where $U_\pm^{(y)}$ and $V_\pm$ are $\g$-valued fields on $\Sigma$ (independent of $z$). One now has to relate these fields to the fundamental field $\gt$ of the model. This is done by applying the boundary condition \eqref{Eq:BC-Gen} on the gauge field $A_\mu$, or more precisely on its jet $\At_\mu \in \g^{[\Zc]}$. We recall that the components $A_\pm$ of the gauge field are related to $\gh$ and $\Lc_\pm$ by $A_\pm = \gh\,\Lc_\pm\,\gh^{-1} - (\p_\pm \gh)\gh^{-1}$. Taking the jet of this equation, we get
\begin{equation}
\At_\pm = \gt\,\Lt_\pm\,\gt^{-1}  - (\p_\pm \gt)\gt^{-1},
\end{equation}
where
\begin{equation}
\Lt_\pm = \Bigl( \p_z^p \Lc_\pm\bigl|_{z=z_r} \Bigr)_{r}^{p\in\lbrace 0,\cdots,m_r-1\rbrace} \;\in \g^{[\Zc]}
\end{equation}
is the jet of the Lax connection. The boundary condition \eqref{Eq:BC-Gen} can thus be rewritten as
\begin{equation}\label{Eq:BC-Jet}
\gt\,\Lt_\pm\,\gt^{-1}  - (\p_\pm \gt)\gt^{-1} \; \in \ff.
\end{equation}
This can be seen as a relation between the fields $U_\pm^{(y)}$ and $V_\pm$ appearing in the Lax connection and the fundamental field $\gt$ of the model. From now on, we suppose that we have solved this relation and thus that we have written the jet $\Lt_\pm(\gt)$ in terms of the field $\gt$.

We now have enough material to express the action of the 2-dimensional integrable model as a functional of $\gh$. More precisely, as shown in~\cite{Benini:2020skc}, it is given by
\begin{equation}
S[\gt] = \iint_\Sigma \left\llangle \Lt(\gt), \gt^{-1} \dd \gt \right\rrangle_{[\Zc]} + I_{\text{WZ},G^{[\Zc]}}[\gt],
\end{equation}
where $I_{\text{WZ},G^{[\Zc]}}[\gt]$ is the Wess-Zumino term of the $G^{[\Zc]}$-valued field $\gt$, defined with respect to the bilinear form $\llangle \cdot,\cdot \rrangle_{[\Zc]}$ on the Lie algebra $\g^{[\Zc]}$. By construction, this action is invariant under the gauge transformations \eqref{Eq:GaugeEdge}.\\

Let us end this paragraph with a remark. The approach we have sketched in this subsection follows the same strategy as in the examples of the PCM and the Yang-Baxter model treated above. In particular, we impose a boundary condition on the gauge field and restrict to gauge transformations $A\mapsto A^u$ that are compatible with this condition. The 2-dimensional field $\gt$ is then extracted from the 4-dimensional field $\gh$, which parametrises the component $A_{\bar z}$ of the gauge field: more precisely, $\gt$ corresponds to the degrees of freedom in $\gh$ that cannot be trivially gauged away. The existence of these surviving 2d degrees of freedom is then due to the restriction on the gauge transformations and thus to the boundary conditions. For completeness and to ease the comparison with other works in the literature, we note that the approach followed in~\cite{Benini:2020skc,Lacroix:2020flf} is slightly different although equivalent. Indeed, in these works, all transformations $A \mapsto A^u$ are considered as true gauge transformations, without any compatibility requirement with boundary conditions. The consequence is that the 4-dimensional action \eqref{Eq:Action4d} in itself is not invariant under all gauge transformations (in particular under certain gauge transformations which are not trivial on the poles defect $\Sigma\times\Zc$): to obtain a fully gauge invariant theory, one has to introduce additional 2d fields $\gt$ living on the poles defect $\Sigma \times \Zc$ and add to the 4-dimensional action a 2-dimensional term depending on this additional field $\gt$, coupled to the gauge field $A$, in a way which ensures that this extended action is gauge-invariant with respect to all transformations $A\mapsto A^u$. These additional degrees of freedom $\gt$ are then called \textit{edge modes}. In this approach, since all transformations $A \mapsto A^u$ are true gauge transformations, one can gauge away all degrees of freedom in the gauge field, leaving in the end an effective theory depending only on the edge modes $\gt$. Although the origin of the 2-dimensional field $\gt$ is different in this formulation with edge modes compared to the approach followed in these notes, these two formulations are in the end equivalent (we will not detail further the edge modes approach here and refer to~\cite{Benini:2020skc} for more details).
 
\paragraph{Towards integrable $\mathcal{E}$-models.} The work~\cite{Benini:2020skc}, summarised in the previous paragraphs, treats a very general class of 4d-CS setups corresponding to arbitrary 1-forms $\omega$. The last step needed to explicitly describe the resulting 2-dimensional integrable models is to solve the boundary condition \eqref{Eq:BC-Jet} and express the jet $\Lt_\pm$ in terms of the fundamental 2d field $\gt$: indeed, in the above discussion, we have supposed that such an expression $\Lt_\pm(\gt)$ was given but did not describe how to find it explicitly. This last step was performed in~\cite{Lacroix:2020flf}, at least under a certain technical assumption on the 1-form $\omega$, namely that it possesses a double pole at $z=\infty$. Let us briefly summarise the results of this work. For this class of $\omega$, there is a defect term corresponding to the pole at infinity, which can be treated by simply imposing the boundary condition $A_\mu|_{z=\infty}=0$, similarly to the cases of the PCM and the Yang-Baxter model. Once this pole at infinity is treated, one can focus on the finite poles $\Zc' = \Zc \cap \C$. In particular, we can drop the components of the jet $\At_\mu$ corresponding to the infinite pole: we then obtain a quantity that we shall denote as $\Ab_\mu$ and which encodes the jet at the finite poles defect $\Sigma \times \Zc'$. This quantity is valued in the defect algebra $\df=\g^{[\Zc']}$ associated to these finite poles, whereas the full jet $\At_\mu$ belongs to $\g^{[\Zc]}$. Having already treated the boundary condition at infinity, the choice of an appropriate boundary condition at the finite poles amounts to the choice of a maximal isotropic subalgebra $\kf$ of $\df$ (in analogy with the choice of a maximal isotropic subalgebra $\ff$ of $\g^{[\Zc]}$ when including the pole at infinity).

The 2-dimensional field $\gt$ contains degrees of freedom extracted from $\gh$ at the defect $\Sigma \times \lbrace \infty \rbrace$. The gauge transformations compatible with the boundary conditions $A_\mu|_{z=\infty}=0$ allow one to gauge away all these degrees of freedom except for $\gh|_{z=\infty}$. However, we can also eliminate the latter using the other freedom in $\gh$, namely the right multiplication $\gh\mapsto \gh v$ by $v:\Sigma \rightarrow G$ independent of $z$. Finally, we are left with degrees of freedom $\gb$ associated purely with the finite poles $\Zc'$, and which are naturally valued in the corrresponding defect group $D=G^{[\Zc']}$. The residual gauge transformations on these fields are then
\begin{equation}\label{Eq:GaugeE}
\gb \mapsto \ub \gb, \qquad \text{ with } \quad \ub \in K,
\end{equation}
where $K$ is the subgroup of $D$ corresponding to the isotropic subalgebra $\kf$ (to be compared with Equation \eqref{Eq:GaugeEdge} for the case before treating the pole at infinity). Moreover, the boundary condition \eqref{Eq:BC-Gen} now becomes
\begin{equation}
\gb\,\Lb_\pm\,\gb^{-1}  - (\p_\pm \gb)\gb^{-1} \; \in \kf.
\end{equation}
This is the constraint that is solved explicitly in~\cite{Lacroix:2020flf}. Reinserting the corresponding solution $\Lb_\pm(\gb)$ in the action of the model, one gets an explicit expression of the 2-dimensional action $S[\gb]$, which is invariant under the gauge transformations \eqref{Eq:GaugeE} by $K$. This then defines an integrable $\sigma$-model on the coset $K \!\setminus\! D$. As shown in~\cite{Lacroix:2020flf}, this model belongs to the class of so-called $\mathcal{E}$-models, which was introduced in~\cite{Klimcik:1995ux, Klimcik:1995dy, Klimcik:1996nq} to make manifest properties of (Poisson-Lie) T-dualities of certain $\sigma$-models. It was observed that many integrable $\sigma$-models, including for instance Yang-Baxter and $\lambda$-deformations, belong to this class~\cite{Klimcik:2015gba,Klimcik:2017ken,Severa:2017kcs,Klimcik:2019kkf,Hoare:2020mpv} (see also Ben Hoare's lectures \textit{``Integrable deformations of $\sigma$-models''}~\cite{LectureHoare}). In this context, the results of~\cite{Lacroix:2020flf} then construct a very general class of integrable $\mathcal{E}$-models, which contains these previously known examples but also new ones, by deriving them from 4d-CS.

\section{Conclusion and perspectives}
\label{Sec:Conclusion}

In these lectures, we have reviewed how one can systematically construct integrable 2-dimensional field theories (and in particular integrable $\sigma$-models) from the 4-dimensional semi-holomorphic Chern-Simons theory with disorder defects. The main concepts and results discussed in these lectures are summarised in the diagrams \ref{Fig:Space} and \ref{Fig:Summary}: we will not repeat this summary here and invite the reader to revisit these diagrams for a schematic overview. We have illustrated this approach on two main examples: the Principal Chiral Model and its Yang-Baxter deformation. Moreover, we have discussed some other examples in exercises and have sketched a more general and systematic treatment of the integrable models obtained by this approach at the end of the lectures. Let us conclude with a few perspectives on further developments.

\subsection{Exploring the panorama of integrable field theories}

Armed with the systematic approach provided by the 4-dimensional Chern-Simons theory, a natural perspective is to explore the panorama of integrable 2-dimensional field theories that one can obtain through this approach. 

\paragraph{Integrable coset $\bm{\sigma}$-models and superstrings.} The main two examples discussed in these lectures are integrable $\sigma$-models on a Lie group $G$ or its deformation. Other important examples of integrable $\sigma$-models include models on coset spaces $G/H$ such as symmetric spaces (which contain for instance spheres and Anti-de Sitter spaces in any dimension). These integrable coset models~\cite{Eichenherr:1979ci,Metsaev:1998it,Bena:2003wd,Young:2005jv,Arutyunov:2020sdo} and their integrable deformations~\cite{Delduc:2013fga,Delduc:2013qra,Hollowood:2014rla,Hollowood:2014qma} can also be obtained from the 4-dimensional Chern-Simons theory, as shown in the works~\cite{Costello:2019tri,Fukushima:2020dcp,Costello:2020lpi,Tian:2020ryu,Tian:2020meg,Fukushima:2021eni}, following a strategy similar to the one reviewed in these lectures\footnote{More precisely, the works mentioned here study integrable coset $\sigma$-models using the introduction of disorder defects in the 4-dimensional Chern-Simons theory. Let us also mention for completeness that in the case of integrable $\sigma$-models on Kähler manifolds, there exists an alternative construction of these models from the 4-dimensional Chern-Simons theory based on order defects built from $\beta\gamma$-systems: we refer to~\cite{Costello:2019tri,Bykov:2019vkf,Bykov:2020llx,Bykov:2020nal,Bykov:2020tao,Affleck:2021ypq} for more details.} for non-coset models but supplemented with an averaging process over the action of a finite order automorphism of the underlying Lie group (we will not enter into more details about this approach here). We additionally note that some of these references start with an extension of the theory considered in these lectures where the Lie algebra $\g$ underlying the construction is replaced by a Lie superalgebra, resulting in $\sigma$-models which also possess fermionic degrees of freedom. In particular, this allows to obtain integrable $\sigma$-models on semi-symmetric supercosets.

An important example of an integrable model based on such a semi-symmetric supercoset is the Green-Schwarz superstring on AdS$_5 \times$S$^5$, built from the Lie superalgebra $\mathfrak{psu}(2,2|4)$. In contrast with a standard $\sigma$-model, this theory possesses a dynamical worldsheet metric and exhibits diffeomorphism invariance as well as a local fermionic $\kappa$-symmetry (see~\cite{Arutyunov:2009ga,Beisert:2010jr} for a review of various aspects of this model and its integrability properties). The 4-dimensional Chern-Simons origin of this integrable superstring was discussed in the recent work~\cite{Costello:2020lpi}. In particular, it was shown in this reference that the diffeomorphism invariance and $\kappa$-symmetry arise from the introduction of an additional ingredient in the 4-dimensional setup, called the Beltrami differential. We refer to the lectures \textit{``Integrability, holography and Chern-Simons theory''} by B. Stefa\'nski~\cite{LecturesStefanski} for more details on this subject.

\paragraph{Towards a classification.} Many known integrable field theories have been shown to fit in the framework of 4-dimensional Chern-Simons theory and this approach has also allowed the construction of many new ones. This raises two important open questions: does this approach describe all possible 2-dimensional integrable field theories and is it possible to classify all the models obtained from it? Concerning the first question, although the 4-dimensional Chern-Simons theory has been shown so far to recover many examples, let us mention for completeness that the Sine-Gordon model and more generally affine Toda field theories have for the moment not been included in this formalism.

A very broad class of integrable $\sigma$-models arising from 4-dimensional Chern-Simons theory was discussed in~\cite{Benini:2020skc,Lacroix:2020flf}, as briefly sketched in the last subsection \ref{Sec:GenBC} of these lectures. In this approach, one of the main ingredient entering the defining data of these integrable models is the choice of a maximally isotropic subalgebra of the so-called defect Lie algebra (whose definition depends on the nature of the poles of the 1-form $\omega$). In view of classifying the integrable $\sigma$-models arising from 4-dimensional Chern-Simons theory, a natural derived mathematical question is thus to classify all these maximally isotropic subalgebras, which is at the moment an open problem.

\paragraph{Higher genus Riemann surfaces.} The integrable $\sigma$-models discussed above possess a spectral parameter valued in the Riemann sphere $\CP$. The 4-dimensional Chern-Simons theory also allows to construct integrable $\sigma$-models with spectral parameter valued in other Riemann surfaces and in particular higher genus ones. These models, which appear to be new, are yet to be fully understood. Some of them have been studied in~\cite{Costello:2019tri} and the recent work~\cite{Derryberry:2021rne}, where their target space has been identified with a moduli space of bundles over the underlying Riemann surface.

\subsection{Quantisation}

In these lectures, we have focused on the classical aspects of the 4-dimensional Chern-Simons approach to integrable field theories. A natural and important perspective is the question of the quantisation of these theories.

\paragraph{Quantum integrability.} Given a classical integrable system, characterised by conserved charges in involution, a natural question is whether there exists a quantisation of this model which is integrable at the quantum level, with the conserved charges in involution being promoted to quantum commuting conserved operators. In particular, this question naturally arises for the integrable systems obtained from the 4-dimensional Chern-Simons approach.

For the case of integrable lattice systems, the quantum integrable structure is controlled by the so-called $R$-matrix, solution of the Yang-Baxter equation, which allows the resolution of the theory by the Quantum Inverse Scattering Method / Bethe ansatz (see Ana L. Retore's lectures \textit{``Introduction to classical and quantum integrability''}~\cite{LectureRetore}). As mentioned in the introduction, one of the main motivation behind the introduction of the 4-dimensional Chern-Simons theory is the fact that it naturally reproduces the Yang-Baxter equation (as a generalisation of the Reidemeister move of 3-dimensional Chern-Simons theory which now includes the spectral parameter). This is at the basis of the works~\cite{Costello:2013zra,Costello:2013sla,Witten:2016spx,Costello:2017dso,Costello:2018gyb} in which integrable lattice systems and their $R$-matrices are studied from the point of view of 4-dimensional Chern-Simons theory, both at the classical and the quantum level, and at all order in the quantum parameter $\hbar$. The quantum integrable structure of this class of systems is thus well understood.

In these lectures, we have focused our attention on 2-dimensional integrable field theories rather than integrable lattice models. The question of the quantisation and the quantum integrability of classically integrable field theories has also been studied extensively in the literature. In particular, for certain classes of these theories, one can consider a quantisation procedure that ensures the existence of quantum commuting conserved charges, based on a $R$-matrix solution of the Yang-Baxter equation and which then allows the application of results stemming from the Quantum Inverse Scattering Method / Bethe ansatz framework, similar to the case of integrable lattice systems.

The question of the quantum integrability of the classically integrable field theories constructed from 4-dimensional Chern-Simons theory is an important problem which is in general still open at the moment. First results in this direction were announced in~\cite{YamazakiTalk} for the integrable field theories obtained from the 4-dimensional Chern-Simons theory on $\Sigma\times C$ where $C$ is a Riemann surface endowed with a 1-form $\omega$ without zeroes. As mentioned in the introduction of these lectures, this is the precise setup which is used to describe integrable lattice models in~\cite{Costello:2013zra,Costello:2013sla,Witten:2016spx,Costello:2017dso,Costello:2018gyb} as well as integrable field theories obtained from order defects in~\cite{Costello:2019tri}. Schematically, the quantum integrable structure of these models with order defects can be seen as a continuum limit of the lattice case mentioned above and should then be accessible from the 4-dimensional Chern-Simons approach. In particular, the fact that the 1-form $\omega$ has no zeroes plays an important role in this quantisation: indeed, as argued in~\cite{Costello:2013zra,Costello:2013sla,Witten:2016spx,Costello:2017dso,Costello:2018gyb}, this condition ensures that the 4-dimensional Chern-Simons theory can be quantised perturbatively in the quantum parameter $\hbar$.

By contrast, we have focused in these lectures on the integrable field theories arising from the 4-dimensional Chern-Simons theory with disorder defects, which correspond to zeroes of $\omega$. According to the above discussion, the quantisation of these models is more subtle. This in fact echoes some of the known results about integrable $\sigma$-models and their quantisation. Indeed, these models, such as for instance the Principal Chiral Model considered in these lectures, generally suffer from the so-called problem of non-ultralocality~\cite{Maillet:1985fn,Maillet:1985ek}, which prevents a straightforward use of techniques coming from the Quantum Inverse Scattering Method to these models (their quantum properties are then studied through other methods, such as the Factorised Scattering Theory, which has led to many important results in particular in the context of the AdS/CFT correspondence). More generally, the analysis of~\cite{Vicedo:2019dej} shows that all integrable models obtained from 4-dimensional Chern-Simons theory with disorder defects are faced with this problem of non-ultralocality (see the Discussion section of~\cite{Vicedo:2019dej} for further comments on this aspect). The study of the quantum integrability of these models form first principle is thus at the moment an interesting open question.

Let us finally mention that in the work~\cite{Gaiotto:2020fdr}, it was shown that certain integrable systems called Kondo problems can also be related to the 4-dimensional Chern-Simons theory. Quantum aspects of these systems are discussed in~\cite{Gaiotto:2020fdr}, in particular in the framework of the so-called ODE/IQFT correspondence~\cite{Voros:1994,Dorey:1998pt,Bazhanov:1998wj,Bazhanov:2003ni} (see also~\cite{Lukyanov:2006gv} for previous results on the ODE/IQFT correspondence for Kondo problems and~\cite{Gaiotto:2020dhf,Kotousov:2021vih,Wu:2021jir} for further recent developments). This correspondence relates certain observables of an Integrable Quantum Field Theory (IQFT), in particular the spectrum of its commuting operators, to the properties of some well-chosen Ordinary Differential Equations (ODE). In the context of the Kondo problems and their relation to 4-dimensional Chern-Simons theory, these ODE take particular form which depends on the 1-form $\omega$ underlying the 4-dimensional setup. This is in agreement with the fact, mentioned in Subsection \ref{Sec:CommentsLax}, that the 4-dimensional Chern-Simons theory is related to the so-called affine Gaudin models: indeed, it was proposed in~\cite{Feigin:2007mr} that these models can serve as a natural framework for studying the ODE/IQFT correspondence (see also~\cite{Lacroix:2018fhf,Lacroix:2018itd,Gaiotto:2020dhf,Kotousov:2021vih} for further developments on the quantisation of affine Gaudin models and their relation to the ODE/IQFT correspondence).

\paragraph{Renormalisation.} Another important aspect for the quantisation of the integrable field theories obtained from the 4-dimensional Chern-Simons theory is the question of their renormalisation. For certain models obtained from order defects, this question was discussed in~\cite{Costello:2019tri}, where it is mentioned that the renormalisation group flow of the theory amounts to a flow of the positions of the order defects on the Riemann surface $C$.

The models considered in these lectures, and which are obtained from disorder defects, are $\sigma$-models, for which renormalisation is controlled by the geometrical properties of the target space (and in particular its curvature). It is believed that classically integrable $\sigma$-models, such as the ones considered in these lectures, are renormalisable: this question has been the subject of many works, including recently, and this belief has been checked on various examples (the subject being quite vast, it would be difficult to do justice to these references in this single opening paragraph). For the models considered in these lectures, the continuous parameters of the theory are the positions of the poles and zeroes of the meromorphic 1-form $\omega=\vp(z)\dd z$. If these models are renormalisable, we then expect these positions to run with the renormalisation group flow. Based on various examples, it was conjectured in~\cite{Delduc:2020vxy} that this flow at one-loop can be rewritten compactly in terms of the function $\vp(z)$ defining $\omega$ as
\begin{equation}
\frac{\dd\;}{\dd \tau} \vp(z) = -c_{\g} \frac{\dd\;}{\dd z} \Bigl( \vp(z)f(z) \Bigr),
\end{equation}
where $\tau=\log(\mu)/4\pi$ is the ``renormalisation group time'', $c_\g$ is the so-called dual Coxeter number of the Lie algebra $\g$ and $f(z)$ is a function of the spectral parameter which is also defined in terms of the zeroes and poles of $\vp(z)$ (see~\cite{Delduc:2020vxy} for details). In the case of models with $\omega=\vp(z)\dd z$ possessing a double pole at infinity and an arbitrary number of simple poles, this conjecture was proven recently in~\cite{Hassler:2020xyj}, using the interpretation~\cite{Lacroix:2020flf} of these models as $\mathcal{E}$-models. In view of the importance of the 1-form $\omega=\vp(z)\dd z$ in the 4-dimensional Chern-Simons theory and the geometrical interpretation of the spectral parameter $z\in\CP$ in this setup, it would be natural to study the origin and/or interpretation of the above RG-flow in this context. Some first results in this direction, in the form of a conjecture closely related to the one above for the flow of $\omega$, have been discussed recently in the work~\cite{Derryberry:2021rne}.

\subsection{Relation to the 6-dimensional holomorphic twistor Chern-Simons theory}

As a final comment, let us mention that, in the recent works~\cite{Bittleston:2020hfv,Penna:2020uky}, the 4-dimensional Chern-Simons theory has been related to a 6-dimensional holomorphic Chern-Simons theory on twistor space first proposed in~\cite{CostelloTwistor}. The relevance of twistor space for the description of integrable field theories has been already put forward through another unifying principle proposed to describe integrability, which is based on reductions of anti-self-dual Yang-Mills equations (see for instance the textbook~\cite{BookTwistor} for a review and references) and which also provides a geometrical interpretation of the spectral parameter, similarly to the 4-dimensional Chern-Simons approach. In this context, the works~\cite{Bittleston:2020hfv,Penna:2020uky} then show that these two unifying approaches are in fact related and thus offer some new perspectives in the understanding of integrable systems.

\section*{Acknowledgements}

These notes were written for the lectures delivered at the school ``Integrability, Dualities and Deformations'', that ran from 23 to 27 August 2021 in Santiago de Compostela and virtually (for more informations, see \href{https://indico.cern.ch/e/IDD2021}{https://indico.cern.ch/e/IDD2021}). I am grateful to Riccardo Borsato and Fedor Levkovich-Maslyuk for valuable discussions and comments on the draft of these lectures, as well as to Ben Hoare, Nikita Nekrasov, Ana L. Retore and Beno\^it Vicedo for useful discussions. I would also like to thank the organisers of the school for the opportunity to present these lectures, as well as the participants, for making the school a quite enjoyable experience. This work was funded by the Deutsche Forschungsgemeinschaft (DFG, German Research Foundation) under Germany's Excellence Strategy - EXC 2121 ``Quantum Universe'' - 390833306.

\appendix

\section{Meromorphic functions and \texorpdfstring{$\bm{\delta}$}{delta}-distributions}
\label{App:Delta}

In this appendix, we prove the identity \eqref{Eq:DerPoleDirac}. By a translation, it is enough to prove it for $y=0$: we thus want to show
\begin{equation}\label{Eq:DerPoleApp}
\p_{\zb} \left( \frac{1}{z} \right) = -2i\pi \, \delta^{(2)}(z),
\end{equation}
where $\delta^{(2)}(z)$ is the Dirac-distribution defined with respect to the volume form $\dd z \wedge \dd \zb$. It is clear that the derivative $\p_{\zb}(1/z)$ is zero outside of $z=0$, since the function $z \mapsto 1/z$ is holomorphic on $\C\setminus\lbrace 0 \rbrace$. However, due to the pole at $z=0$, this derivative should be treated as a distribution on the entire complex plane $\C$. There are various ways of computing this distribution: here we will follow a rather direct computational way. We introduce the real and imaginary parts $a=(z+\zb)/2$ and $b=(z-\zb)/2i$ of $z$, such that $z=a+ib$ and $\zb=a-ib$. We will now identify the point $z\in \C$ with the vector $\vec r=(a,b) \in \R^2$. In particular, the derivative with respect to $\zb$ reads
\begin{equation}
\p_{\zb} = \frac{1}{2}\left( \frac{\p\;}{\p a} + i \frac{\p\;}{\p b} \right).
\end{equation}
A direct computation shows that
\begin{equation}
\p_{\zb} \left( \frac{1}{z} \right) = \frac{1}{2}\left( \frac{\p\;}{\p a} + i \frac{\p\;}{\p b} \right) \left( \frac{a-ib}{a^2+b^2} \right) = \frac{1}{2} \left( \frac{\p\;}{\p a} \frac{a}{a^2+b^2} + \frac{\p\;}{\p b} \frac{b}{a^2+b^2} \right) + \frac{i}{2} \left( -\frac{\p\;}{\p a} \frac{b}{a^2+b^2} + \frac{\p\;}{\p b} \frac{a}{a^2+b^2} \right).
\end{equation}
We introduce the vector fields
\begin{equation}
\vec v = \left( \frac{a}{a^2+b^2}, \frac{b}{a^2+b^2} \right) \qquad \text{ and } \qquad \vec w = \left( -\frac{b}{a^2+b^2}, \frac{a}{a^2+b^2} \right)
\end{equation}
on $\R^2$, such that
\begin{equation}\label{Eq:Div}
\p_{\zb} \left( \frac{1}{z} \right) = \frac{1}{2} \,\text{div}\,\vec v + \frac{i}{2} \,\text{div}\,\vec w.
\end{equation}
One easily checks by a direct computation that
\begin{equation}
\text{div}\,\vec v = \text{div}\,\vec w = 0, \qquad \forall\,\vec r \neq \vec 0,
\end{equation}
in agreement with our observation that the derivative $\p_{\zb}(1/z)$ vanishes for $z\neq 0$. To take into account the potential distribution terms in these divergences, we will use the (2-dimensional) Gauss theorem. Indeed, denoting by $D_R \subset \R^2$ the disk of radius $R$ centred at $(0,0)$ and $C_R$ its boundary, which is then a circle, the Gauss theorem tells us that
\begin{equation}
\iint_{D_R} \text{div}\,\vec v\;\dd a \wedge \dd b = \int_{C_R} \vec v \cdot \dd \ell\,\vec e_r.
\end{equation}
The right-hand side of this equation is the flux of $\vec v$ through the circle, with $\dd \ell$ the line element and $\vec e_r$ the radial unit vector, normal to the circle. A simple computation shows that $\vec v \cdot \vec e_r = 1/\sqrt{a^2+b^2}$. Since $\dd \ell = \sqrt{a^2+b^2}\, \dd\theta$, where $\theta\in [0,2\pi]$ is the polar angle, we find
\begin{equation}
\iint_{D_R} \text{div}\,\vec v\;\dd a \wedge \dd b = 2\pi.
\end{equation}
Thus the integral of $\text{div}\,\vec v$ over the disk $D_R$ is independent of $R$ and in particular tends to the non-zero quantity $2\pi$ when $R \to 0$. This translates to the fact that $\text{div}\,\vec v$ is proportional (with factor $2\pi$) to a Dirac-distribution centred at $\vec r=\vec 0$. There is a simple physical analogy behind this result: the vector field $\vec v / 2\pi$ is the electric field created by a punctual charge at the origin in 2 dimensions (for $\epsilon_0=1$): by Gauss law, the divergence of this electric field is the corresponding charge density, which is nothing but the Dirac-distribution as we are considering a point source (the above computation is then a check of the integral Gauss law, stating that the flux over the circle is equal to the charge inside the disk, which is always 1 independently of the radius). In conclusion, we then have
\begin{equation}
\text{div}\,\vec v = 2\pi\,\tilde{\delta}^{(2)}(\vec r).
\end{equation}
There is a slight subtlety to take into account here: indeed, the Dirac-distribution appearing in the above equation is by construction defined with respect to the volume form $\dd a \wedge \dd b$ and not $\dd z \wedge \dd \zb$, which is why we denoted it as $\tilde{\delta}^{(2)}$ and not $\delta^{(2)}$. A direct computation shows that $\dd z \wedge \dd \zb = -2i\, \dd a \wedge \dd b$. We thus have $\tilde{\delta}^{(2)}=-2i\,\delta^{(2)}$, hence
\begin{equation}
\text{div}\,\vec v = -4i\pi\,\delta^{(2)}(z).
\end{equation}
One can apply the same argument to the vector field $\vec w$. In particular we find
\begin{equation}
\iint_{D_R} \text{div}\,\vec w\;\dd a \wedge \dd b = \int_{C_R} \vec w \cdot \dd \ell\,\vec e_r = 0,
\end{equation}
since $\vec w$ is orthogonal to the radial vector $\vec e_r$. Thus $\text{div}\,\vec w$ does not contain a Dirac-distribution and is genuinely vanishing on all $\R^2$. Combining these results with Equation \eqref{Eq:Div}, we finally get the desired identity \eqref{Eq:DerPoleApp}.

\newpage

\bibliographystyle{JHEP}

\providecommand{\href}[2]{#2}\begingroup\raggedright

\end{document}